\documentclass[12pt]{article}
\usepackage{fullpage}
\usepackage[english]{babel}
\usepackage[latin1]{inputenc}
\usepackage{epsfig}
\usepackage{color,graphicx,graphics,psfrag}
\usepackage{amsmath,amstext,amssymb,amsfonts, amscd}

\def\Box{\leavevmode\vbox{\hrule
     \hbox{\vrule\kern4pt\vbox{\kern4pt}%
           \vrule}\hrule}}
\def\blackbox{\leavevmode\vrule height 5pt width 4pt depth 0pt\relax}
\def\endproof{\null\hfill {$\blackbox$}\bigskip}

\newcounter{appendix}
\def\appendix{\advance\c@appendix by 1
   \def\thesection{\Alph{section}}
   \ifnum\c@appendix=1 \setcounter{section}{-1} \fi
   \@startsection {section}{1}{\z@}{-3.5ex plus -1ex minus 
   -.2ex}{2.3ex plus .2ex}{\Large\bf}}

\def\paragraph#1{{\bf #1\ }}

\newtheorem{lemma}{Lemma}[section]

\newtheorem{definition}[lemma]{Definition}

\newtheorem{proposition}[lemma]{Proposition}

\newtheorem{remark}{Remark}[section]

\title{Numerical approximation of the Euler-Maxwell model in the quasineutral limit} 
\author{P. Degond, F. Deluzet, D. Savelief} 
\date{} 
\begin{document}

\maketitle

\vspace{0. cm}

\begin{center}
1-Université de Toulouse; UPS, INSA, UT1, UTM ;\\ 
Institut de Mathématiques de Toulouse ; \\
F-31062 Toulouse, France. \\
2-CNRS; Institut de Mathématiques de Toulouse UMR 5219 ;\\ 
F-31062 Toulouse, France.\\
email: pierre.degond@math.univ-toulouse.fr; fabrice.deluzet@math.univ-toulouse.fr; dominique.savelief@math.univ-toulouse.fr \\
\end{center}

\vspace{0.5 cm}
\begin{abstract}
We derive and analyze an Asymptotic-Preserving scheme for the Euler-Maxwell system in the quasi-neutral limit. We prove that the linear stability condition on the time-step is independent of the scaled Debye length $\lambda$ when $\lambda \to 0$. Numerical validation performed on Riemann initial data and for a model Plasma Opening Switch device show that the AP-scheme is convergent to the Euler-Maxwell solution when $\Delta x/ \lambda \to 0$ where $\Delta x$ is the spatial discretization. But, when $\lambda /\Delta x \to 0$, the AP-scheme is consistent with the quasi-neutral Euler-Maxwell system. The scheme is also perfectly consistent with the Gauss equation. The possibility of using large time and space steps leads to several orders of magnitude reductions in computer time and storage.
\end{abstract}

\medskip
\noindent
{\bf Acknowledgements:} This work has been supported by the french magnetic fusion programme 
'fédération de recherche sur la fusion par confinement magnétique', 
in the frame of the contract 'APPLA' (Asymptotic-Preserving schemes for 
Plasma Transport) and by the 'Fondation Sciences et Technologies pour 
l'Aéronautique et l'Espace', in the frame of the project 'Plasmax'.


\vskip 0.4cm


\setcounter{equation}{0}
\section{Introduction}
\label{EM:sec_intro}

The goal of this paper is to derive, analyze and validate a new Asymptotic-Preserving (AP) scheme for the Euler-Maxwell (EM) system of plasma physics in the quasi-neutral limit. The Euler-Maxwell system provides a fluid description of a plasma interacting with an electromagnetic wave. In the one-fluid setting where the plasma ions are supposed immobile (sections \ref{sec_one_fluid} to \ref{sec_1F_space}), the electron fluid obeys a system of isentropic gas dynamics equations subjected to the Lorentz force. The electromagnetic field is a solution of the Maxwell equations coupled to the fluid equations through the electrical charge and current. In the two-fluid case (section \ref{sec_2F}), each electron or ion species obey its own system of isentropic gas dynamics equations. The restriction to the isentropic case is for simplicity only: all concepts extend straightforwardly to full Euler systems including energy equations. 

When scaled to dimensionless variables (see section \ref{sec_one_fluid}), the EM system depends on the scaled Debye length $\lambda$ which is the ratio of the physical Debye length $\lambda_D$ to a typical dimension of the system $x_0$. The Debye length $\lambda_D$ is the characteristic length scale associated to the coupling between the particles and the electromagnetic waves and is one of the most important parameters in plasma physics \cite{Chen, Krall_Trivelpiece}. It is usually small because the electrostatic interaction occurs at spatial scales which are much smaller than the usual scales of interest. However, there are situations, for instance in boundary layers, or at the plasma-vacuum interface, where the electrostatic interaction scale must be taken into account. This means that the choice of the relevant scale $x_0$ may depend on the location inside the system and that in general, the parameter $\lambda$ may vary by orders of magnitude from one part of the domain to another one. 

In the scaled EM system, $\lambda$ appears both in the Ampere and Gauss equations. Therefore, when $\lambda$ is very small, a quasi-neutral regime, where the local electric charge is everywhere close to zero, appears. Simultaneously, $\lambda \ll 1$ implies that the speed of light is very large compared to the hydrodynamic speeds. In the limit $\lambda \to 0$, the scaled EM system formally converges to a system consisting of the Faraday equation for the Magnetic field, of the magnetostatics Ampere equation (i.e. without the displacement current) and of a stationary elliptic equation for the electric field (but which is not the usual Poisson equation). This system, later on referred to as the Quasi-Neutral Euler-Maxwell (QN-EM) system, bears analogies with the so-called Electron-MagnetoHydrodynamics equations (EMH) \cite{Gor_Kin_Rud_94}.   

This paper proposes a suitable numerical scheme for both the  $\lambda = O(1)$ and $\lambda \ll 1$ regimes. Physically, $\lambda^{-1}$ measures the temporal and spatial frequencies of plasma oscillations and electromagnetic waves. When $\lambda \ll 1$, they are very large and impose strong constraints on numerical discretizations. For classical explicit schemes, the time and space steps $\Delta t, \, \Delta x$ must resolve these frequencies and be of order $O(\lambda)$ to prevent the onset of numerical instabilities. For this reason, most studies are based on quasi-neutral models \cite{Cri_Deg_Vig_05, DiPeso_JCP_111_237, Hewett_JCP_29_219, Joyce_JCP_138_540, Lyster_JCP_102_180, Mankofsky_JCP_70_89, Rambo_JCP_118_152, Sle_Ste}. However, when $\lambda$ varies from one region to the other, quasi-neutral models lead to the wrong solution where $\lambda = O(1)$. A possible way to handle such situations is to decompose the simulation domain and to use the full EM or the QN-EM models according to whether $\lambda \ll 1$ or $\lambda = O(1)$ \cite{Deg_Par_Vig_MCM, Deg_Par_Vig_SiamMMS, Fra_Ock, Ha_Sle, Slemrod_1, Slemrod_3, Sternberg_JCP_111_347}. However, this domain decomposition approach suffers from many drawbacks. The coupling between the EM and QN-EM at the interfaces is not well-defined, which questions the physical reliability of the any particular strategy. Additionally, the domain decomposition must often be updated with time, which introduces a costly mesh adaptation strategy. Therefore, methods which are able to handle both regimes and are free of time and space step constraints related to $\lambda$ are much more flexible, versatile  and robust. This is the route which is followed in the present work. 

More specifically, we look for Asymptotic-Preserving (AP) schemes for the EM model with respect to the limit $\lambda \to 0$.  The AP property can be defined as follows. Consider a singular perturbation problem $P^\lambda$ whose solutions converge to those of a limit problem $P^0$ when $\lambda \to 0$ (here $P^\lambda$ is the EM model and $P^0$ is the QN-EM model). A scheme $P^\lambda_{\delta,h}$ for problem $P^\lambda$ with time-step $\delta$ and space-step $h$ is called Asymptotic Preserving (or AP) if it is stable independently of the value of $\lambda$ when $\lambda \to 0$ and if the scheme $P^0_{\delta,h}$ obtained by letting $\lambda \to 0$ in $P^\lambda_{\delta,h}$ with fixed $(\delta,h)$ is consistent with problem $P^0$. This property is illustrated by the commutative diagram  below: 
$$ \begin{CD} 
P^\lambda_{\delta,h} @>{(\delta,h) \to 0}>>  P^\lambda \\
@VV{\lambda \to 0}V  @VV{\lambda \to 0}V\\
P^0_{\delta,h} @>{(\delta,h) \to 0}>> P^0 
\end{CD}
$$
The possibility of letting $\lambda \to 0$ in $P^\lambda_{\delta,h}$ with fixed $(\delta,h)$ implicitly assumes that the stability condition on $(\delta,h)$ is independent of $\lambda$ when $\lambda \to 0$. This property is referred to as 'Asymptotic Stability'. The concept of an AP scheme has been introduced by S. Jin \cite{Jin} for diffusive limits of kinetic models and has been widely expanded since then \cite{Ben_Lem_Mie_JCP08, Bue_Cor_NumerMath07, Bue_Cor_M2AN02, Buet_Despres_JCP06, Car_Gou_Laf_JCP08, Fil_Jin_JCP10, Gos_Tos_NumerMath04, Klar_SINUM99, Lem_Mie_SISC08, McC_Low_JCP08,  Sea_Kla_JCP06}. 

In order to achieve the AP property, a certain degree of time implicitness must be introduced. In section \ref{sec_1F_time}, we will review various implicit schemes in view of this AP property and show that only one of the proposed schemes does exhibit this property. Specifically, we need a fully implicit discretization of the Maxwell equations together with an implicit current in the Ampere equation as well as an implicit mass flux in the mass conservation equation. A linearized stability analysis in Fourier space shows that the resulting scheme is actually AP. The implicit mass-flux strategy has already been used for the Euler-Poisson problem \cite{Cri_Deg_Vig_07, Deg_Liu_Vig_08, Deg_Del_Liu_Sav_Vig_JSC, Vignal_SIAP10} and Vlasov-Poisson problem \cite{Bel_Cro_Deg_JSC09, Deg_Del_Nav_JCP10} and is also key in the large magnetic-field asymptotics \cite{Deg_Del_Loz_Nar_Neg_CMS, Deg_Del_Neg_MMS10, Deg_Del_San_JCP09} and in the low Mach-number asymptotics \cite{Deg_Tan}. It is a well established fact \cite{Leveque_2} that, in order to enforce stability of the hydrodynamics equations, some numerical viscosity must be added. In section \ref{sec_1F_space}, we show that consistency with the Gauss equation is obtained if corresponding numerical viscosity terms are added to the Ampere equation. The concepts are then extended to the two-fluid EM model in section \ref{sec_2F} and a numerical validation is given in section \ref{EM:sec_num}. The numerical results practically demonstrate the Asymptotic-Preserving character of the AP-scheme. By comparison, in highly under-resolved situations (i.e. when the time and space steps do not resolve the fastest scales) a classical (time-explicit) scheme exhibits a strong instability.
Implicit method have previously been proposed in the context of Particle-In-Cell methods for the Vlasov equation (see \cite{Cohen_JCP_46_15, Langdon_JCP_51_107, Mason_JCP_51_484} for the electrostatic case and \cite{Brackbill_JCP_46_271, Hewett_JCP_72_121, Mason_JCP_71_429, Wallace_JCP_63_434} in the electromagnetic case). For hydrodynamic models, we refer to \cite{Fabre_JCP_101_445, Collela_JCP_149_168, Schneider_IJNM_05_399, Shumlak_JCP_187_620}. However, few of these methods are implicit and none has been analyzed in view of the AP-property. Finally, we refer to \cite{Deg_Cemracs} for a recent review on AP-schemes applied to plasma models.

\setcounter{equation}{0}
\section{The one-fluid Euler-Maxwell model}
\label{sec_one_fluid}

\subsection{General framework}
\label{subsec_1F_EMS}

The one-fluid Euler-Maxwell (EM) system consists of the mass and momentum balance equations for the electron fluid coupled to the Maxwell equations. The mass and momentum balance equations are written:
\begin{eqnarray}
& & \hspace{-1cm} \partial_t n + \nabla \cdot (nu) = 0, \label{1F_n} \\
& & \hspace{-1cm} m ( \partial_t (nu) + \nabla \cdot (nu \otimes u)) + \nabla p = - e n (E+u \times B), \label{1F_u} 
\end{eqnarray}
where $n(x,t) \geq 0$, $u(x,t) \in {\mathbb R}^d $ stand for the electron density and electron velocity respectively. They depend on the space-variable $x \in {\mathbb R}^d$ and on the time $t\geq0$. We denote by $e$ the positive elementary charge and by $m$, the electron mass. The electron pressure $p=p(n)$ is supposed to be a given function of $n$ (isentropic assumption) for simplicity. However, the subsequent analysis would extend straightforwardly to the case where $p$ is determined by an energy balance equation. The operators $\nabla$ and $\nabla \cdot $ are respectively the gradient and divergence operators and $u \otimes u$ denotes the tensor product of the vector $u$ with itself. We assume that the dimension $d=3$ for this presentation. We have neglected electron-ion collisions which otherwise would introduce a friction term in (\ref{1F_u}). This term could be added with no change to the subsequent theory and is omitted for simplicity. 

The electric field $E(x,t) \in {\mathbb R}^d $ and the magnetic field $B(x,t) \in {\mathbb R}^d $ are solutions of the Maxwell equations:
\begin{eqnarray}
& & \hspace{-1cm} \partial_t B + \nabla \times E = 0, \label{1F_B} \\
& & \hspace{-1cm} c^{-2} \partial_t E - \nabla \times B = - \mu_0 j , \label{1F_E} \\
& & \hspace{-1cm} \nabla \cdot B = 0, \label{1F_divB} \\
& & \hspace{-1cm} \nabla \cdot E = \epsilon_0^{-1} \rho, \label{1F_divE} 
\end{eqnarray}
where $\epsilon_0$, $\mu_0$ and $c$ are the vacuum permittivity, permeability and light velocity respectively, which satisfy $\epsilon_0 \mu_0 c^2 = 1$. Eqs (\ref{1F_B}), (\ref{1F_E}) and (\ref{1F_divE}) are the Faraday, Ampere and Gauss equations respectively. The divergence constraints (\ref{1F_divB}), (\ref{1F_divE}) are consequences of (\ref{1F_B}), (\ref{1F_E}), as soon as they are satisfied initially, which we will assume from now on. 

Finally, the electrical charge $\rho(x,t) \in {\mathbb R} $ and the electrical current $j(x,t) \in {\mathbb R}^d $ are given by 
\begin{eqnarray}
& & \hspace{-1cm} \rho = e (n_i - n), \label{1F_rho} \\
& & \hspace{-1cm} j = - e n u , \label{1F_j} 
\end{eqnarray}
where $n_i$ is the background ion density, which is supposed uniform and constant in time. Similarly, the ions are supposed steady, so that their contribution to the electrical current is identically zero.

\subsection{Scaling of the one-fluid Euler-Maxwell system}
\label{subsec_1F_scaling}

To scale this system to dimensionless units, we introduce scaling units $x_0$, $t_0$, $u_0$, $n_0$, $p_0$, $E_0$, $B_0$, $\rho_0$, $j_0$ for space, time, velocity, density, pressure, electric field, magnetic field, charge density and current density respectively. To reduce the number of dimensionless parameters, we make the following hypotheses: 
\begin{enumerate}
\item The spatial and temporal scales are linked by $x_0 = u_0 t_0$.
\item The velocity scale is chosen in such a way that the drift energy and thermal energy scales are the same: $m u_0^2 = p_0 n_0^{-1}$. For convenience, we introduce a temperature scale $T_0$ by $k_B T_0 = p_0 n_0^{-1}$. 
\item The density scale is fixed by the uniform ion background: $n_0 = n_i$.
\item The charge density scale is fixed by the number density scale by $\rho_0 = e n_0$.
\item The current density scale is fixed by the density and velocity scales by $j_0 = e n_0 u_0$.
\item The electric field scale is such that the electrical and thermal (or drift) energy scales are the same: $e E_0 x_0 = p_0 n_0^{-1}$.
\end{enumerate}
Assumptions number 1, 3, 4 and 5 are natural. Assumptions 2 and 6 guarantee that the inertia force, the pressure force and the electric force have the same order of magnitude. With these six relations, there are only three dimensionless parameters, which are: 
\begin{eqnarray}
& & \hspace{-1cm} \alpha = \frac{u_0}{c}, \quad \beta = \left( \frac{u_0 B_0}{E_0} \right)^{1/2} , \quad \lambda = \left( \frac{\epsilon_0 k_B T}{e^2 n_0 x_0^2} \right)^{1/2} .  \label{alpha_beta_lambda} 
\end{eqnarray}
The first one is the ratio of the plasma velocity to the speed of light. The second one is the ratio of the induction electric field to the reference electric field. The third one is the Debye length scaled by the reference space scale. 

In this scaling, the EM system is written (by abuse of notation, we keep the same notations for the dimensionless variables as for the physical variables): 
\begin{eqnarray}
& & \hspace{-1cm} \partial_t n + \nabla \cdot (nu) = 0, \label{s1F_n} \\
& & \hspace{-1cm}  \partial_t (nu) + \nabla \cdot (n u \otimes u) + \nabla p(n) = - n (E+ \beta^2 u \times B), \label{s1F_u}\\
& & \hspace{-1cm} \beta^2 \partial_t B + \nabla \times E = 0, \label{s1F_B} \\
& & \hspace{-1cm} \lambda^2 (\alpha^2 \partial_t E - \beta^2 \nabla \times B) = \alpha^2 n u  , \label{s1F_E} \\
& & \hspace{-1cm} \nabla \cdot B = 0, \label{s1F_divB} \\
& & \hspace{-1cm} \lambda^2 \nabla \cdot E = 1 - n, \label{s1F_divE} 
\end{eqnarray}
We are interested in the limit $\lambda \to 0$ (quasineutral limit). To choose how the remaining parameters $\alpha$ and $\beta$ scale with $\lambda$, we adopt the principle of the least degeneracy, i.e. we choose the scaling which produces the limit system with the largest number of terms. If we examine (\ref{s1F_E}), we notice that whatever the choice of $\alpha$, we have $\lambda^2 \alpha^2 \partial_t E \ll \alpha^2 n u$. So, in the limit $\lambda \to 0$, of these two terms, only $\alpha^2 n u$ remains. The principle of least degeneracy thus imposes that the remaining term of (\ref{s1F_E}) i.e. $\lambda^2 \beta^2 \nabla \times B$ be of the same order of magnitude as $\alpha^2 n u$, which imposes $\lambda^2 \beta^2 = \alpha^2$. Now, the choice $\beta^2 = 1$ is the least degenerate one as regards eqs. (\ref{s1F_u}) and (\ref{s1F_B}) because, either $\beta \ll 1$ or $\beta \gg 1$ will then lead to reduced equations with a smaller number of terms. Based on these considerations, we choose
\begin{eqnarray}
& & \hspace{-1cm} \alpha = \lambda, \quad \beta = 1,  \label{alpha_beta} 
\end{eqnarray}
which leads to the final form of the scaled Euler-Maxwell system: 
\begin{eqnarray}
& & \hspace{-1cm} \partial_t n^\lambda + \nabla \cdot (n^\lambda u^\lambda) = 0, \label{S1F_n} \\
& & \hspace{-1cm}  \partial_t (n^\lambda u^\lambda) + \nabla \cdot (n^\lambda u^\lambda \otimes u^\lambda) + \nabla p(n^\lambda) = - n^\lambda (E^\lambda+ u^\lambda \times B^\lambda), \label{S1F_u}\\
& & \hspace{-1cm} \partial_t B^\lambda + \nabla \times E^\lambda = 0, \label{S1F_B} \\
& & \hspace{-1cm} \lambda^2 \partial_t E^\lambda - \nabla \times B^\lambda =  n^\lambda u^\lambda  , \label{S1F_E} \\
& & \hspace{-1cm} \nabla \cdot B^\lambda = 0, \label{S1F_divB} \\
& & \hspace{-1cm} \lambda^2 \nabla \cdot E^\lambda = 1 - n^\lambda, \label{S1F_divE} 
\end{eqnarray}
where we have highlighted the dependence of the solution upon the parameter $\lambda$.

\subsection{Quasi-neutral limit $\lambda \to 0$}
\label{subsec_1F_QNlim}

In the  limit $\lambda \to 0$, we suppose that $n^\lambda \to n^0$, $u^\lambda \to u^0$, \ldots. Then, formally, the scaled EM system leads to the Quasi-Neutral Euler-Maxwell (QN-EM) system  
\begin{eqnarray}
& & \hspace{-1cm} \nabla \cdot u^0 = 0, \label{QN1F_n} \\
& & \hspace{-1cm}  \partial_t u^0 + \nabla \cdot ( u^0 \otimes u^0)  = - (E^0+ u^0 \times B^0), \label{QN1F_u}\\
& & \hspace{-1cm} \partial_t B^0 + \nabla \times E^0 = 0, \label{QN1F_B} \\
& & \hspace{-1cm}  - \nabla \times B^0 =  u^0  , \label{QN1F_E} \\
& & \hspace{-1cm} \nabla \cdot B^0 = 0, \label{QN1F_divB} \\
& & \hspace{-1cm} n^0=1, \label{QN1F_divE} 
\end{eqnarray}
The divergence free constraint on $u^0$ is a consequence of (\ref{QN1F_E}), while the divergence free constraint on $B^0$ is a consequence of (\ref{QN1F_B}) (and of the divergence free initial data). Finally, $n^0=1$ is no more a dynamical variable of the problem. Therefore, the core three equations of the QN-EM model are (\ref{QN1F_u}), (\ref{QN1F_B}), (\ref{QN1F_E}).

In this model, the time evolutions of $u^0$ and $B^0$ are constrained by (\ref{QN1F_E}). $E^0$ is the Lagrange multiplier of this constraint. To resolve it and find an explicit equation for $E^0$, it suffices to take the curl of (\ref{QN1F_B}), add it to (\ref{QN1F_u}) and use (\ref{QN1F_E}) to cancel the time-derivatives. This leads to: 
\begin{eqnarray}
& & \hspace{-1cm}  \nabla \times ( \nabla \times E^0) + E^0 =  - \nabla \cdot ( u^0 \otimes u^0)   -  u^0 \times B^0, \label{QN1F_E_ref}
\end{eqnarray}
which is a well-posed elliptic equation for $E^0$ (provided suitable boundary conditions are given, such as perfectly conducting or absorbing boundary conditions ; we will treat the question of boundary conditions in relation to the numerical examples). In  the QN-EM model, the hyperbolic character of the Maxwell equations is lost: $E^0$ adjusts to the variations of $B^0$ instantaneously. 

More precisely, the QN-EM model (\ref{QN1F_n})-(\ref{QN1F_divE}) is equivalent to:
\begin{eqnarray}
& & \hspace{-1cm} \nabla \cdot u^0 = 0, \label{FQN1F_n} \\
& & \hspace{-1cm}  \partial_t u^0 + \nabla \cdot ( u^0 \otimes u^0)  = - (E^0+ u^0 \times B^0), \label{FQN1F_u}\\
& & \hspace{-1cm} \partial_t B^0 + \nabla \times E^0 = 0, \label{FQN1F_B} \\
& & \hspace{-1cm}  \nabla \times ( \nabla \times E^0) + E^0 =  - \nabla \cdot ( u^0 \otimes u^0)   -  u^0 \times B^0, \label{FQN1F_E_ref} \\
& & \hspace{-1cm} \nabla \cdot B^0 = 0, \label{FQN1F_divB} \\
& & \hspace{-1cm} n^0=1, \label{FQN1F_divE} 
\end{eqnarray}
if and only if $u^0|_{t=0}$ and $B^0|_{t=0}$ are related by 
\begin{eqnarray}
& & \hspace{-1cm}  - \nabla \times B^0|_{t=0} =  u^0|_{t=0}  . \label{FQN1F_E} 
\end{eqnarray}
Indeed, the 'only if' part of the statement has just been proved. To prove the 'if' part, we take the curl of (\ref{FQN1F_B}), add it to (\ref{FQN1F_u}) and use (\ref{FQN1F_E_ref}) to deduce that 
\begin{eqnarray}
& & \hspace{-1cm}  \partial_t (\nabla \times B^0 +  u^0) = 0  . \label{FQN1F_E2} 
\end{eqnarray}
Then, if (\ref{FQN1F_E}) is satisfied, (\ref{QN1F_E}) is satisfied for all times. We will look for AP schemes which are consistent with the form (\ref{FQN1F_n})-(\ref{FQN1F_divE}) of the QN-EM model. 

If the initial conditions of the EM model do not satisfy (\ref{FQN1F_E}), an initial layer occurs, during which high frequency oscillations are produced. The QN-EM model produces some kind of time averaging of these high frequency oscillations. The AP scheme introduces numerical dissipation which damps out these fast oscillations in order to approach the quasi-neutral dynamics.

\begin{remark}
If we neglect the inertia of the electrons, which amounts to removing the drift term in the momentum equation (\ref{QN1F_u}), the QN-EM model reduces to:  
\begin{eqnarray*}
& & \hspace{-1cm} \partial_t B^0 + \nabla \times E^0 = 0,  \\
& & \hspace{-1cm} u^0 = - \nabla \times B^0   , \\
& & \hspace{-1cm} E^0+ u \times B^0=0, \\
& & \hspace{-1cm} \nabla \cdot B^0 = 0, 
\end{eqnarray*}
which is the so-called Electron-MagnetoHydrodynamics (EMH) system \cite{Gor_Kin_Rud_94}. Here, we do not make any assumption about the electron time scales, which leads to a slightly more complex dynamics. 
\label{rem_emh}
\end{remark}

In the limit $\lambda \to 0$, the type of the equation for the electric field changes completely,  from a hyperbolic equation (the Ampere law (\ref{S1F_E})) to an elliptic one (\ref{QN1F_E_ref}). This is the signature that the EM model is a singularly perturbed problem in the limit $\lambda \to 0$. In the process of building an AP scheme, the first step is to reformulate the problem in such a way that this singular perturbation character appears more explicitly. This task is performed in the next section.

\subsection{Reformulation of the EM model for finite $\lambda$}
\label{subsec_1F_reform}

In this section, we plan to find an equivalent formulation of the scaled EM model in such a way that the electric field equation appears as a singular perturbation of the electric field equation (\ref{QN1F_E_ref}) of the QN-EM model. With this aim, we take the curl of (\ref{S1F_B}), add it to (\ref{S1F_u}), and use (\ref{S1F_E}) to eliminate the time derivatives of $nu$ and $B$. This leads to 
\begin{eqnarray}
& & \hspace{-1cm}  \lambda^2 \partial^2_t E^\lambda + \nabla \times ( \nabla \times E^\lambda) + n^\lambda E^\lambda = - \nabla \cdot (n^\lambda u^\lambda \otimes u^\lambda) - \nabla p(n^\lambda)  - n^\lambda  u^\lambda \times B^\lambda.  \label{S1F_E_ref0}
\end{eqnarray}
In this form, it is clear that, when $\lambda \to 0$ and $n \to 1$, (\ref{S1F_E_ref0}) formally tends to (\ref{QN1F_E_ref}). This equation is a wave equation for $E$ with wave-speed $\lambda^{-1}$. It replaces the Ampere equation (\ref{QN1F_E}) in the reformulated Euler-Maxwell (REM) model: 
\begin{eqnarray}
& & \hspace{-1cm} \partial_t n^\lambda + \nabla \cdot (n^\lambda u^\lambda) = 0, \label{RS1F_n} \\
& & \hspace{-1cm}  \partial_t (n^\lambda u^\lambda) + \nabla \cdot (n^\lambda u^\lambda \otimes u^\lambda) + \nabla p(n^\lambda) = - n^\lambda (E^\lambda + u^\lambda \times B^\lambda), \label{RS1F_u}\\
& & \hspace{-1cm} \partial_t B^\lambda + \nabla \times E^\lambda = 0, \label{RS1F_B_ref} \\
& & \hspace{-1cm}  \lambda^2 \partial^2_t E^\lambda + \nabla \times ( \nabla \times E^\lambda) + n^\lambda E^\lambda = - \nabla \cdot (n^\lambda u^\lambda \otimes u^\lambda) - \nabla p(n^\lambda)  - n^\lambda  u^\lambda \times B^\lambda,  \label{RS1F_E} \\
& & \hspace{-1cm} \nabla \cdot B^\lambda = 0, \label{RS1F_divB} \\
& & \hspace{-1cm} \lambda^2 \nabla \cdot E^\lambda = (1 - n^\lambda), \label{RS1F_divE} 
\end{eqnarray}
We stress the fact that this system is equivalent to the initial EM model, provided that $E$ satisfies (\ref{QN1F_E}) at the initial time. This condition provides the Cauchy datum on $\partial_t E$ requested by this second order problem. 

The use of the REM model preferably to the EM model, in conjunction with an implicit time discretization of (\ref{RS1F_E}), is the key for the build-up of an AP scheme for the EM model in the quasi-neutral limit $\lambda \to 0$.

\subsection{Linearization of the EM model}
\label{subsec_1F_linearization}

The numerical stability analysis will use the Fourier analysis of the linearized system. In this section, we investigate the linearization of the EM and QN-EM models about the uniform stationary state $n^\lambda = 1$, $u^\lambda = 0$, $E^\lambda = 0$, $B^\lambda = 0$. Expanding $n^\lambda = 1 + \varepsilon \tilde n^\lambda$, $u^\lambda = \varepsilon \tilde u^\lambda$, $E^\lambda = \varepsilon \tilde E^\lambda$, $B^\lambda = \varepsilon \tilde B^\lambda$, with $\varepsilon \ll 1$ being the intensity of the perturbation to the stationary state, and retaining only the linear terms in $\varepsilon$, we find the linearized EM model (in scaled units):  
\begin{eqnarray}
& & \hspace{-1cm} \partial_t \tilde n^\lambda + \nabla \cdot \tilde u^\lambda = 0, \label{LS1F_n} \\
& & \hspace{-1cm}  \partial_t \tilde u^\lambda + T \nabla \tilde n^\lambda = - \tilde E^\lambda, \label{LS1F_u}\\
& & \hspace{-1cm} \partial_t \tilde B^\lambda + \nabla \times \tilde E^\lambda = 0, \label{LS1F_B} \\
& & \hspace{-1cm} \lambda^2 \partial_t \tilde E^\lambda - \nabla \times \tilde B^\lambda =  \tilde u^\lambda  , \label{LS1F_E} \\
& & \hspace{-1cm} \nabla \cdot \tilde B^\lambda = 0, \label{LS1F_divB} \\
& & \hspace{-1cm} \lambda^2 \nabla \cdot \tilde E^\lambda =  - \tilde n^\lambda, \label{LS1F_divE} 
\end{eqnarray}
with $T = p'(1)$. 
Introducing $\hat n^\lambda$, $\hat u^\lambda$, $\hat E^\lambda$, $\hat B^\lambda$, the partial Fourier transforms of $\tilde n^\lambda$, $\tilde u^\lambda$, $\tilde E^\lambda$, $\tilde B^\lambda$ with respect to $x$, we are led to the following system of ODE's: 
\begin{eqnarray}
& & \hspace{-1cm} \partial_t \hat n^\lambda + i \xi \cdot \hat u^\lambda = 0, \label{FLS1F_n} \\
& & \hspace{-1cm}  \partial_t \hat u^\lambda + i T \xi \hat n^\lambda = - \hat E^\lambda, \label{FLS1F_u}\\
& & \hspace{-1cm} \partial_t \hat B^\lambda + i \xi \times \hat E^\lambda = 0, \label{FLS1F_B} \\
& & \hspace{-1cm} \lambda^2 \partial_t \hat E^\lambda - i \xi \times \hat B^\lambda =  \hat u^\lambda  , \label{FLS1F_E} \\
& & \hspace{-1cm} i \xi  \cdot \hat B^\lambda = 0, \label{FLS1F_divB} \\
& & \hspace{-1cm} i \lambda^2 \xi \cdot \hat E^\lambda =  - \hat n^\lambda, \label{FLS1F_divE} 
\end{eqnarray}
where $\xi$ is the Fourier dual variable to $x$. We denote the solution of this system by $U^\lambda(\xi,t)=(\hat n^\lambda, \hat u^\lambda, \hat E^\lambda, \hat B^\lambda)$. We look for solutions of the form of a Laplace transform $U^\lambda(\xi,t)= e^{-st} U_0^\lambda(\xi)$. A simple algebra leads to the solution $s=0$ as well as to two non-trivial solutions: 

\begin{enumerate}
\item The electromagnetic mode: 
\begin{eqnarray}
& & \hspace{-1cm} 
s^\lambda_{\pm,em} = \pm \frac{i}{\lambda} (1 + |\xi|^2)^{1/2}
, \label{s_electrommagnetic} 
\end{eqnarray}
associated with the polarization $\hat E^\lambda \, \bot \,  \xi$, 

\item The electrostatic mode:
\begin{eqnarray}
& & \hspace{-1cm} 
s^\lambda_{\pm,es} = \pm \frac{i}{\lambda} (1 + T \lambda^2 |\xi|^2)^{1/2}
, \label{s_electrostatic} 
\end{eqnarray}
associated with the polarization $\hat E^\lambda \, \parallel \, \xi$. 

\end{enumerate}

In the limit $\lambda \to 0$, both $s^\lambda_{\pm,em}$ and $s^\lambda_{\pm,es}$ tend to $\infty$, which corresponds to high frequency oscillations of the solution $U^\lambda(\xi,t)$. The only mode of the QN-EM corresponds to $s=0$. It is indeed easy to see that the linearized QN-EM model 
\begin{eqnarray}
& & \hspace{-1cm} \partial_t \tilde n^0 + \nabla \cdot \tilde u^0 = 0, \label{LQN1F_n} \\
& & \hspace{-1cm}  \partial_t \tilde u^0 + T \nabla \tilde n^0 = - \tilde E^0, \label{LQN1F_u}\\
& & \hspace{-1cm} \partial_t \tilde B^0 + \nabla \times \tilde E^0 = 0, \label{LQN1F_B} \\
& & \hspace{-1cm} - \nabla \times \tilde B^0 =  \tilde u^0  , \label{LQN1F_E} \\
& & \hspace{-1cm} \nabla \cdot \tilde B^0 = 0, \label{LQN1F_divB} \\
& & \hspace{-1cm} 0 =  - \tilde n^0, \label{LQN1F_divE} 
\end{eqnarray}
has only steady-state solutions $n^0 = 0$, $E^0 = 0$ (with adequate boundary conditions), while $\tilde B^0$ is any steady-state field satisfying (\ref{LQN1F_divB}) and $\tilde u^0 = - \nabla \times \tilde B^0$.

\setcounter{equation}{0}
\section{Time-semi-discretization, AP property and linearized stability}
\label{sec_1F_time}

We denote by $ \delta  $ the time step. For any function $g(x,t)$, we denote by
$ g^m (x) $ an approximation of $g(x,t^m)$  with $ t^m = m \delta $. We present different time-semi-discretizations of the problem which are classified according to their level of implicitness.

\subsection{Time-semi-discretizations of the EM system}
\label{subsec_1F_time}

As mentioned in section \ref{EM:sec_intro}, we will consider different levels of time-implicitness.  We recall that we need at least a semi-implicit discretization of the Maxwell equations otherwise the scheme is unconditionally unstable. As a consequence, the Lorentz force in the momentum equation must also be evaluated implicitly. This will be the first level of implicitness. The second level takes the current in the Ampere equation as well as the mass flux in the mass conservation equation implicitly. The third level considers a fully implicit discretization of the Maxwell equations, in addition to the previous levels of implicitness.

All these schemes can be put in a unified framework by considering the following discretization: 
\begin{eqnarray}
& & \hspace{-1cm} \delta^{-1} (n^{\lambda, m+1} - n^{\lambda, m}) + \nabla \cdot (n^{\lambda, m+a} u^{\lambda, m+a}) = 0, \label{DS1F_n} \\
& & \hspace{-1cm}  \delta^{-1} ( n^{\lambda, m+1} u^{\lambda, m+1} - n^{\lambda, m} u^{\lambda, m}) + \nabla \cdot (n^{\lambda, m} u^{\lambda, m} \otimes u^{\lambda, m}) + \nabla p(n^{\lambda, m}) = \nonumber \\
& & \hspace{5cm} - (n^{\lambda, m+1-a} E^{\lambda, m+1}+ n^{\lambda, m} u^{\lambda, m} \times B^{\lambda, m}), \label{DS1F_u}\\
& & \hspace{-1cm} \delta^{-1} (B^{\lambda, m+1} - B^{\lambda, m}) + \nabla \times E^{\lambda, m+b} = 0, \label{DS1F_B} \\
& & \hspace{-1cm} \lambda^2 \delta^{-1} (E^{\lambda, m+1} - E^{\lambda, m}) - \nabla \times B^{\lambda, m+c} =  n^{\lambda, m+a} u^{\lambda, m+a}  , \label{DS1F_E} \\
& & \hspace{-1cm} \nabla \cdot B^{\lambda, m+1} = 0, \label{DS1F_divB} \\
& & \hspace{-1cm} \lambda^2 \nabla \cdot E^{\lambda, m+1} = (1 - n^{\lambda, m+1}), \label{DS1F_divE} 
\end{eqnarray}
with $a$, $b$ and $c$ taking the values $0$ or $1$. The various cases are as follows: 

\begin{enumerate}
\item First level of implicitness: $(a,b,c) = (0,1,0)$ or $(0,0,1)$: the scheme is semi-implicit in the Maxwell equations. The Lorentz force is implicit. The rest is explicit. This is the classical strategy. 
\item Second level of implicitness: $(a,b,c) = (1,1,0)$ or $(1,0,1)$: additionally, the current in the Ampere equation and the mass flux in the mass conservation equations is implicit. 
\item Third level of implicitness: $(a,b,c) = (1,1,1)$: the Maxwell equations are fully implicit as well as the current in the Ampere equations and the mass flux in the mass conservation equation. 
\end{enumerate}
We note that the mass flux in the mass conservation equation and the current in the Ampere equation must have the same degree of implicitness in order to guarantee the consistency with the Gauss equation. The various schemes will be referred to by the value of the triple $(a,b,c)$. For instance the $(1,0,1)$-scheme will refer to the scheme with $(a,b,c) = (1,0,1)$. With this level of implicitness, it is convenient to use an explicit evaluation of the density in the Lorentz force (\ref{DS1F_u}), because this reduces the complexity of the inversion of the implicit scheme. This choice does not restrict the AP-character of the scheme (when applicable) nor does it change its linearized stability properties.

We note that the first level cannot be AP. Indeed, taking the limit $\lambda \to 0$ in the $(0,1,0)$ or $(0,0,1)$ schemes, we find that they do not lead to a valid recursion which allows the computation of the variables at time $m+1$ from the knowledge of those at time $m$. 

The second level could be AP. If we let $\lambda \to 0$ in the $(1,1,0)$ scheme, we find the following recursion: 
\begin{eqnarray}
& & \hspace{-1cm} \delta^{-1} (B^{0, m+1} - B^{0, m}) + \nabla \times E^{0, m+1} = 0, \label{QND110S1F_B} \\
& & \hspace{-1cm}  n^{0, m+1} u^{0, m+1}  = - \nabla \times B^{0, m}  , \label{QND110S1F_E} \\
& & \hspace{-1cm}  \delta^{-1} ( n^{0, m+1} u^{0, m+1} - n^{0, m} u^{0, m}) + \nabla \cdot (n^{0, m} u^{0, m} \otimes u^{0, m}) + \nabla p(n^{0, m}) = \nonumber \\
& & \hspace{5cm} - (n^{0, m} E^{0, m+1} + u^{0, m} u^{0, m} \times B^{0, m}), \label{QND110S1F_u}\\
& & \hspace{-1cm} \nabla \cdot B^{0, m+1} = 0, \label{QND110S1F_divB} 
\end{eqnarray}
Taking the curl of (\ref{QND110S1F_B}) and adding to (\ref{QND110S1F_E}), the third equation can be recast into the following equation for $E^{0, m+1}$: 
\begin{eqnarray}
& & \hspace{-1cm}  n^{0, m} E^{0, m+1} = - \nabla \times ( \nabla \times E^{0, m}) - \nabla \cdot (n^{0, m} u^{0, m} \otimes u^{0, m}) - \nabla p(n^{0, m}) \nonumber \\
& & \hspace{9cm} -  n^{0, m} u^{0, m} \times B^{0, m}, \label{QND110S1F_u_2}
\end{eqnarray}
and the scheme is consistent with the QN-EM model (\ref{FQN1F_B})-(\ref{FQN1F_divB}). It is also obviously a valid recursion. 

If we let $\lambda \to 0$ in the $(1,0,1)$ scheme and we use the same computation, we find the following recursion: 
\begin{eqnarray}
& & \hspace{-1cm} \delta^{-1} (B^{0, m+1} - B^{0, m}) + \nabla \times E^{0, m} = 0, \label{QND101S1F_B} \\
& & \hspace{-1cm}  n^{0, m+1} u^{0, m+1}  = - \nabla \times B^{0, m+1}  , \label{QND101S1F_E} \\
& & \hspace{-1cm}  n^{0, m} E^{0, m+1} = - \nabla \times ( \nabla \times E^{0, m}) - \nabla \cdot (n^{0, m} u^{0, m} \otimes u^{0, m}) - \nabla p(n^{0, m}) \nonumber \\
& & \hspace{9cm} -  n^{0, m} u^{0, m} \times B^{0, m}, \label{QND101S1F_u_2} \\
& & \hspace{-1cm} \nabla \cdot B^{0, m+1} = 0, \label{QND101S1F_divB} 
\end{eqnarray}
and again,the scheme is consistent with the QN-EM model (\ref{FQN1F_B})-(\ref{FQN1F_divB}) and provides a valid recursion formula.

It seems that both the $(1,1,0)$ and the $(1,0,1)$ schemes would be good candidates AP schemes. However, in a forthcoming section, we will see that they are not linearly stable. By contrast, the $(1,1,1)$ scheme will be found linearly stable. It is AP because, if we let $\lambda \to 0$ in the $(1,1,1)$ scheme, we find the following recursion: 
\begin{eqnarray}
& & \hspace{-1cm} \delta^{-1} (B^{0, m+1} - B^{0, m}) + \nabla \times E^{0, m+1} = 0, \label{QND111S1F_B} \\
& & \hspace{-1cm}  n^{0, m+1} u^{0, m+1}  = - \nabla \times B^{0, m+1}  , \label{QND111S1F_E} \\
& & \hspace{-1cm}  n^{0, m} E^{0, m+1} + \nabla \times ( \nabla \times E^{0, m+1}) =  - \nabla \cdot (n^{0, m} u^{0, m} \otimes u^{0, m}) - \nabla p(n^{0, m}) \nonumber \\
& & \hspace{9cm} -  n^{0, m} u^{0, m} \times B^{0, m}, \label{QND111S1F_u_2} \\
& & \hspace{-1cm} \nabla \cdot B^{0, m+1} = 0, \label{QND111S1F_divB} 
\end{eqnarray}
which is obviously consistent with the QN-EM model. It also provides a valid recursion for all the variables.

\subsection{Linearized stability analysis}
\label{subsec_1F_linear_stab}

The goal of this section is to analyze the linearized stability properties of the previous schemes. More precisely, we want to show that only the $(1,1,1)$ scheme has the Asymptotic Stability property when $\lambda \to 0$, under a suitably defined CFL condition {\bf independent of the value of $\lambda$ when $\lambda \to 0$}. We will prove $L^2$-stability uniformly with respect to $\lambda$ for the linearization of the EM model (\ref{LS1F_n})-(\ref{LS1F_divE}).  

In general, time-semi-discretizations of hyperbolic problems are unconditionally unstable.  This is because the skew adjoint operator $\partial_x$ has the same effect as a centered space-differencing. For fully discrete schemes, stability is obtained at the price of adding numerical viscosity. To mimic the effect of this viscosity, in the present section, we will consider the linearized Viscous Euler-Maxwell (VEM) model, which consists of the linearized EM model (\ref{LS1F_n})-(\ref{LS1F_divE}) with additional viscosity terms (in this section, we drop the tildes for notational convenience): 
\begin{eqnarray}
& & \hspace{-1cm} \partial_t  n^\lambda + \nabla \cdot  u^\lambda - \beta \Delta n^\lambda = 0, \label{VLS1F_n} \\
& & \hspace{-1cm}  \partial_t  u^\lambda + T \nabla  n^\lambda - \beta \Delta u^\lambda = -  E^\lambda, \label{VLS1F_u}\\
& & \hspace{-1cm} \partial_t  B^\lambda + \nabla \times  E^\lambda = 0, \label{VLS1F_B} \\
& & \hspace{-1cm} \lambda^2 \partial_t  E^\lambda - \nabla \times  B^\lambda =   u^\lambda - \beta \nabla n^\lambda , \label{VLS1F_E} \\
& & \hspace{-1cm} \nabla \cdot  B^\lambda = 0, \label{VLS1F_divB} \\
& & \hspace{-1cm} \lambda^2 \nabla \cdot  E^\lambda =  -  n^\lambda, \label{VLS1F_divE} 
\end{eqnarray}
where $\beta$ is a numerical viscosity coefficient. We keep in mind that, in the spatially discretized case, $\beta$ is proportional to the mesh size $h$: 
\begin{equation}
\beta = \gamma h , \label{EM:beta}
\end{equation}
with the constant $\gamma$ to be specified later on. To keep the consistency with the Gauss equation, we need to add a numerical viscosity contribution into  the Ampere equation. 

The time-semi-discretization of this model leads to 
\begin{eqnarray}
& & \hspace{-1cm} \delta^{-1} (n^{\lambda, m+1} - n^{\lambda, m}) + \nabla \cdot ( u^{\lambda, m+a}) - \beta \Delta n^{\lambda, m} = 0, \label{DVLS1F_n} \\
& & \hspace{-1cm}  \delta^{-1} (u^{\lambda, m+1} - u^{\lambda, m}) + T \nabla n^{\lambda, m} - \beta \Delta u^{\lambda, m} = - E^{\lambda, m+1}, \label{DVLS1F_u}\\
& & \hspace{-1cm} \delta^{-1} (B^{\lambda, m+1} - B^{\lambda, m}) + \nabla \times E^{\lambda, m+b} = 0, \label{DVLS1F_B} \\
& & \hspace{-1cm} \lambda^2 \delta^{-1} (E^{\lambda, m+1} - E^{\lambda, m}) - \nabla \times B^{\lambda, m+c} =  u^{\lambda, m+a} - \beta \nabla n^{\lambda, m} , \label{DVLS1F_E} \\
& & \hspace{-1cm} \nabla \cdot B^{\lambda, m+1} = 0, \label{DVLS1F_divB} \\
& & \hspace{-1cm} \lambda^2 \nabla \cdot E^{\lambda, m+1} = n^{\lambda, m+1}.  \label{DVLS1F_divE} 
\end{eqnarray}
Passing to Fourier space with $\xi$ being the dual variable to $x$, we find the following recursion relations:  
\begin{eqnarray}
& & \hspace{-1cm} \delta^{-1} (\hat n^{\lambda, m+1} - \hat n^{\lambda, m}) + i \xi \cdot  \hat u^{\lambda, m+a} + \beta |\xi|^2 \hat n^{\lambda, m} = 0, \label{FDVLS1F_n} \\
& & \hspace{-1cm}  \delta^{-1} (\hat u^{\lambda, m+1} - \hat u^{\lambda, m}) + i T \xi \hat n^{\lambda, m} + \beta |\xi|^2 \hat u^{\lambda, m}  = - \hat E^{\lambda, m+1}, \label{FDVLS1F_u}\\
& & \hspace{-1cm} \delta^{-1} (\hat B^{\lambda, m+1} - \hat B^{\lambda, m}) + i \xi \times \hat E^{\lambda, m+b} = 0, \label{FDVLS1F_B} \\
& & \hspace{-1cm} \lambda^2 \delta^{-1} (\hat E^{\lambda, m+1} - \hat E^{\lambda, m}) - i \xi \times \hat B^{\lambda, m+c} =  \hat u^{\lambda, m+a} - i \beta \xi \hat n^{\lambda, m} , \label{FDVLS1F_E} \\
& & \hspace{-1cm} i \xi \cdot \hat B^{\lambda, m+1} = 0, \label{FDVLS1F_divB} \\
& & \hspace{-1cm} i \lambda^2 \xi \cdot \hat E^{\lambda, m+1} = \hat n^{\lambda, m+1}.  \label{FDVLS1F_divE} 
\end{eqnarray}
All solutions of this recursion can be found as linear combinations of elementary solutions of the form $U^{\lambda, m}(\xi) = (q^\lambda(\xi))^m U^{\lambda, 0}(\xi)$ where $q^\lambda(\xi) \in {\mathbb C}$ and $U^{\lambda, m} = (\hat n^{\lambda, m}, \hat u^{\lambda, m}, \hat E^{\lambda, m}, \hat B^{\lambda, m})$. Elementary algebra shows that the characteristic roots $q^\lambda(\xi)$ are the solutions of the two polynomial equations: 
\begin{eqnarray}
& & \hspace{-1.5cm} 
P_\lambda(q) = \lambda^2 (q-1)^2 (q-1 + \beta |\xi|^2 \delta) + q^d \delta^2 |\xi|^2 (q-1 + \beta |\xi|^2 \delta)  + q^{a+1} \delta^2 (q-1) = 0, \label{char_root_em} 
\end{eqnarray}
for the electromagnetic modes and 
\begin{eqnarray}
& & \hspace{-1cm} 
Q_\lambda(q) = \lambda^2 (q-1) (q-1 + \beta |\xi|^2 \delta) + q^{a+1} \delta^2 + T \delta^2 \lambda^2 |\xi|^2 q^a + \nonumber \\
& & \hspace{7cm} + \beta \delta \lambda^2 (q-1 + \beta |\xi|^2 \delta) |\xi|^2 
= 0, 
\label{char_root_es} 
\end{eqnarray}
for the electrostatic ones, where we have defined 
$$ d = b+c.$$
In particular, this shows that whatever choice of the semi-implicitation of the Maxwell equations (either at the level of the Faraday equation or at the level of the Ampere equation), the linearized stability properties of the schemes are the same. 

A necessary and sufficient condition for $L^2$ stability is that $|q^\lambda(\xi)| <1$. However, requesting this condition for all $\xi \in {\mathbb R}$ is too restrictive. To account for the effect of a spatial discretization in this analysis, we must restrict the range of admissible Fourier wave-vectors $\xi$ to the interval $[-\frac{\pi}{h}, \frac{\pi}{h}]$. Indeed, a space discretization of step $h$ cannot represent wave-vectors of magnitude larger than $\frac{\pi}{h}$. This motivates the following definition of stability: 

\begin{definition}
The scheme is stable if and only if
\begin{equation}
|q^\lambda_\pm(\xi)|\leq 1, \quad \forall \xi \quad \mbox{ such that } \quad |\xi| < \frac{\pi}{h}.
\label{EM:stab_cnd}
\end{equation}
\label{EM:def_stab}
\end{definition}

Now, our goal is to find which of the schemes are stable under a sufficient conditions on $\delta$ which is independent of $\lambda$ when $\lambda \to 0$ (Asymptotic Stability). We prove: 

\begin{proposition}
(i) The schemes $(0,1,0)$, $(0,0,1)$, $(1,1,0)$, $(1,0,1)$ are not Asymptotically Stable.

\noindent
(ii) The scheme $(1,1,1)$ is stable under the CFL condition $\delta \leq \Gamma h$ where $\Gamma$ is a constant independent of $\lambda$ and is therefore Asymptotically Stable. 
\label{EM:prop_stab}
\end{proposition}

\medskip
\noindent
{\bf Proof:} (i) Let us examine the $(0,1,0)$ and $(0,0,1)$ schemes first, i.e. with $a=0$ and $d=1$. In either cases, the polynomials (\ref{char_root_em}), (\ref{char_root_es}) can be written:
\begin{eqnarray}
& & \hspace{-1cm} 
P_\lambda(q) = \lambda^2 P_1(q) + P_0(q), \quad Q_\lambda(q) = \lambda^2 Q_1(q) + Q_0(q),   \label{char_root} 
\end{eqnarray}
where $P_0$, $P_1$, $Q_0$, $Q_1$ are independent of $\lambda$. More precisely, we have 
\begin{eqnarray}
& & \hspace{-1.5cm} 
\mbox{deg} P_1 = 3, \quad  \mbox{deg}P_0 = 2, \quad \mbox{deg} Q_1 = 2, \quad  \mbox{deg}Q_0 = 1,  \label{deg_char_root} 
\end{eqnarray}
where deg refers to the degree of the polynomial. Therefore, one of the characteristic roots $q^\lambda$ of either equations tends to infinity and behaves like $\delta^2 (1 + |\xi|^2) / \lambda^2$ when $\lambda \to 0$ for the electromagnetic mode and $\delta^2/(\lambda^2 ( 1+T \delta^2 |\xi|^2 + \beta \delta |\xi|^2))$ for the electrostatic mode. In either cases, an instability develops when $\lambda \to 0$ with fixed $\delta$. 

Let us now examine the $(1,1,0)$ and $(1,0,1)$ schemes, i.e. with $a=1$ and $d=1$. In either cases, we have 
\begin{eqnarray}
& & \hspace{-1cm} 
\mbox{deg} P_1 = 3, \quad  \mbox{deg}P_0 = 3, \quad \mbox{deg} Q_1 = 2, \quad  \mbox{deg}Q_0 = 2,  \label{deg_char_root_2} 
\end{eqnarray}
Let us consider the electromagnetic mode. The two roots of $P_1$ are $1$ and $1-\beta |\xi|^2 \delta$. None of them is a root of $P_0$ as soon as $\beta |\xi|^2 \delta >0$. In these conditions, it is easy to see that the roots of (\ref{char_root_em}) are continuous with respect to $\lambda$ as $\lambda \to 0$. Their limit $q^0(\xi)$ is therefore a solution of $P_0(q) = 0$, which is a cubic equation with obvious root $q=0$. The two remaining roots are easily found to be
$$ q^0_\pm = \frac{1-|\xi|^2}{2} \pm \left( \left(\frac{1-|\xi|^2}{2}  \right)^2 + (1-\beta |\xi|^2 \delta) |\xi|^2 \right)^{1/2}. $$
When $|\xi|$ is large, the negative root becomes less than $-1$, which implies instability of the scheme. Since, when the space step $h \to 0$, the maximal admissible wave-vector tends to infinity, there is no hope to counter-balance this instability by any restriction on the numerical parameters. 

\medskip
\noindent 
(ii) For the $(1,1,1)$ scheme, we have $a=1$ and $d=2$. For the electromagnetic mode, we use the same method as for the case $a=1$ and $d=1$, but now $q=0$ is a double root of $P_0$ and the remaining root is:
$$ q^0 = 1 - \frac{\beta |\xi|^4 \delta}{1+|\xi|^2}. $$
We always have $q\leq 1$ and $q \geq 0$ if and only if $\delta \leq (1+|\xi|^2)/(\beta |\xi|^4)$. With the condition $|\xi| \leq  \pi/h$, and (\ref{EM:beta}), a sufficient condition for stability in the limit $\delta \to 0$ is 
\begin{eqnarray}
& & \hspace{-1cm} 
\delta \leq \frac{1}{\gamma \pi^2} h ( 1 + \frac{h^2}{\pi^2}) \leq \frac{2}{\gamma \pi^2} h, 
\label{stab_0} 
\end{eqnarray}
under the additional restriction $h \leq \pi$ which can always be assumed. For the electrostatic mode, a similar strategy can be developed and we notice that  $q=0$ is a double root of $Q_0$ and no additional stability condition is required. Now, by the continuity of the roots with respect to $\lambda$, there exists $\Gamma$ with $0 < \Gamma < {2}/({\gamma \pi^2})$ and $\lambda_0(\Gamma)>0$ such that under the condition $\delta \leq \Gamma h$, $|\xi| \leq \pi/h$, and $\lambda \leq \lambda_0$, all characteristic roots $q^\lambda$ satisfy $|q^\lambda|<1$. This proves the Asymptotic stability of the scheme. \endproof

\setcounter{equation}{0}
\section{Spatial discretization: enforcing the Gauss law}
\label{sec_1F_space}

\subsection{One-dimensional framework}
\label{subsec_1F_one_dim}

We now concentrate on the $(0,0,1)$ scheme (further on referred to as the 'classical scheme') and 
the $(1,1,1)$ scheme (the 'AP-scheme') and we investigate the spatial discretization. A specific attention will be devoted to the enforcement of Gauss's law. For the sake of the exposition, we restrict ourselves to the one-dimensional case. In this case, all unknowns of the problem only depend upon a one-dimensional spatial coordinate $x \in {\mathbb R}$. The electric field has a longitudinal component $E_x$, and a transverse component. We assume a rectilinear polarization, and choose an orthonormal reference frame $(e_x,e_y,e_z)$ such that $e_x$ is in the $x$ direction and $e_y$ is in the transverse electric field direction. The magnitude of this transverse component is denoted by $E_y$. Finally, the magnetic field is aligned with $e_z$ and its magnitude is denoted by $B_z$. By the divergence free condition, the $x$ component of $B$ must be uniform, and we assume that it vanishes completely. The velocity has components in both the $x$ and $y$ directions, called $u_x$ and $u_y$.

In this geometry, the dimensionless EM model is written:
\begin{eqnarray}
& & \hspace{-1cm} \partial_t n^\lambda + \partial_x (n^\lambda u^\lambda_x) = 0, \label{1F_n_cont} \\
& & \hspace{-1cm}  \partial_t (n^\lambda u^\lambda_x) + \partial_x ( n^\lambda (u^\lambda_x)^2 + p(n^\lambda) ) = - n^\lambda (E^\lambda_x + u^\lambda_y B^\lambda_z), \label{1F_qx_cont} \\
& & \hspace{-1cm}  \partial_t (n^\lambda u^\lambda_y) + \partial_x ( n^\lambda u^\lambda_x u^\lambda_y ) = - n^\lambda (E^\lambda_y - u^\lambda_x B^\lambda_z),  \label{1F_qy_cont} \\
& & \hspace{-1cm}  \partial_t B^\lambda_z + \partial_x E^\lambda_y = 0, \label{1F_faraday_cont} \\
& & \hspace{-1cm}  \lambda^2 \partial_t E^\lambda_x = n^\lambda u^\lambda_x, \label{1F_amperex_cont} \\
& & \hspace{-1cm}  \lambda^2 \partial_t E^\lambda_y + \partial_x B^\lambda_z = n^\lambda u^\lambda_y, \label{1F_amperey_cont} \\
& & \hspace{-1cm}  \lambda^2 \partial_x E^\lambda_x = 1-n^\lambda, \label{1F_gauss_cont}
\end{eqnarray}
The associated QN-EM model is obtained by taking $\lambda \to 0$. We get: 
\begin{eqnarray}
& & \hspace{-1cm}  E^0_x + u^0_y B^0_z = 0, \label{1F_qx_lambda=0_cont} \\
& & \hspace{-1cm}  \partial_t  u^0_y  = - E^0_y ,  \label{1F_qy_lambda=0_cont} \\
& & \hspace{-1cm}  \partial_t B^0_z + \partial_x E^0_y = 0, \label{1F_faraday_lambda=0_cont} \\
& & \hspace{-1cm}  u^0_x=0, \label{1F_amperex_lambda=0_cont} \\
& & \hspace{-1cm}  \partial_x B^0_z = u^0_y, \label{1F_amperey_lambda=0_cont} \\
& & \hspace{-1cm}  n^0=1, \label{1F_gauss_lambda=0_cont}
\end{eqnarray}
Taking the $x$-derivative of (\ref{1F_faraday_lambda=0_cont}), adding to (\ref{1F_qy_lambda=0_cont}) and using (\ref{1F_amperey_lambda=0_cont}) leads to 
\begin{eqnarray}
& & \hspace{-1cm}  - \partial^2_x E^0_y + E^0_y = 0, \label{1F_amperey_lambda=0_cont2} 
\end{eqnarray}
Conversely, the QN-EM model obtained by replacing (\ref{1F_amperey_lambda=0_cont}) by (\ref{1F_amperey_lambda=0_cont2}) is equivalent to the original one provided that $\partial_x B^0_z|_{t=0} =  u^0|_{t=0}$. The proof is similar to the full 3D case in section \ref{subsec_1F_QNlim}.

The time discretization of the one-dimensional EM model is given by (omitting the exponent $\lambda$): 
\begin{eqnarray}
& & \hspace{-1cm}  \delta^{-1} (n^{m+1} - n^m) + \partial_x (n^{m+a} u_x^{m+a}) = 0, \label{1F_n_d} \\
& & \hspace{-1cm}  \delta^{-1} (n^{m+1} u_x^{m+1} - n^m u_x^m) + \partial_x ( n^m (u_x^m)^2 + p(n^m) ) = \nonumber \\
& & \hspace{5cm}  =  - (n^{m+1-a} E_x^{m+1} + n^m u_y^m B_z^m), \label{1F_qx_d} \\
& & \hspace{-1cm}  \delta^{-1} (n^{m+1} u_y^{m+1} - n^m u_y^m)  + \partial_x ( n^m u_x^m u_y^m ) = - (n^{m+1-a} E_y^{m+1} - n^m u_x^m B_z^m),  \label{1F_qy_d} \\
& & \hspace{-1cm}  \delta^{-1} (B_z^{m+1} - B_z^m) + \partial_x E_y^{m+b} = 0, \label{1F_faraday_d} \\
& & \hspace{-1cm}  \lambda^2 \delta^{-1} (E_x^{m+1} - E_x^m) = n^{m+a} u_x^{m+a}, \label{1F_amperex_d} \\
& & \hspace{-1cm}  \lambda^2 \delta^{-1} (E_y^{m+1} - E_y^m) + \partial_x B_z^{m+c} = n^{m+a} u_y^{m+a}, \label{1F_amperey_d} \\
& & \hspace{-1cm}  \lambda^2 \partial_x E_x^{m+1} = 1-n^{m+1}, \label{1F_gauss_d}
\end{eqnarray}
again, with $(a,b,c)=(0,0,1)$ (classical scheme) or $(1,1,1)$ (AP scheme).

\subsection{Spatial discretization}
\label{subsec_1F_spatial}

Now, we introduce a spatial discretization with a uniform mesh of step $ h $ and we denote by $C_k$ the cell $[(k-1/2) h, (k+1/2)h]$ and $x_k = kh$, with $k \in {\mathbb Z}$. Like in usual first-order shock capturing schemes, the fluid unknowns $n$ and $u$ are approximated by piecewise constant functions within the cell $C_k$ and represented by cell-centered values $ n|_{k}^{m}$, $ u|_{k}^{m}$ at time $t^m = m \delta$. The electric field $E_x$ and the magnetic field $B_z$ are approximated at the interfaces \, $x_{k+1/2}= (k+1/2)h$ \, by \, $ E_x|_{k+1/2}^{m} $ and $ B_z|_{k+1/2}^{m}$, while $E_y$ is approximated by cell-centered quantities $ E_y|_{k}^{m}$. The discretization of the hydrodynamic part is performed by means of a first order shock capturing scheme. 
We denote by $f_{n}|_{k+1/2}^{m} $, $f_{u_{x}}|_{k+1/2}^{m}$, $f_{u_{y}}|_{k+1/2}^{m} $ the numerical fluxes for the mass and $x$ and $y$-components of the momentum conservation equations respectively, at time $ t^m = m \delta $ and at the cell interface $x_{k+1/2}$. 

The fully discretized scheme is written: 
\begin{eqnarray}
& & \hspace{-1cm}  \delta^{-1} (n|_k^{m+1} - n|_k^m) + h^{-1} (f_n|_{k+1/2}^{m+a} - f_n|_{k-1/2}^{m+a})  = 0, \label{1F_n_fd} \\
& & \hspace{-1cm} \delta^{-1}  ((nu_x)|_k^{m+1} - (nu_x)|_k^m) + h^{-1} (f_{u_x}|_{k+1/2}^{m} - f_{u_x}|_{k-1/2}^{m})  = \nonumber \\
& & \hspace{5cm}  =  - n|_k^{m+1-a} \tilde E_x|_k^{m+1}  - (nu_y)|_k^m \,  \tilde B_z|_k^m, \label{1F_qx_fd} \\
& & \hspace{-1cm} \delta^{-1}  ((nu_y)|_k^{m+1} - (nu_y)|_k^m) + h^{-1} (f_{u_y}|_{k+1/2}^{m} - f_{u_y}|_{k-1/2}^{m})  = \nonumber \\
& & \hspace{5cm}  =  - n|_k^{m+1-a} E_y|_k^{m+1}  + (nu_x)|_k^m \, \tilde  B_z|_k^m, \label{1F_qy_fd} \\
& & \hspace{-1cm}  \delta^{-1} (B_z|_{k+1/2}^{m+1} - B_z|_{k+1/2}^m) + h^{-1} (E_y|_{k+1}^{m+b} - E_y|_{k}^{m+b}) = 0, \label{1F_faraday_fd} \\
& & \hspace{-1cm}  \lambda^2 \delta^{-1} (E_x|_{k+1/2}^{m+1} - E_x|_{k+1/2}^m) = f_n|_{k+1/2}^{m+a}, \label{1F_amperex_fd} \\
& & \hspace{-1cm}  \lambda^2 \delta^{-1} (E_y|_k^{m+1} - E_y|_k^m) + h^{-1} (B_z|_{k+1/2}^{m+c} - B_z|_{k-1/2}^{m+c}) = (n u_y)|_k^{m+a}, \label{1F_amperey_fd}  
\end{eqnarray}
where 
\begin{eqnarray}
& & \hspace{-1cm}   \tilde B_z|_k^m  = \frac{1}{2} (B_z|_{k+1/2}^{m} + B_z|_{k-1/2}^m) , \quad \tilde E_x|_k^{m+1} = \frac{1}{2} (E_x|_{k+1/2}^{m+1} + E_x|_{k-1/2}^{m+1}) .  \label{interp_Bz_Ex} 
\end{eqnarray}

The numerical hydrodynamic fluxes $f_n|_{k+1/2}^{m}$, $f_{u_x}|_{k+1/2}^{m}$ and $f_{u_y}|_{k+1/2}^{m}$ are computed using a Local Lax-Friedrichs (LLF) scheme \cite{Leveque_2} (also known as the Rusanov scheme \cite{Rusanov}; we note that this scheme enters the class of polynomial solvers of \cite{Deg_Pey_Rus_Vil}: it corresponds to the case of a degree $ 0 $ polynomial). In the case $(a,b,c)=(0,0,1)$ (classical scheme), the numerical fluxes are given by:
\begin{eqnarray}
& & \hspace{-1cm}  f_n|_{k+1/2}^{m} = \frac{1}{2} \left[ (nu_x)|_k^m + (nu_x)|_{k+1}^m + \mu_{k+1/2}^m \, \left( n|_k^m - n|_{k+1}^m \right) \right], 
\label{1F_fn_fd} \\
& & \hspace{-1cm}  f_{u_x}|_{k+1/2}^{m} = \frac{1}{2} \left[ \left(nu_x^2 + p(n) \right)|_k^m + \left(nu_x^2 + p(n) \right)|_{k+1}^m + \right.\nonumber \\
& & \hspace{6cm} \left. + \mu_{k+1/2}^m \, \left( (nu_x)|_k^m - (nu_x)|_{k+1}^m \right)  \right], 
\label{1F_fux_fd} \\
& & \hspace{-1cm}  f_{u_y}|_{k+1/2}^{m} = \frac{1}{2} \left[ (n u_x u_y)|_k^m + (n u_x u_y)|_{k+1}^m + \mu_{k+1/2}^m \, \left( (nu_y)|_k^m - (nu_y)|_{k+1}^m \right)  \right] . 
\label{1F_fuy_fd} 
\end{eqnarray}
In the case $(a,b,c)=(1,1,1)$ (AP scheme) the momentum fluxes (\ref{1F_fux_fd}) and (\ref{1F_fuy_fd}) are unchanged. The mass flux $\tilde f_n|_{k+1/2}^{m+1}$ is given by: 
\begin{eqnarray}
& & \hspace{-1cm}  \tilde f_n|_{k+1/2}^{m+1} = \frac{1}{2} \left[ (nu_x)|_k^{m+1} + (nu_x)|_{k+1}^{m+1} + \mu_{k+1/2}^m \, \left( n|_k^m - n|_{k+1}^m \right) \right]. 
\label{1F_fn_m+1_fd} 
\end{eqnarray}
Only the central discretization part of the flux is implicit, while the numerical viscosity term (in factor of $\mu_{k+1/2}^m$) is kept explicit. The tilde is there to make a typographic distinction from the explicit flux (\ref{1F_fn_fd}). Indeed, using the momentum balance equation (\ref{1F_qx_fd}), we can relate the implicit flux (\ref{1F_fn_m+1_fd}) to the explicit one (\ref{1F_fn_fd}) by the following relation: 
\begin{eqnarray}
& & \hspace{-1cm}  \tilde f_n|_{k+1/2}^{m+1} = f_n|_{k+1/2}^{m} - \frac{\delta}{4} \left[ n|_{k+1}^m \, E_x|_{k+3/2}^{m+1} + (n|_{k+1}^m + n|_{k}^m)  \, E_x|_{k+1/2}^{m+1} + \frac{1}{4} n|_{k}^m  \, E_x|_{k-1/2}^{m+1}  \right] \nonumber \\
& & \hspace{0cm} - \frac{\delta h^{-1}}{2} (f_{u_x}|_{k+3/2}^{m} - f_{u_x}|_{k-1/2}^{m}) - \frac{\delta}{2} \left[ (nu_y)|_k^m \,  \tilde B_z|_k^m + (nu_y)|_{k+1}^m \,  \tilde B_z|_{k+1}^m \right] 
. \label{1F_fn_m+1_fd2} 
\end{eqnarray}
This flux involves an average of $E_x^{m+1}$ over three neighbouring mesh points which is too diffusive and poorly accurate. In order to reduce numerical diffusion, we replace (\ref{1F_fn_m+1_fd2}) by the following expression: 
\begin{eqnarray}
& & \hspace{-1cm}  \tilde f_n|_{k+1/2}^{m+1} = f_n|_{k+1/2}^{m} - \frac{\delta}{2} (n|_{k+1}^m + n|_{k}^m)  \, E_x|_{k+1/2}^{m+1}  \nonumber \\
& & \hspace{0cm} - \frac{\delta h^{-1}}{2} (f_{u_x}|_{k+3/2}^{m} - f_{u_x}|_{k-1/2}^{m}) - \frac{\delta}{2} ( (nu_y)|_k^m  + (nu_y)|_{k+1}^m ) \, B_z|_{k+1/2}^m  
. \label{1F_fn_m+1_fd2_mod} 
\end{eqnarray}
This implicit flux can be viewed as an order $O(\delta)$ modification of the explicit flux. This simple modification is crucial in making the scheme AP. 

We now specify the numerical viscosity $\mu_{k+1/2}^m$. In the LLF scheme, the quantity $\mu_{k+1/2}^m$ is an evaluation of the local maximal wave speed at the interface $x_{k+1/2}$. It is computed as follows: we introduce
$$ \nu^{+}|_{k+1/2}^{m} = \max \left( \nu^*|_{k+1/2}^m , \nu^*|_{k+1}^m \right)  \quad \mbox{ and } \quad  \nu^{-}|_{k+1/2}^{m} = \min \left( \nu_*|_{k+1/2}^m , \nu_*|_{k}^m \right) , $$
where $\nu^*|_{k}^m$ and $\nu_*|_{k}^m$ (respectively $\nu^*|_{k+1/2}^m$ and $\nu_*|_{k+1/2}^m$) denote the largest and smallest characteristic speeds of the hydrodynamic systems associated to the state $(n|_k^m, \, (nu_x)|_k^m,$ $(nu_y)|_k^m)$ (respectively to the state  $(n|_{k+1/2}^m, \, (nu_x)|_{k+1/2}^m, \, (nu_y)|_{k+1/2}^m)$, with 
$$ n|_{k+1/2}^m = \frac{1}{2} (n|_k^m + n|_{k+1}^m) , \quad (nu_x)|_{k+1/2}^m = \frac{1}{2} ((nu_x)|_k^m + (nu_x)|_{k+1}^m),$$
and similarly for $(nu_y)|_{k+1/2}^m$). Then, 
$$ \mu_{k+1/2}^m =  \max \left\{ \left| \, \nu^{+}|_{k+1/2}^{m}   \, \right| , \left|  \,  \nu^{-}|_{k+1/2}^{m}  \,  \right| \right\}.  $$ 
The time step $ \delta $ must satisfy the CFL condition
$ \delta \leq (\max_{k \in {\mathbb Z}} \mu_{k+1/2}^m) \, h $
to ensure stability of the hydrodynamic part of the scheme.

An important feature of the scheme is that the current $(nu)_x$ in the $x$-component of the Ampere equation (\ref{1F_amperex_fd}) is evaluated by using the mass flux $f_n|_{k+1/2}^{m+a}$. At the level of the continuous problem, these two quantities are identical. Therefore, this approximation is consistent. However, using the mass flux rather than the current allows us to guarantee a perfect consistency with the Gauss equation. Indeed, 
taking the difference of (\ref{1F_amperex_fd}) evaluated at $x_{k+1/2}$ and $x_{k-1/2}$ and using (\ref{1F_n_fd}), we easily check that:
\begin{eqnarray}
& & \hspace{-1cm}  \lambda^2  h^{-1} (E_x|_{k+1/2}^{m+1} - E_x|_{k-1/2}^{m+1}) + n|_{k}^{m+1} = \lambda^2  h^{-1} (E_x|_{k+1/2}^{m} - E_x|_{k-1/2}^{m}) + n|_{k}^{m}.  \label{1F_gauss_fd}
\end{eqnarray}
We deduce that the Gauss equation is exactly satisfied at any time, provided that it is exactly satisfied at initialization. In the $y$-component of the Ampere equation (\ref{1F_amperex_fd}), the current is evaluated using the usual approximation $(n u_y)|_k^{m+a}$ because, in a one-dimensional problem, the $y$-component of the mass flux is independent of $x$ and does not enter the mass balance. In a 2 or 3-dimensional problem, one should evaluate all components of the current using the corresponding components of the mass flux, to ensure consistency with the Gauss equation. 

We now consider the sequence of updates for the two schemes separately.

\subsection{Classical scheme $(a,b,c)=(0,0,1)$: time update}
\label{subsec_1F_class_time_update}

In this case, the time update goes as follows: first the mass conservation eq. (\ref{1F_n_fd}) is used to compute $n^{m+1}$. Then the Faraday eq. (\ref{1F_faraday_fd}) allows us to find $B_z^{m+1}$, immediately followed by the Ampere eqs. (\ref{1F_amperex_fd}), (\ref{1F_amperey_fd}) to find $E_x|_{k+1/2}^{m+1}$ and $E_y|_k^{m+1}$. Finally, with the momentum balance eqs. (\ref{1F_qx_fd}), (\ref{1F_qy_fd}), we find the values of $(nu_x)|_k^{m+1}$ and $(nu_y)|_k^{m+1}$.

\subsection{AP-scheme $(a,b,c)=(1,1,1)$: time update and AP character}
\label{subsubsec_1F_AP_time_update}

The time update follows a different sequence. We first solve for the implicit Maxwell equations. We begin by computing $E_y^{m+1}$. To this aim, inserting (\ref{1F_faraday_fd}) and (\ref{1F_qy_fd}) into (\ref{1F_amperey_fd}) to eliminate $B_z|_{k+1/2}^{m+1}$, and $(nu_y)|_k^{m+1}$ respectively, we find that (\ref{1F_amperey_fd}) is equivalent to
\begin{eqnarray}
& & \hspace{-1cm}  (\lambda^2  + \delta^{2} n|_k^m) E_y|_k^{m+1} - \delta^{2} h^{-2} (E_y|_{k+1}^{m+1} - 2 E_y|_k^{m+1} + E_y|_{k-1}^{m+1}) = \lambda^2 E_y|_k^m  + \delta (nu_y)|_k^m \nonumber \\
& & \hspace{-1cm}   - \delta h^{-1} (B_z|_{k+1/2}^{m} - B_z|_{k-1/2}^{m})  - \delta^{2} h^{-1}  (f_{u_y}|_{k+1/2}^{m} - f_{u_y}|_{k-1/2}^{m}) + \delta^{2} (nu_x)|_k^m \, \tilde  B_z|_k^m
, \label{1F_amperey_ref_fd}  
\end{eqnarray}
or, using (\ref{1F_amperey_fd}) again, between time steps $t^{m-1}$ and $t^m$, to
\begin{eqnarray}
& & \hspace{-1cm}  (\lambda^2 + \delta^{2} n|_k^m ) E_y|_k^{m+1} - \delta^{2} h^{-2} (E_y|_{k+1}^{m+1} - 2 E_y|_k^{m+1} + E_y|_{k-1}^{m+1}) = \lambda^2 (2 E_y|_k^m - E_y|_k^{m-1})   \nonumber \\
& & \hspace{4.5cm}  - \delta^{2} h^{-1} (f_{u_y}|_{k+1/2}^{m} - f_{u_y}|_{k-1/2}^{m}) + \delta^{2} (nu_x)|_k^m \, \tilde  B_z|_k^m
. \label{1F_amperey_ref2_fd}  
\end{eqnarray}
This equation is clearly consistent with the reformulated Ampere eq. (\ref{RS1F_E}). Both (\ref{1F_amperey_ref_fd}) and (\ref{1F_amperey_ref2_fd}) are discrete elliptic equations for $E_y|_k^{m+1}$ which are well-posed provided that suitable boundary conditions are defined.

We now examine the computation of $E_x^{m+1}$. We insert the expression (\ref{1F_fn_m+1_fd2_mod}) of the mass flux into (\ref{1F_amperex_fd}). This yields: 
\begin{eqnarray}
& & \hspace{-1cm}  (\lambda^2  +  \frac{\delta^2}{2} (n|_{k+1}^m + n|_{k}^m) )  \, E_x|_{k+1/2}^{m+1}  = \lambda^2 E_x|_{k+1/2}^m + \delta f_{n}|_{k+1/2}^{m}  \nonumber \\
& & \hspace{0cm} - \frac{\delta^2 h^{-1}}{2} (f_{u_x}|_{k+3/2}^{m} - f_{u_x}|_{k-1/2}^{m}) - \frac{\delta^2}{2}( (nu_y)|_k^m  + (nu_y)|_{k+1}^m ) \, B_z|_{k+1/2}^m . 
\label{1F_amperex_fd_mod} 
\end{eqnarray}
This expression provides an explicit evaluation of $E_x|_{k+1/2}^{m+1}$. 

Once $E_y^{m+1}$ and $E_x^{m+1}$ are known, we can compute $B_z^{m+1}$ using (\ref{1F_faraday_fd}), then $u_x^{m+1}$, $u_y^{m+1}$ and $n^{m+1}$ using (\ref{1F_qx_fd}), (\ref{1F_qy_fd}) and (\ref{1F_n_fd}) respectively. 

Finally, we show that the fully discrete scheme is AP. Indeed, when $\lambda \to 0$, (\ref{1F_gauss_fd}), gives $n|_k^{m+1} = n|_k^m$ exactly. Since we assume consistency with the Gauss equation at time $t=0$, which, in the case $\lambda = 0$, amounts to assuming that $n|_k^0 = 1$ for all $k$, we deduce that $n|_k^m=1$ for all $k$ and $m$. Then, the remaining equations yield, in the limit $\lambda \to 0$: 
\begin{eqnarray}
& & \hspace{-1cm}  \delta E_x|_{k+1/2}^{m+1}  = f_{n}|_{k+1/2}^{m} - \frac{\delta h^{-1}}{2} (f_{u_x}|_{k+3/2}^{m} - f_{u_x}|_{k-1/2}^{m}) - \frac{\delta}{2} ( u_y|_k^m + u_y|_{k+1}^m ) \, B_z|_{k+1/2}^m  , \nonumber \\
& & \hspace{-1cm} \label{1F_amperex_lambda=0_fd_mod} \\
& & \hspace{-1cm}  \delta E_y|_k^{m+1} - \delta h^{-2} (E_y|_{k+1}^{m+1} - 2 E_y|_k^{m+1} + E_y|_{k-1}^{m+1}) =   \nonumber \\
& & \hspace{5cm}   - h^{-1} (B_z|_{k+1/2}^{m} - B_z|_{k-1/2}^{m})  +  u_y|_k^m  \nonumber \\
& & \hspace{5cm}   - \delta h^{-1} (f_{u_y}|_{k+1/2}^{m} - f_{u_y}|_{k-1/2}^{m}) + \delta u_x|_k^m \, \tilde  B_z|_k^m
, \label{1F_amperey_lambda=0_fd}  \\
& & \hspace{-1cm}  \delta^{-1} (B_z|_{k+1/2}^{m+1} - B_z|_{k+1/2}^m) + h^{-1} (E_y|_{k+1}^{m+1} - E_y|_{k}^{m+1}) = 0, \label{1F_faraday_lambda=0_fd} \\
& & \hspace{-1cm} \delta^{-1}  (u_x|_k^{m+1} - u_x|_k^m) + h^{-1} (f_{u_x}|_{k+1/2}^{m} - f_{u_x}|_{k-1/2}^{m})  = \nonumber \\
& & \hspace{5cm}  =  - \frac{1}{2} ( E_x|_{k+1/2}^{m+1} + E_x|_{k-1/2}^{m+1})  - u_y|_k^m \,  \tilde B_z|_k^m, \label{1F_qx_lambda=0_fd} \\
& & \hspace{-1cm} \delta^{-1}  (u_y|_k^{m+1} - u_y|_k^m) + h^{-1} (f_{u_y}|_{k+1/2}^{m} - f_{u_y}|_{k-1/2}^{m})  = \nonumber \\
& & \hspace{5cm}  =  -  E_y|_k^{m+1}  + u_x|_k^m \, \tilde  B_z|_k^m, \label{1F_qy_lambda=0_fd} 
\end{eqnarray}
with the fluxes 
\begin{eqnarray}
& & \hspace{-1cm}  f_n|_{k+1/2}^{m} = \frac{1}{2} ( u_x|_k^m + u_x|_{k+1}^m ), 
\label{1F_fn_lambda=0_fd} \\
& & \hspace{-1cm}  f_{u_x}|_{k+1/2}^{m} = \frac{1}{2} \left[ u_x^2|_k^m + u_x^2|_{k+1}^m +  \mu_{k+1/2}^m \, \left( u_x|_k^m - u_x|_{k+1}^m \right)  \right], 
\label{1F_fux_lambda=0_fd} \\
& & \hspace{-1cm}  f_{u_y}|_{k+1/2}^{m} = \frac{1}{2} \left[ (u_x u_y)|_k^m + (u_x u_y)|_{k+1}^m + \mu_{k+1/2}^m \, \left( u_y|_k^m - u_y|_{k+1}^m \right)  \right] . 
\label{1F_fuy_lambda=0_fd} 
\end{eqnarray}
Since the pressure $p(1)$ is now a constant, it has be removed from (\ref{1F_fux_lambda=0_fd}), because fluxes are defined up to a constant in space. 
Inserting (\ref{1F_fn_lambda=0_fd}) into (\ref{1F_amperex_lambda=0_fd_mod}) and the result into (\ref{1F_qx_lambda=0_fd}), we find 
\begin{eqnarray*}
& & \hspace{-1cm}  
u_x|_k^{m+1} = - \frac{1}{4} (u_x|_{k+1}^m - 2 u_x|_k^m + u_x|_{k-1}^m) + \\
& & \hspace{1cm}  \frac{\delta h^{-1}}{4} ( f_{u_x}|_{k+3/2}^{m} - 3 f_{u_x}|_{k+1/2}^{m} + 3 f_{u_x}|_{k-1/2}^{m} - f_{u_x}|_{k-3/2}^{m}) \\
& & \hspace{1cm} + \frac{\delta}{4} ( u_y|_{k+1}^{m} B_z|_{k+1/2}^{m} - u_y|_{k}^{m} ( B_z|_{k+1/2}^{m}+ B_z|_{k-1/2}^{m}) + u_y|_{k-1}^{m} B_z|_{k-1/2}^{m} ) \\
& & \hspace{0.3cm}  = O( h^2 (1 + \delta)), 
\end{eqnarray*}
which is consistent with (\ref{1F_amperex_lambda=0_cont}). From there, we deduce that
$$ f_{u_x}|_{k+1/2}^{m} = O( h^3 (1 + \delta)), \quad f_{u_y}|_{k+1/2}^{m} = O( h).$$ 
It follows that (\ref{1F_qy_lambda=0_fd}) can be written
\begin{eqnarray*}
& & \hspace{-1cm}  \delta^{-1}  (u_y|_k^{m+1} - u_y|_k^m) =  -  E_y|_k^{m+1}  + O( h ), 
\end{eqnarray*}
which is consistent with (\ref{1F_qy_lambda=0_cont}). A similar computation, inserting (\ref{1F_qx_lambda=0_fd}) into (\ref{1F_fn_lambda=0_fd}) and using (\ref{1F_amperex_lambda=0_fd_mod}) shows that 
$$ f_{n}|_{k+1/2}^{m} = O(\delta h^2). $$ 
Therefore, (\ref{1F_amperex_lambda=0_cont}) is such that 
\begin{eqnarray*}
& & \hspace{-1cm}  E_x|_{k+1/2}^{m+1}  = - \frac{1}{2} ( u_y|_k^m + u_y|_{k+1}^m ) \, B_z|_{k+1/2}^m + O(h^2) ,  
\end{eqnarray*}
which is consistent with (\ref{1F_qx_lambda=0_cont}). The consistency of (\ref{1F_faraday_lambda=0_fd}) with (\ref{1F_faraday_lambda=0_cont}) is obvious. Finally, inserting (\ref{1F_faraday_lambda=0_fd}) into (\ref{1F_amperey_lambda=0_fd}) leads to 
\begin{eqnarray*}
& & \hspace{-1cm}  h^{-1} (B_z|_{k+1/2}^{m+1} - B_z|_{k-1/2}^{m+1}) =  u_y|_k^m  + O(\delta)
, 
\end{eqnarray*}
which is consistent with (\ref{1F_amperey_lambda=0_cont}). This proves that the fully discrete scheme is AP.

\setcounter{equation}{0}
\section{Two-fluid case}
\label{sec_2F}

\subsection{Euler-Maxwell system}
\label{subsec_2F}

The two-fluid Euler-Maxwell (EM) system consists of the mass and momentum balance equations for both the electron and ion fluids coupled to the Maxwell equations. The mass and momentum balance equations are written:
\begin{eqnarray}
& & \hspace{-1cm} \partial_t n_i + \nabla \cdot (n_i u_i) = 0, \label{2F_ni} \\
& & \hspace{-1cm} m_i ( \partial_t (n_i u_i) + \nabla \cdot (n_i u_i \otimes u_i)) + \nabla p_i =  e n_i (E + u_i \times B), \label{2F_ui} \\
& & \hspace{-1cm} \partial_t n_e + \nabla \cdot (n_e u_e) = 0, \label{2F_ne} \\
& & \hspace{-1cm} m_e ( \partial_t (n_e u_e) + \nabla \cdot (n_e u_e \otimes u_e)) + \nabla p_e =  - e n_e (E + u_e \times B), \label{2F_ue} 
\end{eqnarray}
where the indices $i$ and $e$ refer to the ions and electrons respectively. The meaning of the variables is the same as in the one-fluid case, section (\ref{subsec_1F_EMS}). The Maxwell equations (\ref{1F_B})-(\ref{1F_divE}) are unchanged but the definition of the charge and current densities is now given by:
\begin{eqnarray}
& & \hspace{-1cm} \rho = e (n_i - n_e), \label{2F_rho} \\
& & \hspace{-1cm} j = e (n_i u_i - n_e u_e) ,  \label{2F_j} 
\end{eqnarray}
where we assume for simplicity that the ions are singly charged.

In the scaling, the same density and velocity scales for the ions and the electrons are chosen. The thermal energy scale is chosen equal to the ion drift energy scale i.e.  $m_i u_0^2 = p_0 n_0^{-1}$ and an additional dimensionless parameter $\varepsilon$ corresponding to the electron to ion mass ratio appears: 
\begin{eqnarray}
& & \hspace{-1cm} \varepsilon^2 = \frac{m_e}{m_i} .  \label{2F_eps} 
\end{eqnarray}
Apart from this, we use similar scaling hypotheses as in the one-fluid case, section \ref{subsec_1F_scaling}, and find the dimensionless two-fluid EM model: 
\begin{eqnarray}
& & \hspace{-1cm} \partial_t n^\lambda_i + \nabla \cdot (n^\lambda_i u^\lambda_i) = 0, \label{S2F_ni} \\
& & \hspace{-1cm}  \partial_t (n^\lambda_i u^\lambda_i) + \nabla \cdot (n^\lambda_i u^\lambda_i \otimes u^\lambda_i) + \nabla p(n^\lambda_i) = n^\lambda_i (E^\lambda+ u^\lambda_i \times B^\lambda), \label{S2F_ui}\\
& & \hspace{-1cm} \partial_t n^\lambda_e + \nabla \cdot (n^\lambda_e u^\lambda_e) = 0, \label{S2F_ne} \\
& & \hspace{-1cm}  \varepsilon^2 [ \partial_t (n^\lambda_e u^\lambda_e) + \nabla \cdot (n^\lambda_e u^\lambda_e \otimes u^\lambda_e)] + \nabla p(n^\lambda_e) = - n^\lambda_e (E^\lambda+ u^\lambda_e \times B^\lambda), \label{S2F_ue}\\
& & \hspace{-1cm} \partial_t B^\lambda + \nabla \times E^\lambda = 0, \label{S2F_B} \\
& & \hspace{-1cm} \lambda^2 \partial_t E^\lambda - \nabla \times B^\lambda = - j^\lambda := - (n^\lambda_i u^\lambda_i - n^\lambda_e u^\lambda_e ) , \label{S2F_E} \\
& & \hspace{-1cm} \nabla \cdot B^\lambda = 0, \label{S2F_divB} \\
& & \hspace{-1cm} \lambda^2 \nabla \cdot E^\lambda = \rho^\lambda := n^\lambda_i - n^\lambda_e, \label{S2F_divE} 
\end{eqnarray}
where we have introduced the dimensionless charge and current densities $\rho^\lambda$ and $j^\lambda$. 

The quasineutral $\lambda \to 0$ limit leads to the two-fluid QN-EM model, in which only the Ampere and Gauss equations (\ref{S2F_E}), (\ref{S2F_divE}) are formally modified: 
\begin{eqnarray}
& & \hspace{-1cm} \partial_t n^0_i + \nabla \cdot (n^0_i u^0_i) = 0, \label{QN2F_ni} \\
& & \hspace{-1cm}  \partial_t (n^0_i u^0_i) + \nabla \cdot (n^0_i u^0_i \otimes u^0_i) + \nabla p(n^0_i) = n^0_i (E^0+ u^0_i \times B^0), \label{QN2F_ui}\\
& & \hspace{-1cm} \partial_t n^0_e + \nabla \cdot (n^0_e u^0_e) = 0, \label{QN2F_ne} \\
& & \hspace{-1cm}  \varepsilon^2 [ \partial_t (n^0_e u^0_e) + \nabla \cdot (n^0_e u^0_e \otimes u^0_e)] + \nabla p(n^0_e) = - n^0_e (E^0+ u^0_e \times B^0), \label{QN2F_ue}\\
& & \hspace{-1cm} \partial_t B^0 + \nabla \times E^0 = 0, \label{QN2F_B} \\
& & \hspace{-1cm} - \nabla \times B^0 =  - j^0 := - (n^0_i u^0_i - n^0_e u^0_e)  , \label{QN2F_E} \\
& & \hspace{-1cm} \nabla \cdot B^0 = 0, \label{QN2F_divB} \\
& & \hspace{-1cm} 0 = \rho^0 := n^0_i - n^0_e. \label{QN2F_divE} 
\end{eqnarray}
We note that we keep $\varepsilon$ fixed and finite. Taking the difference of (\ref{QN2F_ni}) and (\ref{QN2F_ne}) and using (\ref{QN2F_divE}) shows that the current should be divergence free: 
\begin{eqnarray}
& & \hspace{-1cm} \nabla \cdot j^0 =  0, \label{QN2F_divj} 
\end{eqnarray}
which is consistent with (\ref{QN2F_E}). Taking the difference of (\ref{QN2F_ui}) with $\varepsilon^{-2} \times$ (\ref{QN2F_ue}), we find 
\begin{eqnarray}
& & \hspace{-1cm} \partial_t j^0 + \nabla \cdot \phi^0 =  (n^0_i + \varepsilon^{-2} n^0_e) E + (n^0_i u^0_i + \varepsilon^{-2} n^0_e u^0_e) \times B, \label{QN2F_j} 
\end{eqnarray}
with
\begin{eqnarray}
& & \hspace{1cm} \phi^0 = n^0_i u^0_i \otimes u^0_i + p(n^0_i) \mbox{Id} - (  n^0_e u^0_e \otimes u^0_e + \varepsilon^{-2} p(n^0_e) \mbox{Id} ) , \label{QN2F_flj} 
\end{eqnarray}
the current flux. Then, $B^0$ and $j^0$ both satisfy evolution equations ((\ref{QN2F_B}) and (\ref{QN2F_j})) and are related by the constraint (\ref{QN2F_E}). $E^0$ is the Lagrange multiplier of this constraint. To find it, we take the curl of (\ref{QN2F_B}), subtract it to (\ref{QN2F_j}) and use (\ref{QN2F_E}). We find
\begin{eqnarray}
& & \hspace{-1cm} \nabla \times (\nabla \times E^0)  + (n^0_i + \varepsilon^{-2} n^0_e) E  = \nabla \cdot \phi^0 - (n^0_i u^0_i + \varepsilon^{-2} n^0_e u^0_e) \times B, \label{QN2F_E_ref} 
\end{eqnarray}
which is a well-posed elliptic equation for $E$. An equivalent form of the two-fluid QN-EM model is therefore obtained by replacing the Ampere equation (\ref{QN2F_E}) by its reformulation (\ref{QN2F_E_ref}) provided that 
\begin{eqnarray}
& & \hspace{-1cm}  - \nabla \times B^0|_{t=0} =  -j^0|_{t=0}  . \label{FQN2F_E} 
\end{eqnarray}
The proof is similar as in the one-fluid case. 

We finally note that we can reformulate the Ampere equation in the original EM model by using a similar manipulation. The current equation (\ref{QN2F_j}) and its associated flux (\ref{QN2F_flj}) have the same expression at finite $\lambda$. Then, taking the curl of (\ref{S2F_B}), subtracting it to (\ref{QN2F_j}) (with finite $\lambda$) and using (\ref{QN2F_E}), we find: 
\begin{eqnarray}
& & \hspace{-1cm} \lambda^2 \partial^2_t E^\lambda + \nabla \times (\nabla \times E^\lambda)  + (n^\lambda_i + \varepsilon^{-2} n^\lambda_e) E  = \nabla \cdot \phi^\lambda - (n^\lambda_i u^\lambda_i + \varepsilon^{-2} n^\lambda_e u^\lambda_e) \times B.  \label{S2F_E_ref} 
\end{eqnarray}
The reformulated EM model (REM) which consists of the original EM model in which the Ampere equation (\ref{QN2F_E}) is replaced by (\ref{S2F_E_ref}) is equivalent to the original one provided that the Ampere equation is satisfied at the initial time. Again, our AP-scheme for the two-fluid EM model will be consistent with the REM model.

\subsection{Discrete equations}
\label{subsec_2F_discrete}

We skip the step of the time-semi-discretization as it is similar as in the one-fluid case. The linearized stability analysis of the two-fluid model is left to future work. We provide the final spatio-temporal discretization in the one-dimensional setting for reference. The one-dimensional equations are not recalled. They are similar to the one-fluid case, but simply consist in a duplicate of the mass and momentum balance equations for each species, with the appropriate changes in the sign of the Lorentz force. The final discretization is as follows (the notations are the same as in the one-fluid case):
\begin{eqnarray}
& & \hspace{-1cm}  \delta^{-1} (n_i|_k^{m+1} - n_i|_k^m) + h^{-1} (f_{n_i}|_{k+1/2}^{m+a} - f_{n_i}|_{k-1/2}^{m+a})  = 0, \label{2F_ni_fd} \\
& & \hspace{-1cm} \delta^{-1}  ((n_i u_{ix})|_k^{m+1} - (n_i u_{ix})|_k^m) + h^{-1} (f_{u_{ix}}|_{k+1/2}^{m} - f_{u_{ix}}|_{k-1/2}^{m})  = \nonumber \\
& & \hspace{5cm}  =   n_i|_k^{m+1-a} \tilde E_x|_k^{m+1}  + (n_i u_{iy})|_k^m \,  \tilde B_z|_k^m, \label{2F_qxi_fd} \\
& & \hspace{-1cm} \delta^{-1}  ((n_i u_{iy})|_k^{m+1} - (n_i u_{iy})|_k^m) + h^{-1} (f_{u_{iy}}|_{k+1/2}^{m} - f_{u_{iy}}|_{k-1/2}^{m})  = \nonumber \\
& & \hspace{5cm}  =   n_i|_k^{m+1-a} E_y|_k^{m+1}  - (n_i u_{ix})|_k^m \, \tilde  B_z|_k^m, \label{2F_qyi_fd} \\
& & \hspace{-1cm}  \delta^{-1} (n_e|_k^{m+1} - n_e|_k^m) + h^{-1} (f_{n_e}|_{k+1/2}^{m+a} - f_{n_e}|_{k-1/2}^{m+a})  = 0, \label{2F_ne_fd} \\
& & \hspace{-1cm} \varepsilon^2 \delta^{-1}  ((n_e u_{ex})|_k^{m+1} - (n_e u_{ex})|_k^m) + h^{-1} (f_{u_{ex}}|_{k+1/2}^{m} - f_{u_{ex}}|_{k-1/2}^{m})  = \nonumber \\
& & \hspace{5cm}  =   - n_e|_k^{m+1-a} \tilde E_x|_k^{m+1}  -  (n_e u_{ey})|_k^m \,  \tilde B_z|_k^m, \label{2F_qxe_fd} \\
& & \hspace{-1cm} \varepsilon^2 \delta^{-1}  ((n_e u_{ey})|_k^{m+1} - (n_e u_{ey})|_k^m) + h^{-1} (f_{u_{ey}}|_{k+1/2}^{m} - f_{u_{ey}}|_{k-1/2}^{m})  = \nonumber \\
& & \hspace{5cm}  =   - n_e|_k^{m+1-a} E_y|_k^{m+1}  + (n_e u_{ex})|_k^m \, \tilde  B_z|_k^m, \label{2F_qye_fd} \\
& & \hspace{-1cm}  \delta^{-1} (B_z|_{k+1/2}^{m+1} - B_z|_{k+1/2}^m) + h^{-1} (E_y|_{k+1}^{m+b} - E_y|_{k}^{m+b}) = 0, \label{2F_faraday_fd} \\
& & \hspace{-1cm}  \lambda^2 \delta^{-1} (E_x|_{k+1/2}^{m+1} - E_x|_{k+1/2}^m) = - (f_{n_i}|_{k+1/2}^{m+a} - f_{n_e}|_{k+1/2}^{m+a}), \label{2F_amperex_fd} \\
& & \hspace{-1cm}  \lambda^2 \delta^{-1} (E_y|_k^{m+1} - E_y|_k^m) + h^{-1} (B_z|_{k+1/2}^{m+c} - B_z|_{k-1/2}^{m+c}) = \nonumber \\
& & \hspace{5cm} = - ( (n_i u_{iy})|_k^{m+a} - (n_e u_{ie})|_k^{m+a} ), \label{2F_amperey_fd}  
\end{eqnarray}
with $(a,b,c)=(0,0,1)$ (classical scheme) or $(a,b,c)=(1,1,1)$ (AP scheme) and where $\tilde B_z|_k^m$ and $\tilde E_x|_k^{m+1}$ are given by (\ref{interp_Bz_Ex}). 

In the case $(a,b,c)=(0,0,1)$ (classical scheme), the numerical fluxes are given by:
\begin{eqnarray}
& & \hspace{-1cm}  f_{n_i}|_{k+1/2}^{m} = \frac{1}{2} \left[ (n_i u_{ix})|_k^m + (n_i u_{ix})|_{k+1}^m + \mu_i|_{k+1/2}^m \, \left( n_i|_k^m - n_i|_{k+1}^m \right) \right], 
\label{2F_fni_fd} \\
& & \hspace{-1cm}  f_{u_{ix}}|_{k+1/2}^{m} = \frac{1}{2} \left[ \left(n_i u_{ix}^2 + p(n_i) \right)|_k^m + \left(n_i u_{ix}^2 + p(n_i \right)|_{k+1}^m + \right.\nonumber \\
& & \hspace{6cm} \left. + \mu_i|_{k+1/2}^m \, \left( (n_i u_{ix})|_k^m - (n_i u_{ix})|_{k+1}^m \right)  \right], 
\label{2F_fuxi_fd} \\
& & \hspace{-1cm}  f_{u_{iy}}|_{k+1/2}^{m} = \frac{1}{2} \left[ (n_i u_{ix} u_{iy})|_k^m + (n_i u_{ix} u_{iy})|_{k+1}^m + \right.\nonumber \\
& & \hspace{6cm} \left. + \mu_i|_{k+1/2}^m \, \left( (n_i u_{iy})|_k^m - (n_i u_{iy})|_{k+1}^m \right)  \right] , 
\label{2F_fuyi_fd} \\
& & \hspace{-1cm}  f_{n_e}|_{k+1/2}^{m} = \frac{1}{2} \left[ (n_e u_{ex})|_k^m + (n_e u_{ex})|_{k+1}^m + \mu_e|_{k+1/2}^m \, \left( n_e|_k^m - n_e|_{k+1}^m \right) \right], 
\label{2F_fne_fd} \\
& & \hspace{-1cm}  f_{u_{ex}}|_{k+1/2}^{m} = \frac{1}{2} \left[ (\varepsilon^2 n_e u_{ex}^2 + p(n_e) )|_k^m + (\varepsilon^2 n_e u_{ex}^2 + p(n_e) ) |_{k+1}^m + \right.\nonumber \\
& & \hspace{6cm} \left. + \mu_e|_{k+1/2}^m \, \left( (n_e u_{ex})|_k^m - (n_e u_{ex})|_{k+1}^m \right)  \right], 
\label{2F_fuxe_fd} \\
& & \hspace{-1cm}  f_{u_{ey}}|_{k+1/2}^{m} = \frac{1}{2} \left[ \varepsilon^2  ( (n_e u_{ex} u_{ey})|_k^m + (n_e u_{ex} u_{ey})|_{k+1}^m )  + \right.\nonumber \\
& & \hspace{6cm} \left. + \mu_e|_{k+1/2}^m \, \left( (n_e u_{ey})|_k^m - (n_e u_{ey})|_{k+1}^m \right)  \right] . 
\label{2F_fuye_fd} 
\end{eqnarray}
The numerical viscosities are computed separately for each species with the same method as in section \ref{subsec_1F_spatial}.  In the case $(a,b,c)=(1,1,1)$ (AP scheme) the momentum fluxes (\ref{2F_fuxi_fd}),  (\ref{2F_fuyi_fd}), (\ref{2F_fuxe_fd}),  (\ref{2F_fuye_fd}) are unchanged. Using the same assumptions as in section \ref{subsec_1F_spatial}, the implicit mass fluxes $\tilde f_{n_i}|_{k+1/2}^{m+1}$, $\tilde f_{n_e}|_{k+1/2}^{m+1}$ (where the tildes distinguish them from the explicit ones) are given by: 
\begin{eqnarray}
& & \hspace{-1cm}  \tilde f_{n_i}|_{k+1/2}^{m+1} = f_{n_i}|_{k+1/2}^{m} + \frac{\delta}{2} (n_i|_{k+1}^m + n_i|_{k}^m)  \, E_x|_{k+1/2}^{m+1}  \nonumber \\
& & \hspace{0cm} - \frac{\delta h^{-1}}{2} (f_{u_{ix}}|_{k+3/2}^{m} - f_{u_{ix}}|_{k-1/2}^{m}) + \frac{\delta}{2} ( (n_i u_{iy})|_k^m  + (n_i u_{iy})|_{k+1}^m ) \, B_z|_{k+1/2}^m  
. \label{2F_fni_m+1_fd2_mod} \\
& & \hspace{-1cm}  \tilde f_{n_e}|_{k+1/2}^{m+1} = f_{n_e}|_{k+1/2}^{m} - \frac{\delta}{2 \varepsilon^2} (n_e|_{k+1}^m + n_e|_{k}^m)  \, E_x|_{k+1/2}^{m+1}  \nonumber \\
& & \hspace{0cm} - \frac{\delta h^{-1}}{2 \varepsilon^2} (f_{u_{ex}}|_{k+3/2}^{m} - f_{u_{ex}}|_{k-1/2}^{m}) - \frac{\delta}{2 \varepsilon^2} ( (n_e u_{ey})|_k^m  + (n_e u_{ey})|_{k+1}^m ) \, B_z|_{k+1/2}^m  
. \label{2F_fne_m+1_fd2_mod} 
\end{eqnarray}
With the same computation as in section \ref{subsec_1F_spatial}, we find that this scheme satisfies the discrete Gauss equation exactly: 
\begin{eqnarray}
& & \hspace{-1cm}  \lambda^2  h^{-1} (E_x|_{k+1/2}^{m+1} - E_x|_{k-1/2}^{m+1}) - ( n_i|_{k}^{m+1} - n_e|_{k}^{m+1})  = \nonumber \\
& & \hspace{4cm} = \lambda^2  h^{-1} (E_x|_{k+1/2}^{m} - E_x|_{k-1/2}^{m}) - ( n_i|_{k}^{m} - n_e|_{k}^{m}).  \label{2F_gauss_fd}
\end{eqnarray}

The sequence of updates for the classical scheme $(a,b,c)=(0,0,1)$ is a simple generalization of section \ref{subsec_1F_class_time_update}. For the AP-scheme $(a,b,c)=(1,1,1)$, we first realize that (\ref{2F_amperey_fd}) is equivalent to
\begin{eqnarray}
& & \hspace{-1cm}  (\lambda^2  + \delta^{2} (n_i|_k^m + \varepsilon^{-2} n_e|_k^m)) E_y|_k^{m+1} - \delta^{2} h^{-2} (E_y|_{k+1}^{m+1} - 2 E_y|_k^{m+1} + E_y|_{k-1}^{m+1}) = \lambda^2 E_y|_k^m  \nonumber \\
& & \hspace{1cm}  - \delta ( (n_i u_{iy})|_k^m - (n_e u_{ey})|_k^m )  - \delta h^{-1} (B_z|_{k+1/2}^{m} - B_z|_{k-1/2}^{m}) + \nonumber \\
& & \hspace{1cm}   + \delta^{2} h^{-1}  ( (f_{u_{iy}}|_{k+1/2}^{m} - \varepsilon^{-2} f_{u_{ey}}|_{k+1/2}^{m} ) - (f_{u_{iy}}|_{k-1/2}^{m} - \varepsilon^{-2} f_{u_{ey}}|_{k-1/2}^{m} )) + \nonumber \\
& & \hspace{1cm} + \delta^{2} ((n_i u_{ix})|_k^m + \varepsilon^{-2} (n_e u_{ex})|_k^m)\, \tilde  B_z|_k^m
, \label{2F_amperey_ref_fd}  
\end{eqnarray}
or, using (\ref{1F_amperey_fd}) again, between time steps $t^{m-1}$ and $t^m$, to
\begin{eqnarray}
& & \hspace{-1cm}  (\lambda^2  + \delta^{2} (n_i|_k^m + \varepsilon^{-2} n_e|_k^m)) E_y|_k^{m+1} - \delta^{2} h^{-2} (E_y|_{k+1}^{m+1} - 2 E_y|_k^{m+1} + E_y|_{k-1}^{m+1}) = \nonumber \\
& & \hspace{1cm} = \lambda^2 (2 E_y|_k^m - E_y|_k^{m-1})  + \nonumber \\
& & \hspace{1cm}   + \delta^{2} h^{-1}  ( (f_{u_{iy}}|_{k+1/2}^{m} - \varepsilon^{-2} f_{u_{ey}}|_{k+1/2}^{m} ) - (f_{u_{iy}}|_{k-1/2}^{m} - \varepsilon^{-2} f_{u_{ey}}|_{k-1/2}^{m} )) + \nonumber \\
& & \hspace{1cm} + \delta^{2} ((n_i u_{ix})|_k^m + \varepsilon^{-2} (n_e u_{ex})|_k^m)\, \tilde  B_z|_k^m
. \label{2F_amperey_ref2_fd}  
\end{eqnarray}
This equation is clearly consistent with the reformulated Ampere eq. (\ref{S2F_E_ref}). Both (\ref{2F_amperey_ref_fd}) and (\ref{2F_amperey_ref2_fd}) are discrete elliptic equations for $E_y|_k^{m+1}$ which are well-posed provided that suitable boundary conditions are defined. Similarly, (\ref{2F_amperex_fd}) is equivalent to: 
\begin{eqnarray}
& & \hspace{-1cm}  ( \lambda^2  + \frac{\delta^2}{2} ((n_i|_{k+1}^m + \varepsilon^{-2} n_e|_{k+1}^m) + (n_i|_k^m + \varepsilon^{-2} n_e|_k^m)) )  \, E_x|_{k+1/2}^{m+1}  = \nonumber \\
& & \hspace{0cm} = \lambda^2 E_x|_{k+1/2}^m - \delta ( f_{n_i}|_{k+1/2}^{m} - f_{n_e}|_{k+1/2}^{m} ) \nonumber \\
& & \hspace{0cm} + \frac{\delta^2 h^{-1}}{2} ( (f_{u_{ix}}|_{k+3/2}^{m} - \varepsilon^{-2} f_{u_{ex}}|_{k+3/2}^{m}) - (f_{u_{ix}}|_{k-1/2}^{m} - \varepsilon^{-2} f_{u_{ex}}|_{k-1/2}^{m}) ) \nonumber \\
& & \hspace{0cm} - \frac{\delta^2 }{2} ( ( (n_i u_{iy})|_k^m  + \varepsilon^{-2} (n_e u_{ey})|_k^m ) + ( (n_i u_{iy})|_{k+1}^m  + \varepsilon^{-2} (n_e u_{ey})|_{k+1}^m ) ) \, B_z|_{k+1/2}^m . 
\label{2F_amperex_fd_mod} 
\end{eqnarray}
The remaining updates are processed in a similar way as section \ref{subsec_1F_class_time_update}. The proof that this scheme is AP, i.e. consistent with the one-dimensional version of the QN-EM model, is left to the reader.

\setcounter{equation}{0}
\section{Numerical results}
\label{EM:sec_num}


In this section, we provide a numerical validation of the AP-methodology. We will consider two different one-dimensional test problems. The first one is a simple Riemann problem, where the initial condition is piecewise constant with a discontinuity at the origin. Two different initial conditions will be used, respectively giving rise to shock and rarefaction waves. 
The second test problem corresponds to a more realistic physical situation: it is a one-dimensional model for a Plasma Opening Switch (POS) device. Both tests will be run in the one-fluid and two-fluid cases. We will see that, while the classical scheme develops instabilities in under-resolved situations (when the time or space steps are bigger than the finest time or space scales), the AP-methodology provides a consistent approximation of the solution of the limit quasineutral model. We will also show that in resolved situations, both the classical and AP-schemes have optimal order, i.e. $1/2$ for discontinuous solutions (in the $L^1$ norm) and $1$ for smooth solutions.

\subsection{Riemann problem}
\label{EM:subsec_riemann}

The most general initial conditions for the Riemann problem are given by:
\begin{equation}
n = \left\{ \begin{array}{lll} n_l, & \mbox{ if } & x \leq 0, \\
n_r, & \mbox{ if } & x \geq 0, \end{array} \right. \hspace{1cm}
u_x = \left\{ \begin{array}{lll} u_{xl}, & \mbox{ if } & x \leq 0, \\
u_{xr}, & \mbox{ if } & x \geq 0, \end{array} \right.
\label{eq_ini_riemann}
\end{equation}
where $n_l \not = n_r$ and $u_{xl} \not = u_{xr}$. In the two-fluid case, initial conditions like (\ref{eq_ini_riemann}) are prescribed for $n_e$, $n_i$, $u_{ex}$, $u_{ix}$. In our examples though, we will make $n_l = n_r = 1$ (in dimensionless units) and assume that the initial discontinuity applies only to the velocity with $u_{xl} = -u_{xr}$. Indeed, in this configuration and in the absence of coupling with the electromagnetic field, the solution is particularly simple. Since there is no analytical solution of the system when the coupling with the electromagnetic field is turned on, it is easier to qualitatively interpret the results if the solution without coupling is simple. 

Indeed, in the absence of coupling, and if $u_{xl} > 0$, i.e. if the initial velocity configuration is towards a compression of the fluid, two outgoing shock waves starting at the origin propagate in opposite directions at the same speed and encompass a region of higher density at rest (i.e. with zero velocity). If, on the other hand, $u_{xl} < 0$, i.e. if the initial velocity configuration is that of an expansion, two outgoing rarefaction waves starting at the origin propagate in opposite directions at the same speed and encompass a region of lower density at rest. 

When turning on the electromagnetic field, we will consider two situations. In the first one, the initial values of $E_y$, $B_z$ and $u_y$ are identically zero. Then, they identically vanish at all times and the quantities of interest are $n$, $u_x$ and $E_x$. In the second one, we suppose that the initial $B_z$ is non-zero and uniform. In this case, non-zero values of $E_y$ and $u_y$ are generated. 

In the one-fluid case and in the quasi-neutral limit $\lambda = 0$, the solution corresponds to a fluid at rest (i.e. $u_x = 0$) with uniform density $n=1$. Then, the  behavior of the scheme in the quasi-neutral limit can be compared to this analytical solution. As $\lambda$ decreases, the numerical solution should get closer and closer to this analytical solution.

In the forthcoming simulation, the computational domain is chosen to be $ [ -0.1;0.1] $ and in the two-fluid case, the electron to ion mass ratio is taken to be  $\varepsilon^2 =  10^{-4} $.

\subsubsection{One-fluid outgoing shock waves; zero initial magnetic field}
\label{subsubsec_num_onefluid_shock}

\begin{figure}[hbtp]
 \begin{minipage}[c]{.46\linewidth}
 \includegraphics[scale=0.4]{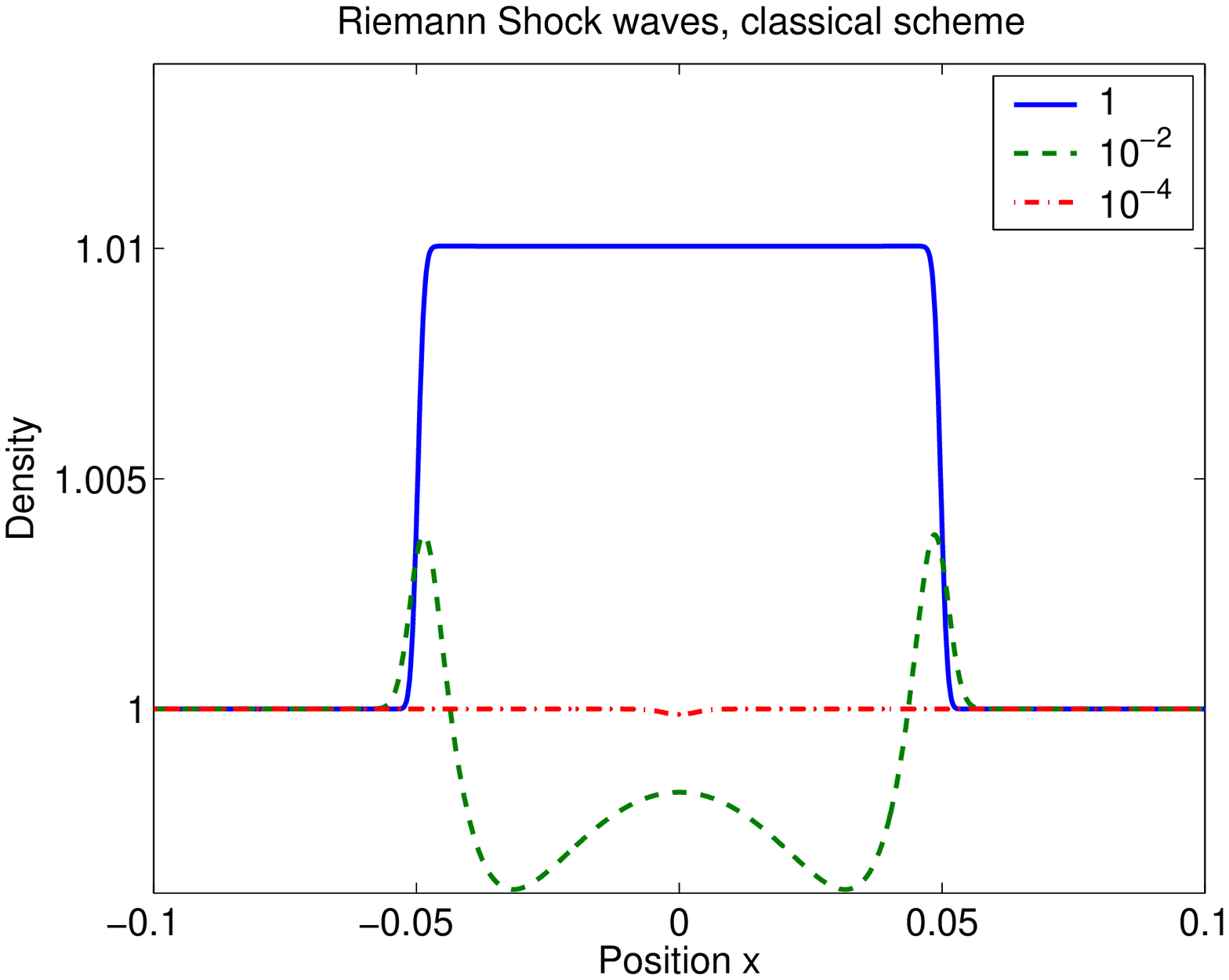}
 \end{minipage}
 \begin{minipage}[c]{.46\linewidth}
  \includegraphics[scale=0.4]{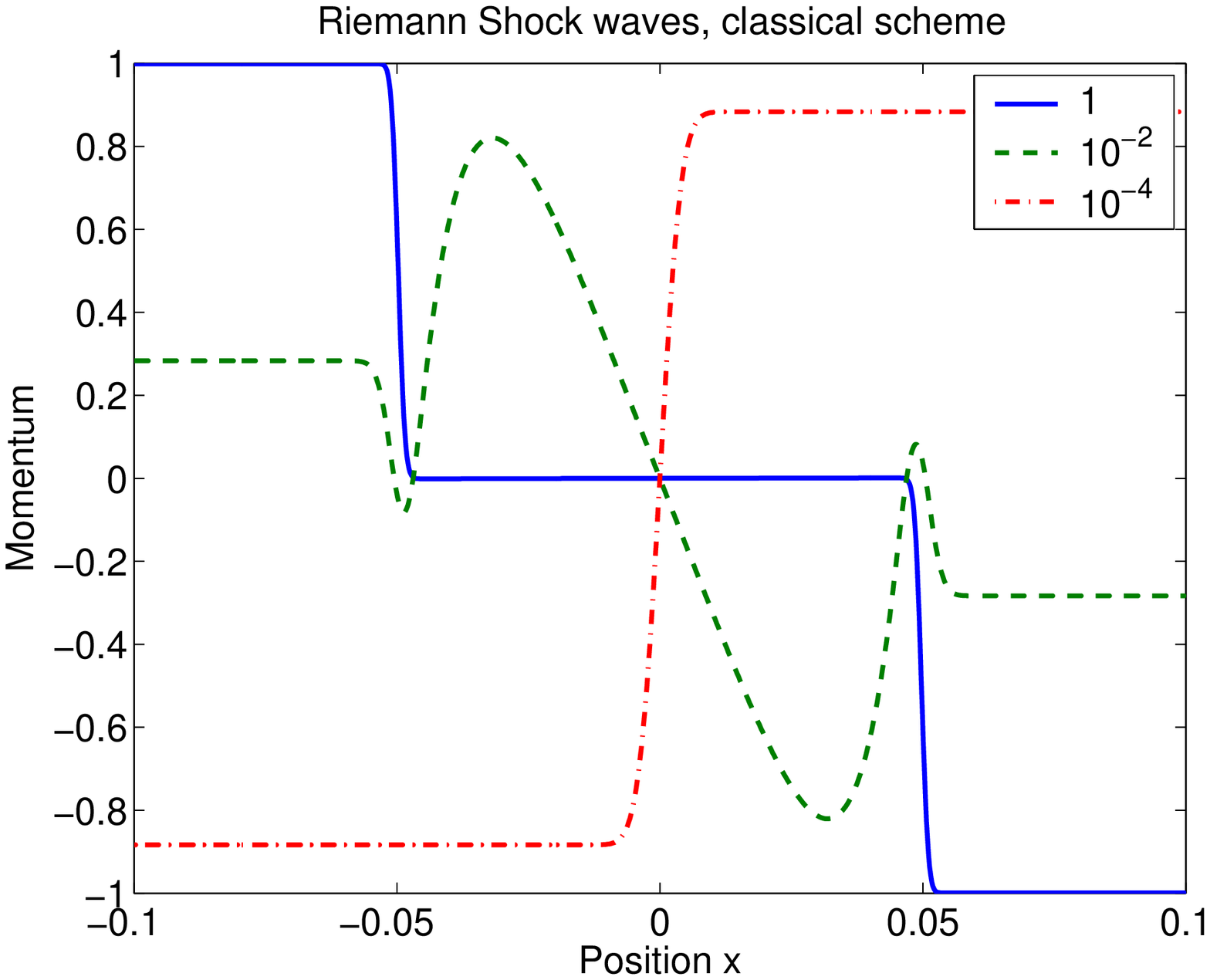}
 \end{minipage}
 \caption{\label{chocs_mono_clas_1000_024} One-fluid shock wave test case with zero initial magnetic field. $n$ (left panel) and $nu_x$ (right panel) as functions of $x$ at time $ t = 5 \times 10^{-4} $ with $ \lambda =1 $, $ \lambda = 10^{-2} $ and $ \lambda = 10^{-4} $ computed with the classical scheme on $N_x= 10^{3} $ space cells.
}
\end{figure}

\begin{figure}[hbtp]
 \begin{minipage}[c]{.46\linewidth}
 \includegraphics[scale=0.4]{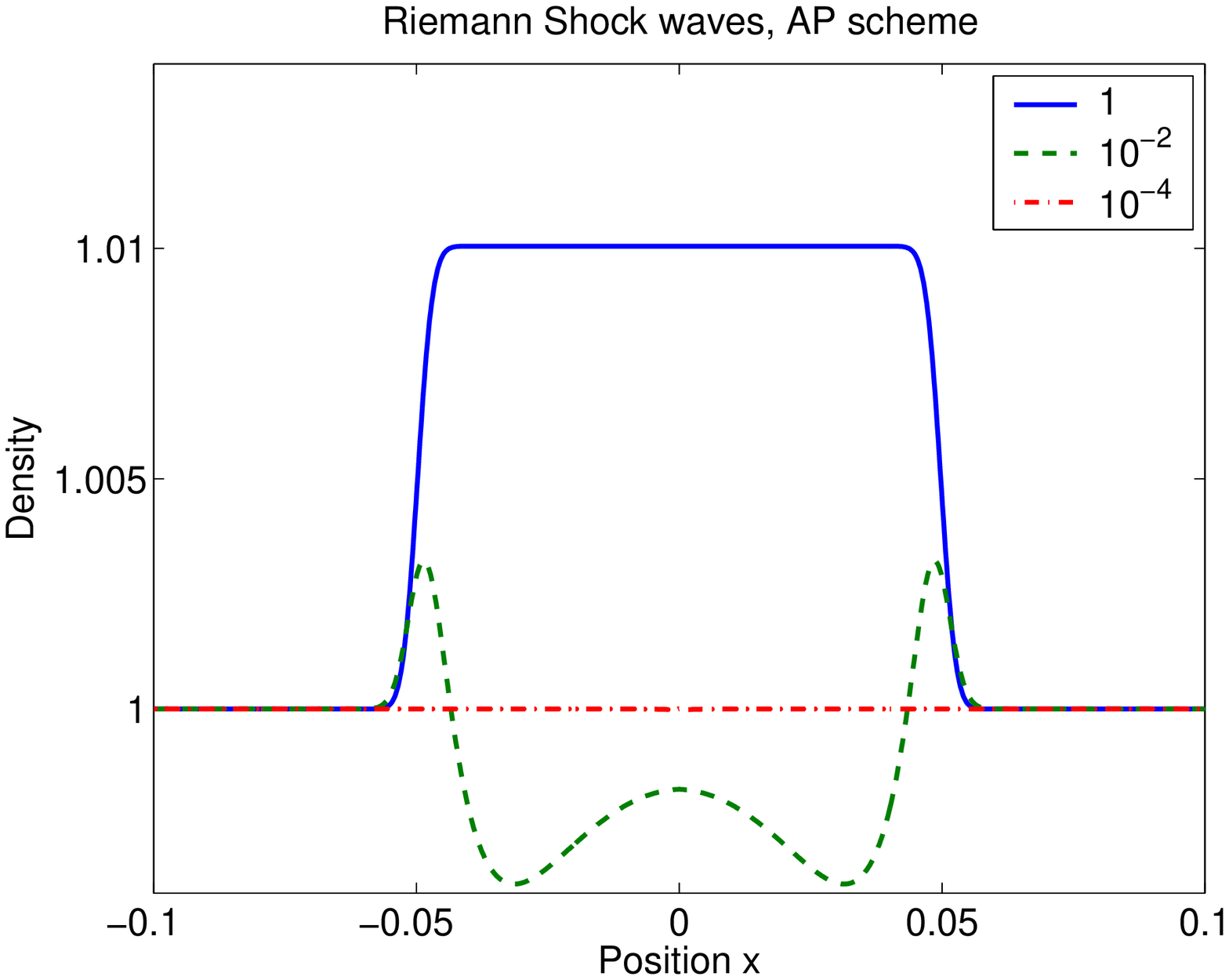}
 \end{minipage}
 \begin{minipage}[c]{.46\linewidth}
  \includegraphics[scale=0.4]{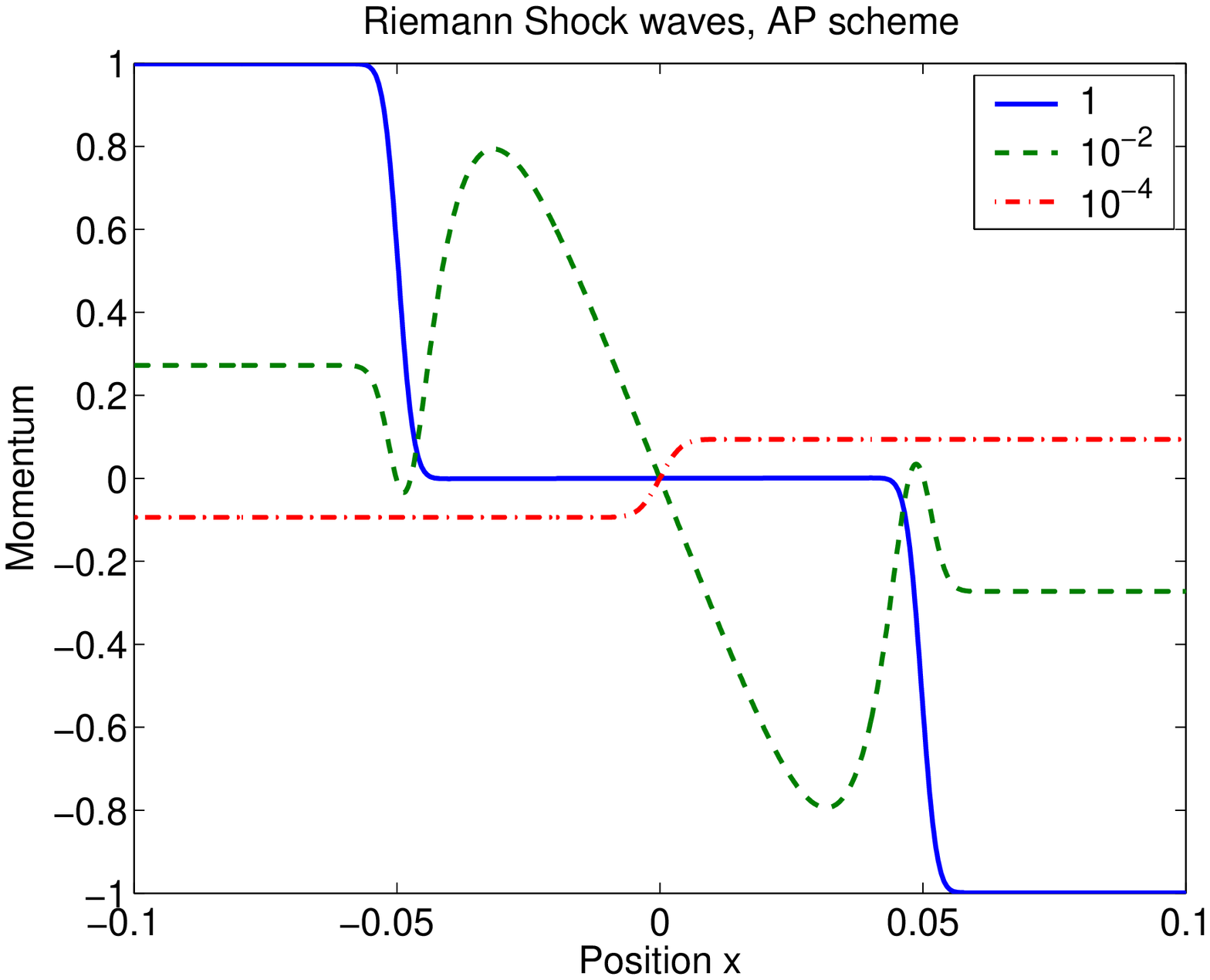}
 \end{minipage}
 \caption{\label{chocs_mono_AP_1000_024} One-fluid shock wave test case with zero initial magnetic field. $n$ (left panel) and $nu_x$ (right panel) as functions of $x$ at time $ t = 5 \times 10^{-4} $ with $ \lambda =1 $, $ \lambda = 10^{-2} $ and $ \lambda = 10^{-4} $ computed with the AP-scheme on $N_x= 10^{3} $ space cells.
}
\end{figure}

In this test case, the initial velocities are $ u_{L} = +1 $ and $ u_{R} = -1 $. We first investigate how the schemes behave as the coupling with the electromagnetic field is turned on, i.e. as $\lambda$ is gradually decreased. Figs. \ref{chocs_mono_clas_1000_024} and \ref{chocs_mono_AP_1000_024} shows how the classical and AP schemes behave when $\lambda$ successively takes the values $\lambda =1$, $\lambda =10^{-2}$ and $\lambda =10^{-4}$. These figures display the density $n$ and momentum $q_x$ as functions of space at a given time $t=5 \times 10^{-4}$ for the classical scheme (Fig. \ref{chocs_mono_clas_1000_024}) and for the AP-scheme (Fig. \ref{chocs_mono_AP_1000_024}). We observe that, when $\lambda = 1$, the coupling is weak and the solution is close to that of the Euler equations with zero Lorentz force. On the other hand, if $ \lambda =10^{-4}$, the solution is close to the corresponding quasi-neutral limit, i.e. $n=1$ and $u_x = 0$. If $\lambda =10^{-2}$, the Debye length is in an intermediate regime and the solution lies in between these two extremes. When $\lambda$ is small, the boundary values of the momentum are different from those of the initial conditions. This is because a very fast wave has crossed the domain and has changed the boundary values of the momentum. This change is allowed by the Neumann boundary conditions which are imposed on the fluid quantities at the domain boundaries. Let us now compare the magnitudes of the momentum between the final and initial times for $\lambda = 10^{-4}$. For the classical scheme, these magnitudes are of the same order of magnitude (Fig. \ref{chocs_mono_clas_1000_024}), whereas for the AP scheme the magnitude at the final time is very small compared to that at the initial time (Fig. \ref{chocs_mono_AP_1000_024}). Therefore, the behavior of the AP scheme is consistent with the quasi-neutral limit while that of the classical scheme is not. The AP scheme therefore ensures a correct transition from the Euler shock to the quasi-neutral fluid when $\lambda$ decreases. 

\begin{figure}[hbtp]
 \begin{minipage}[c]{.46\linewidth}
 \includegraphics[scale=0.4]{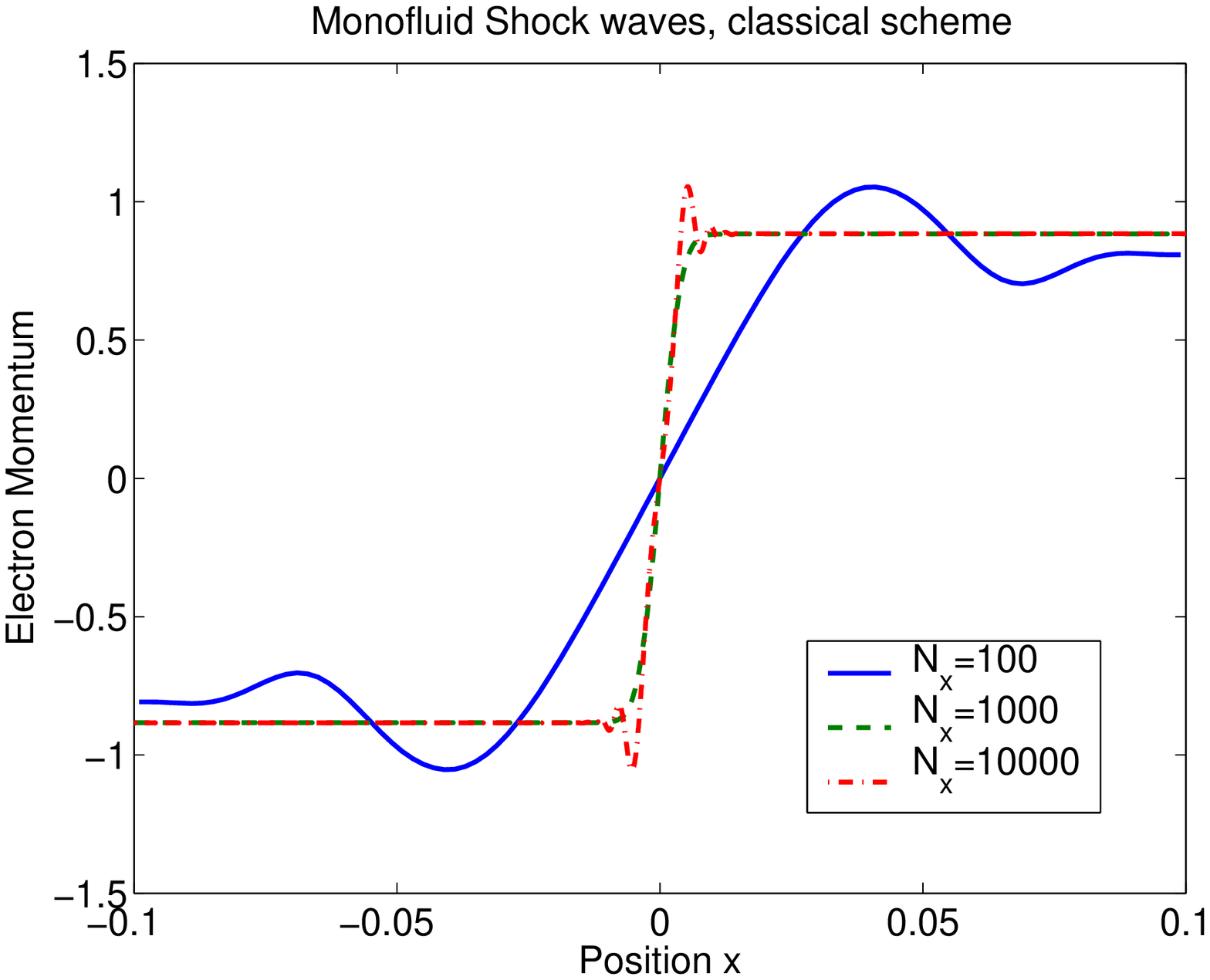}
 \end{minipage}
 \begin{minipage}[c]{.46\linewidth}
  \includegraphics[scale=0.4]{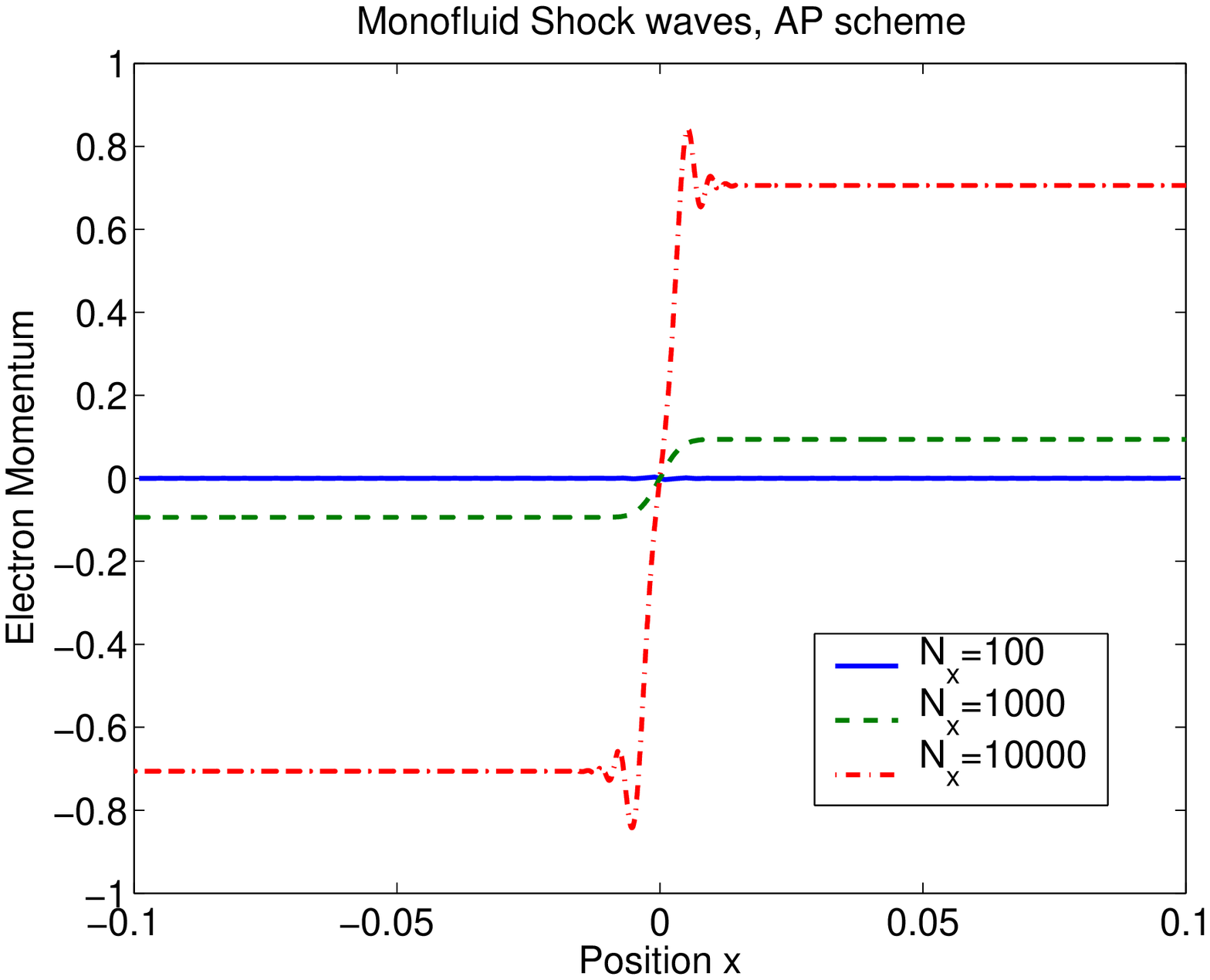}
 \end{minipage}
 \caption{\label{chocs_mono_multx_04_momx} One-fluid shock wave test case with zero initial magnetic field. $n u_x$ as a function of $x$ at time $ t = 5 \times 10^{-4} $ with  $ \lambda = 10^{-4} $  for the classical scheme (left panel) and AP-scheme (right panel).
$\Delta x / \lambda$ respectively equals  $ 0.2 $, $ 2 $ and $ 20 $ for $ N_{x} = 10000 $, $ N_{x} = 1000 $ and  $ N_{x} = 100 $ discretization cells.
}
\end{figure}

\begin{figure}[hbtp]
 \begin{minipage}[c]{.46\linewidth}
 \includegraphics[scale=0.4]{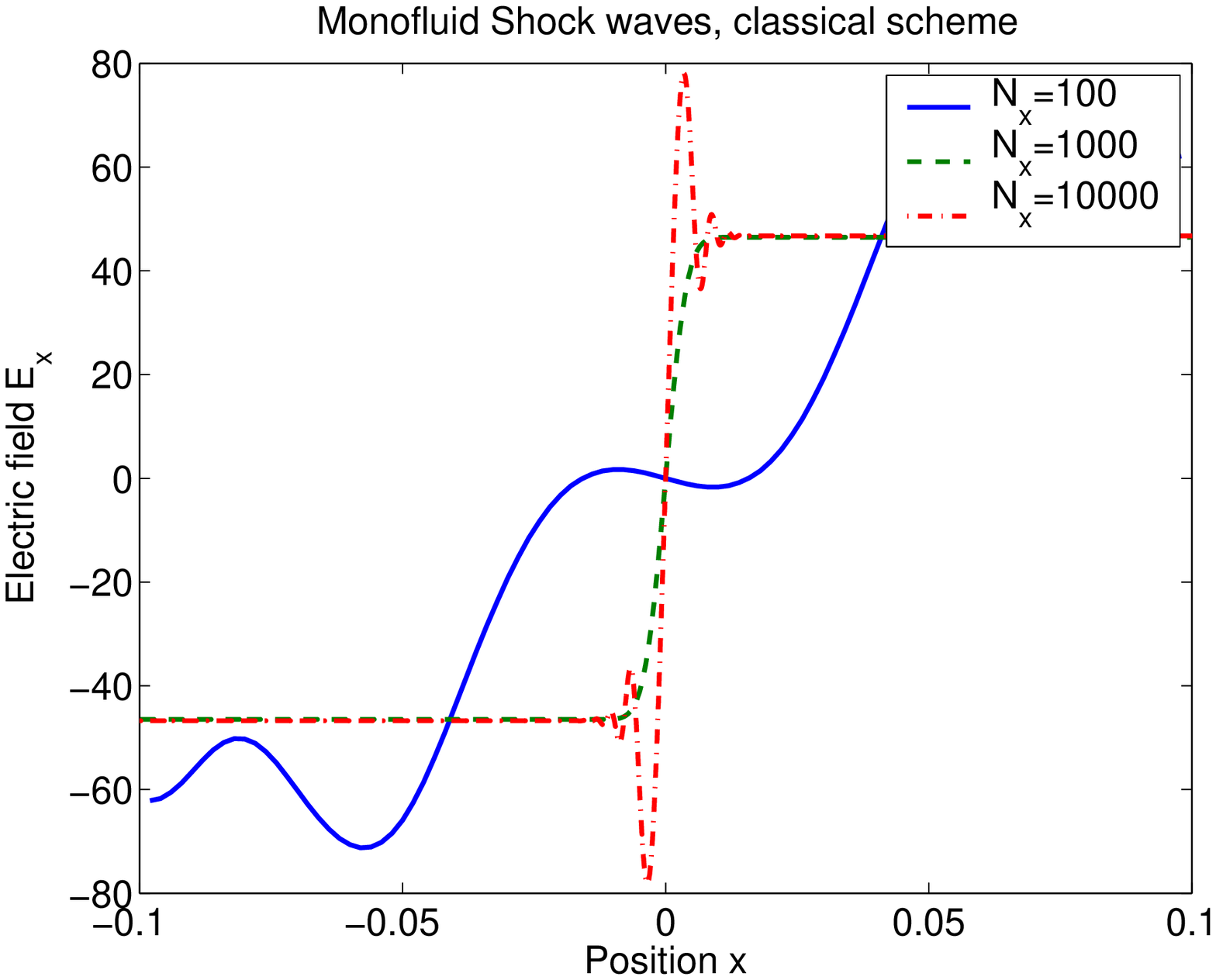}
 \end{minipage}
 \begin{minipage}[c]{.46\linewidth}
  \includegraphics[scale=0.4]{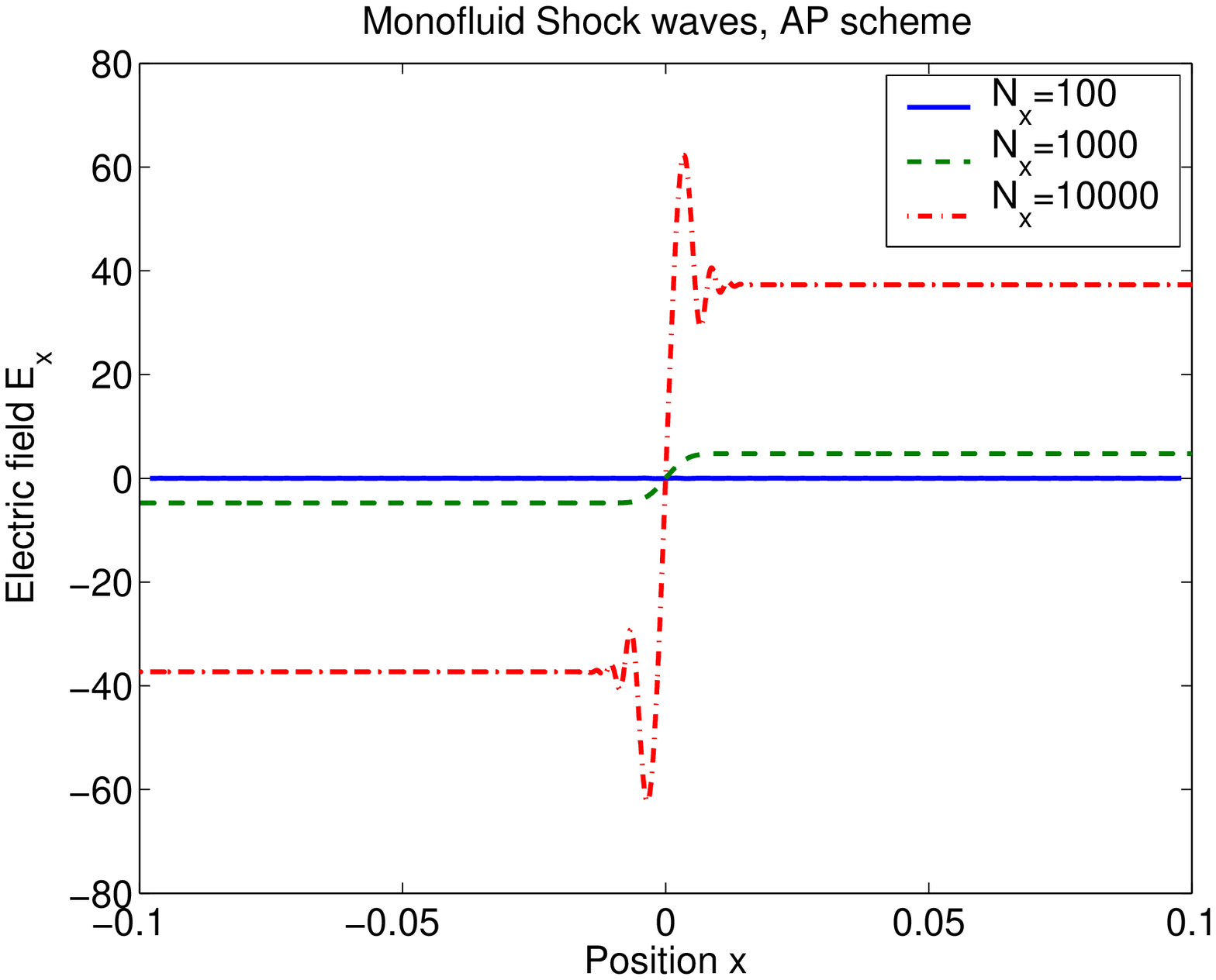}
 \end{minipage}
 \caption{\label{chocs_mono_multx_04_Ex} One-fluid shock wave test case with zero initial magnetic field. $E_x$ as a function of $x$ at time $ t = 5 \times 10^{-4}$ with $ \lambda = 10^{-4} $ for the classical scheme (left panel) and AP-scheme (right panel).
$\Delta x / \lambda$ respectively equals $ 0.2 $, $ 2 $ and $ 20 $ for $ N_{x} = 10000 $, $ N_{x} = 1000 $ and  $ N_{x} = 100 $ discretization cells.
}
\end{figure}

Another way to highlight the consistency of the AP-scheme with the quasi-neutral limit and the corresponding inconsistency of the classical scheme is to investigate how the results depend on the ratio $\Delta x / \lambda$ of  the space step to the Debye length. Figs \ref{chocs_mono_multx_04_momx} and \ref{chocs_mono_multx_04_Ex} display the momentum and electric field (respectively) as functions of $x$ at the same final time as before. The left and right-hand pictures correspond to the classical and AP- schemes respectively. The value of $\lambda$ is kept fixed at $\lambda = 10^{-4}$ but the number of discretization points is decreased from $N_x = 10^4$ to $N_x = 10^3$ and finally $N_x = 10^2$, leading to correspondingly increasing ratios $\Delta x / \lambda = 0.2$, $2$ and $20$ respectively. On the pictures, we observe that the AP-scheme provides a neat transition from a shock wave solution for $\Delta x / \lambda <1$ to the quasi-neutral uniform solution for $\Delta x / \lambda \gg 1$. At variance, the classical scheme provides large magnitude momenta or electric fields, in contradiction to the quasi-neutral solution. However, these solutions are not correct solutions of the problem with finite $\lambda$ either, since the wave number of the oscillations of the solutions have nothing to do with those obtained in the resolved situation $\Delta x / \lambda = 0.2$. Therefore, in the under-resolved situation, the classical scheme  is neither good for the problem with finite $\lambda$ nor for the quasi-neutral limit.

\begin{figure}[hbtp]
 \begin{minipage}[c]{.46\linewidth}
 \includegraphics[scale=0.4]{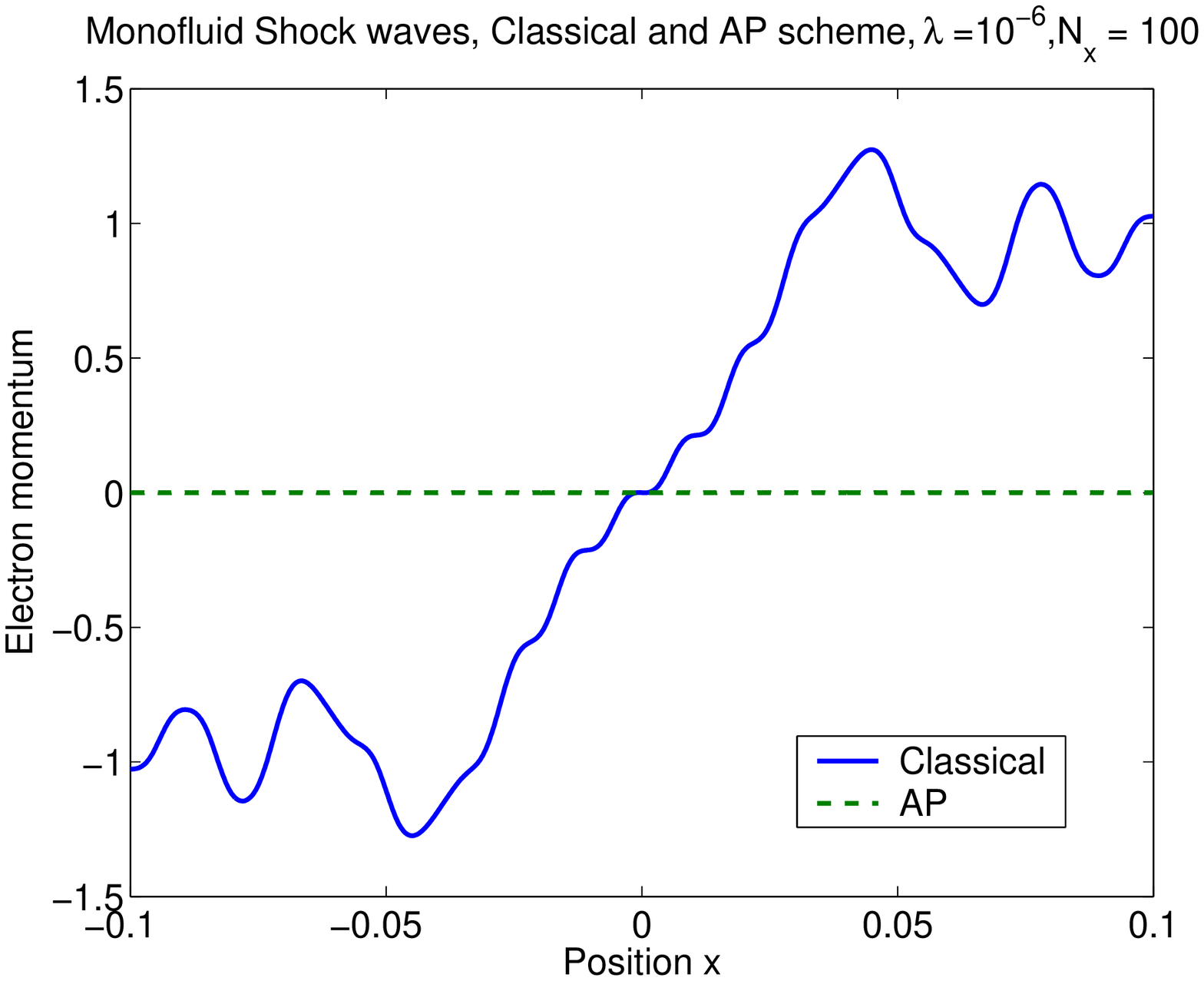}
 \end{minipage}
 \begin{minipage}[c]{.46\linewidth}
  \includegraphics[scale=0.4]{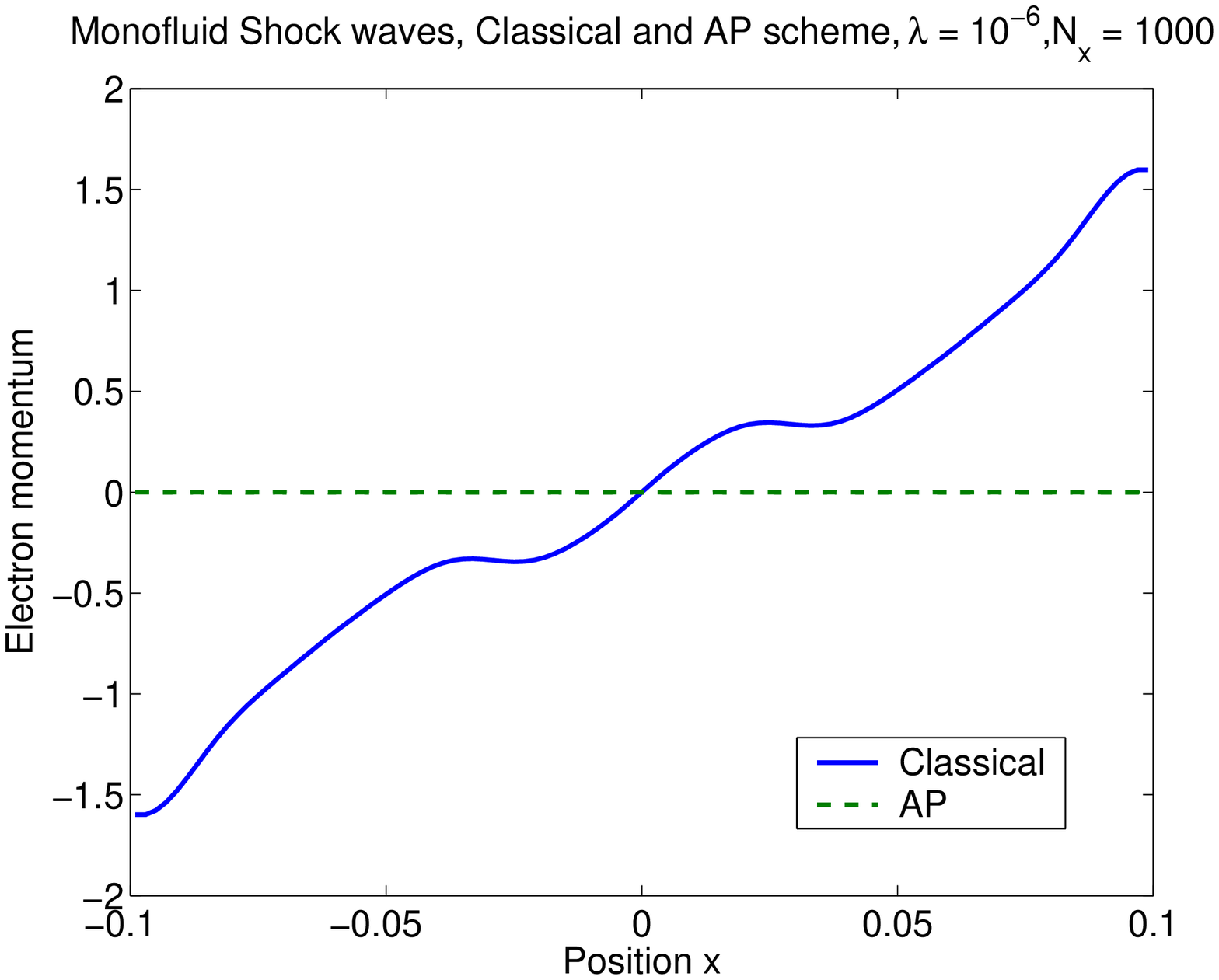}
 \end{minipage}
 \caption{\label{chocs_mono_multx_06} One-fluid shock wave test case with zero initial magnetic field. $nu_x$ as a function of $x$ at time $ t = 5 \times 10^{-4} $ with  $ \lambda = 10^{-6} $ for both the classical and AP schemes. $\Delta x / \lambda$ respectively equals $ 2000 $ and $ 200 $ for $ N_{x} = 100 $ cells (left panel) and  $ N_{x} = 1000 $ cells (right panel). }
\end{figure}

Fig.  \ref{chocs_mono_multx_06} displays the electron momentum as a function of $x$ at the time $t=5 \times 10^{-4}$ in the case $ \lambda = 10^{-6} $, for $N_x = 100$ (left figure) and $N_x = 1000$ (right figure) discretization points and for both the classical and AP- schemes. We see that, for this value of $\lambda$, the momentum computed by the AP scheme is identically zero for both choices of space discretization $\Delta x$, while that computed by the classical scheme keeps an $O(1)$ magnitude. In these cases, $\Delta x / \lambda$ have values respectively equal to $2 10^2$ and $2 10^3$, which shows the ability of the AP-scheme to handle extremely under-resolved situations.

\begin{figure}[hbtp]
 \begin{minipage}[c]{.46\linewidth}
 \psfrag{varepsilon ne}{$\varepsilon(n_e)$}
 \psfrag{Delta x}{$\Delta x$}
 \includegraphics[scale=0.4]{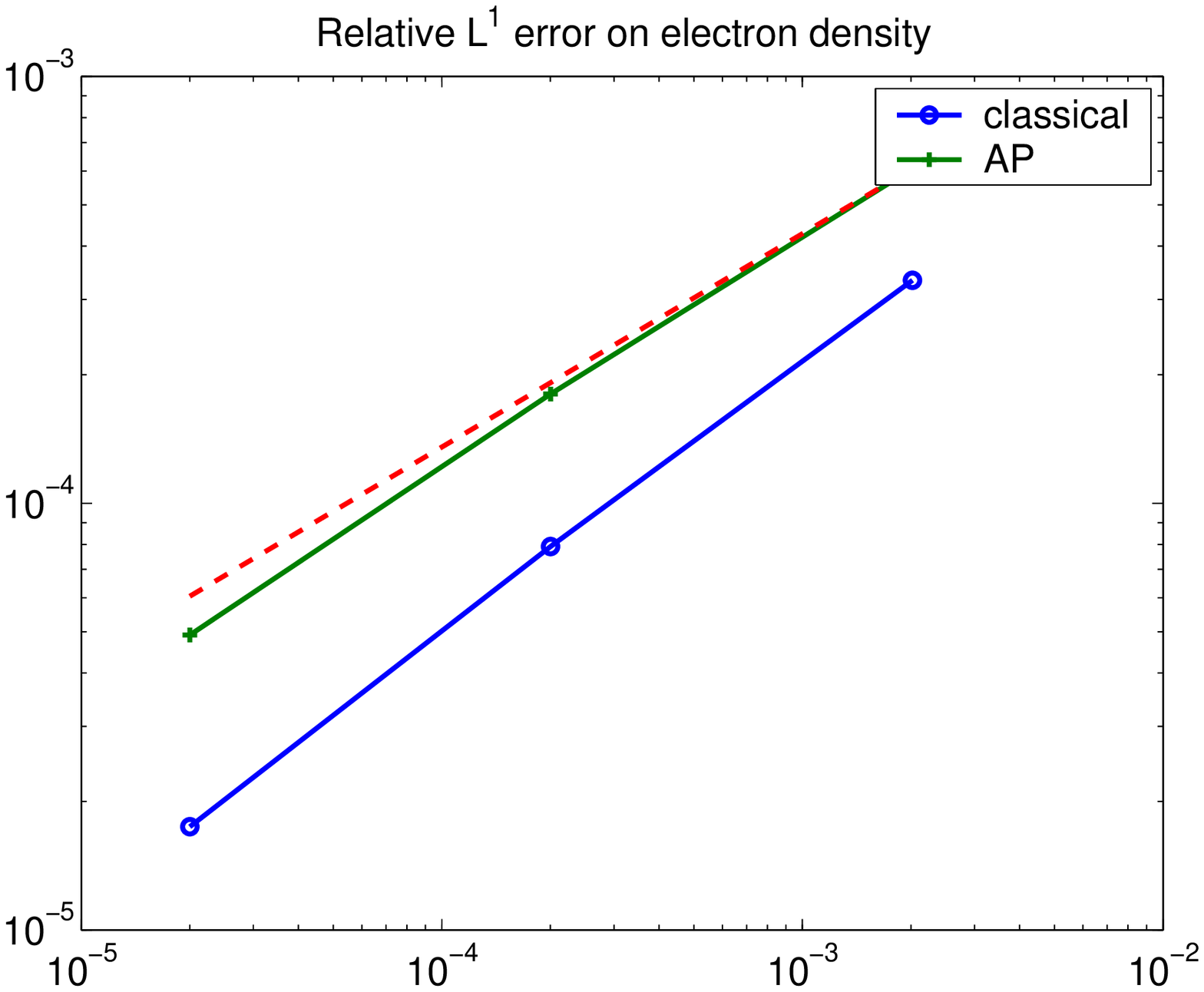}
 \end{minipage}
 \begin{minipage}[c]{.46\linewidth}
 \psfrag{varepsilon ni}{$\varepsilon(n_i)$}
 \psfrag{Delta x}{$\Delta x$}
  \includegraphics[scale=0.4]{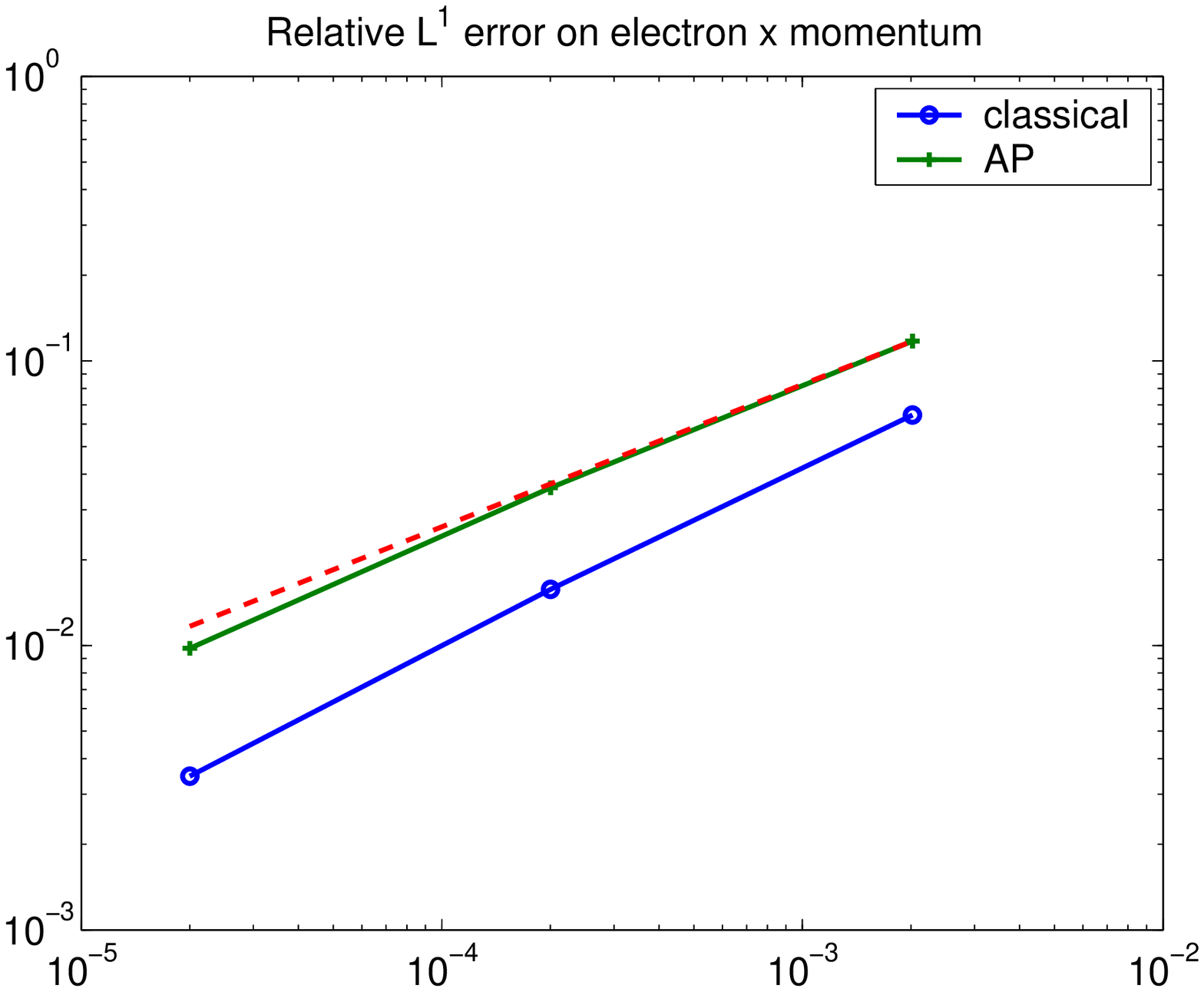}
 \end{minipage}
 \caption{\label{chocs_mono_error_dens_L1} One-fluid shock wave test case with zero initial magnetic field. Relative $L^1$ errors on $n$ (left panel) and $n u_x$ (right panel) as functions of $\Delta x$ at time $t=5\times 10^{-4}$ with $ \lambda=1$ for the classical and AP- schemes. The dashed line represents the theoretical error, with a slope equal to $ \sqrt{\Delta x}$.}
\end{figure}

For Debye lengths $ \lambda = 1 $ and $ \lambda = 10^{-2} $, an approximate reference solution can be computed on a grid such that $ \Delta_{x}^{\text{ref}} < \lambda $.
The grid to compute this reference solution is made of $ 10^{5} $ cells. This grid would be suitable to compute a reference solution for the case $ \lambda = 10^{-4} $, but it cannot be done at reasonable computational cost because the Courant-Friedrichs-Levy condition on the Maxwell equations requires too small time steps. Indeed, 
the reference solution is computed with the classical scheme. This computation is accurate since all physical space and time scales are resolved by the space and time steps. The reference solution can be used to perform a numerical convergence study for the cases $ \lambda = 1 $ and $ \lambda = 10^{-2} $. We compute relative errors in the $L^1$ norm. For instance, the density error is defined by:
$$ \varepsilon_{clas}(n) = \frac{ \max || n_{\text{num}} - n_{\text{ref}} ||_{1} }{ || n_{\text{ref}} ||_{1} },
$$
where $ n_{\text{ref}} $ is the density of the reference solution and $ n_{\text{num}} $ is the density of the approximate solution to test.
The $ L^{1} $ norm is chosen because of the discontinuities involved in the solution of the Riemann problem. It is shown in the literature that the best convergence rate for the numerical approximation of discontinuous solutions of conservation laws is obtained in the $L^1$ norm and that the corresponding order is $1/2$, i.e. $O( \sqrt{\Delta x}) $. Such error indicators are applied to both the classical and AP schemes, and for the $x$-components of the momentum and electric field. These relative errors are plotted in Fig. \ref{chocs_mono_error_dens_L1} as functions of $\Delta x$. We can see that the classical scheme is slightly more precise than the AP-scheme, and that both verify the theoretical order of convergence of $O( \sqrt{\Delta x}) $. Indeed, the slope of the error curve is compared to a straight line of slope $1/2$ and the match is almost perfect. This shows that the AP-scheme is consistent with the problem with finite $\lambda$ in the resolved case. 

The studies performed on this particular test case confirm that the AP-scheme is consistent with the quasi-neutral solution in under-resolved situations, and with the problem with finite $\lambda$ in the resolved situation, as an AP-scheme should do. They also show that the classical scheme does not capture the correct quasi-neutral regime in under-resolved situations. We will now confirm these trends in the forthcoming test problems. 

\subsubsection{One-fluid outgoing rarefaction waves; zero initial magnetic field}
\label{subsubsec_num_onefluid_rarefaction}

\begin{figure}[hbtp]
 \begin{minipage}[c]{.46\linewidth}
 \includegraphics[scale=0.4]{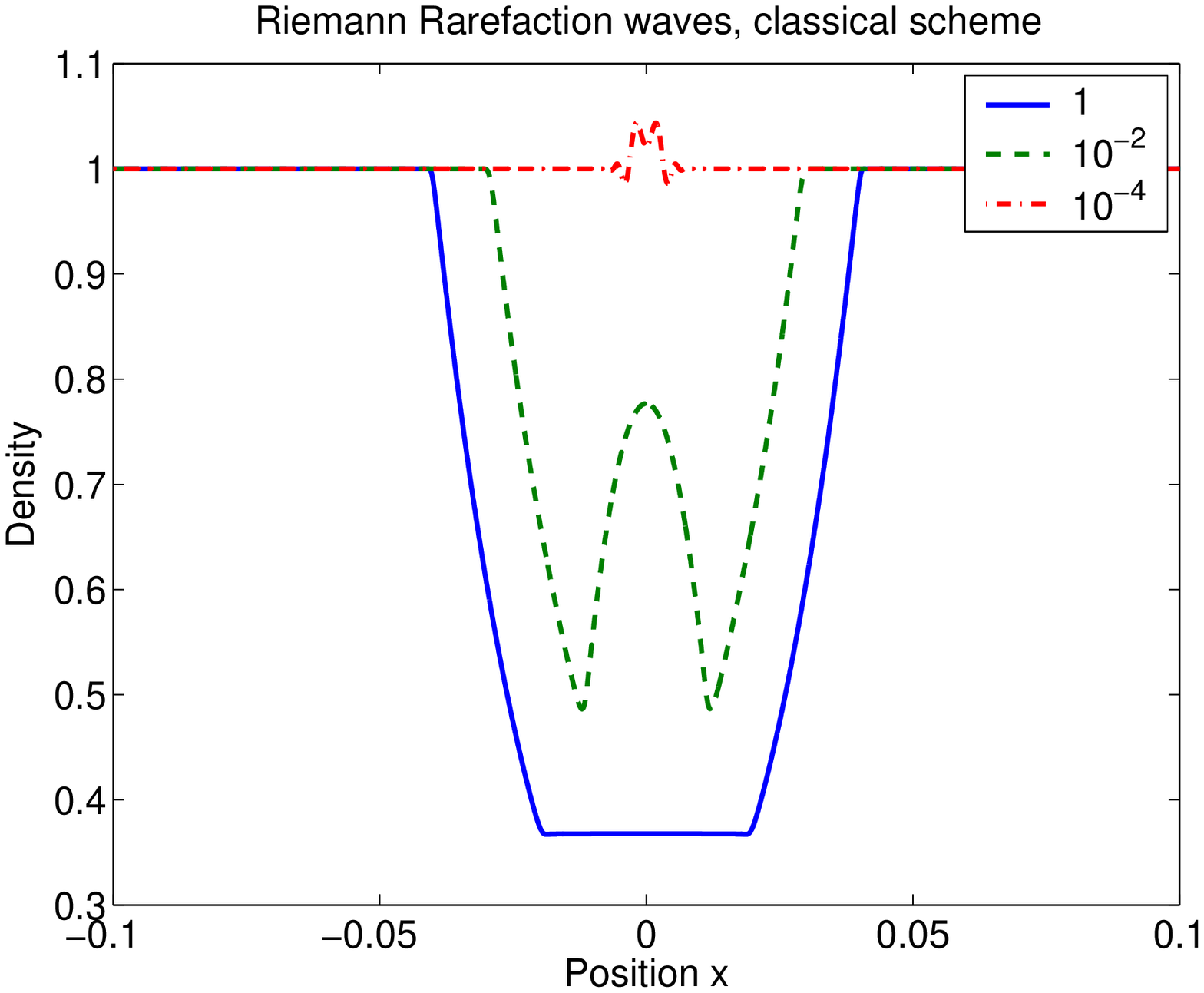}
 \end{minipage}
 \begin{minipage}[c]{.46\linewidth}
  \includegraphics[scale=0.4]{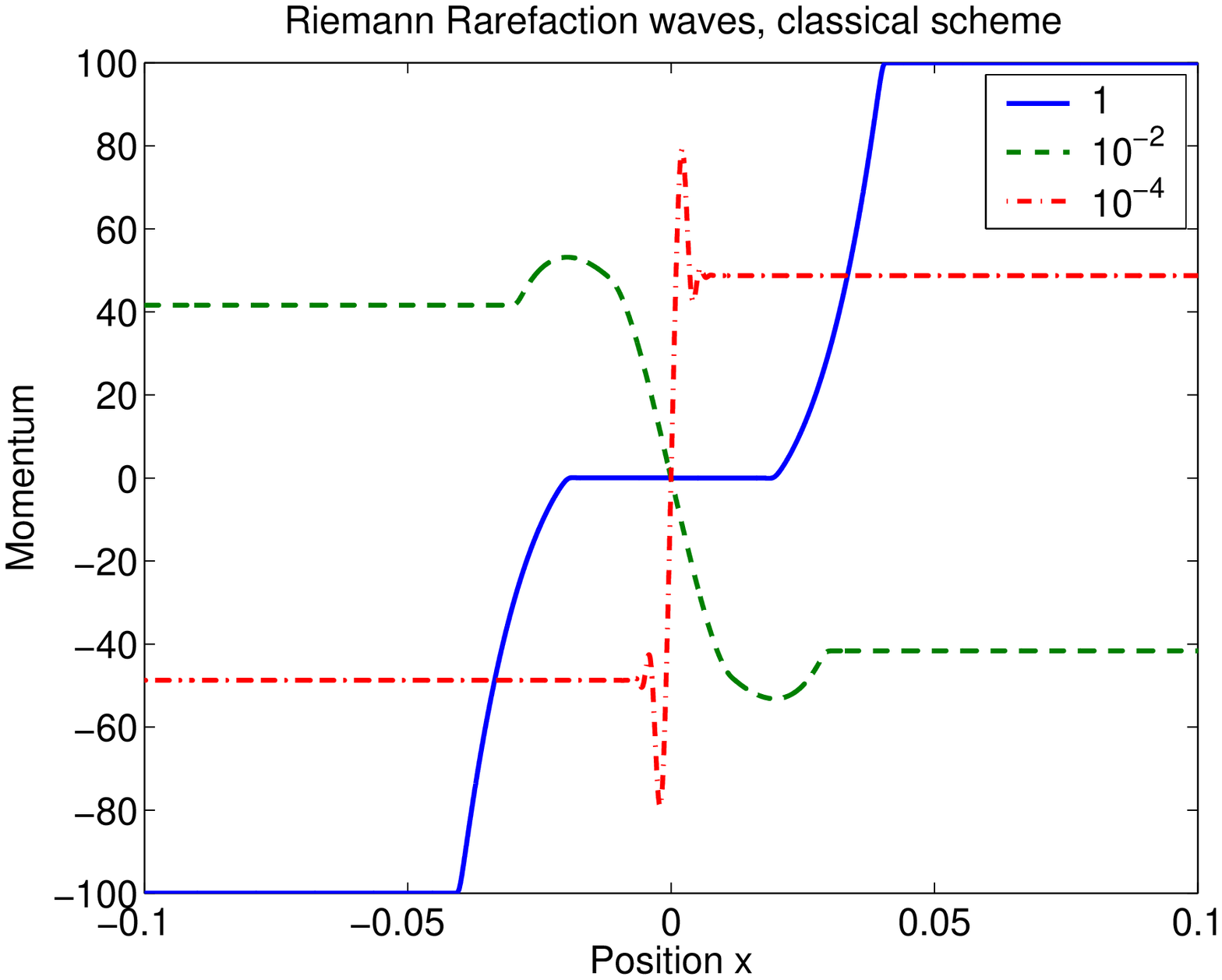}
 \end{minipage}
 \caption{\label{detentes_mono_clas_10000_024} One-fluid rarefaction wave test case with zero initial magnetic field. $n$ (left panel) and $nu_x$ (right panel) as functions of $x$ at time $ t = 2 \times 10^{-4} $ with $ \lambda =1 $, $ \lambda = 10^{-2} $ and $ \lambda = 10^{-4} $ computed with the classical scheme on $ 10^{4} $ discretization cells.
}
\end{figure}

\begin{figure}[hbtp]
 \begin{minipage}[c]{.46\linewidth}
 \includegraphics[scale=0.4]{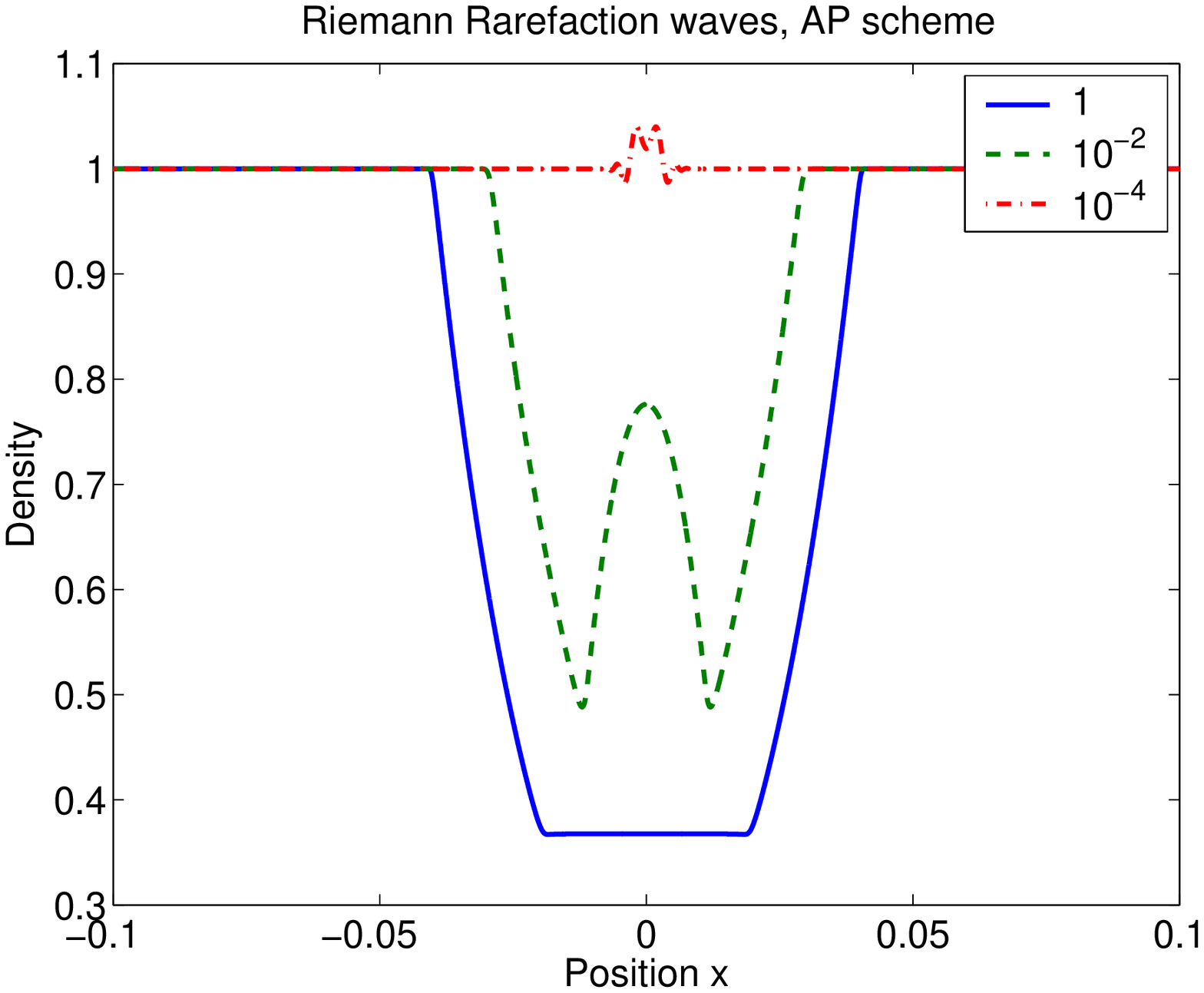}
 \end{minipage}
 \begin{minipage}[c]{.46\linewidth}
  \includegraphics[scale=0.4]{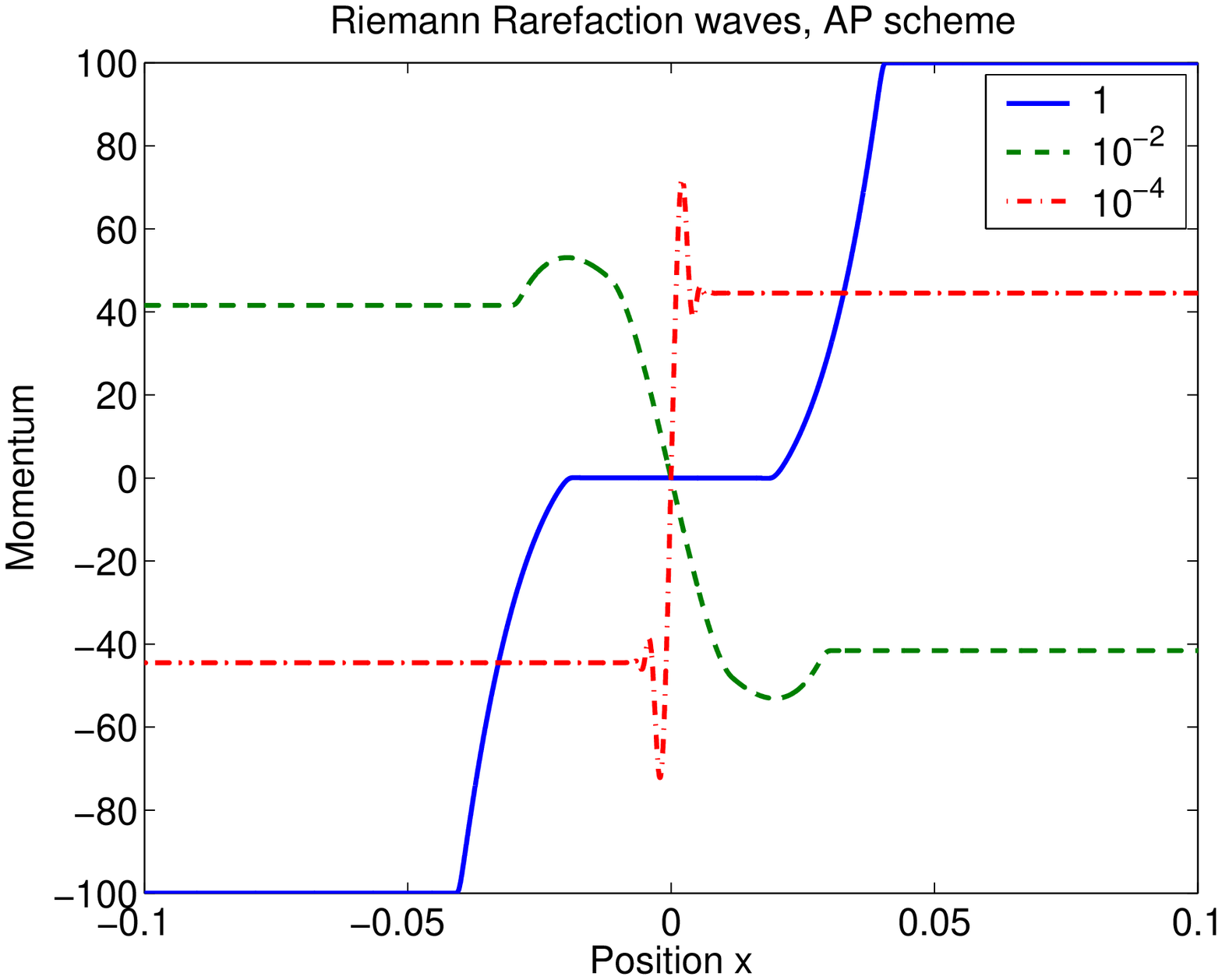}
 \end{minipage}
 \caption{\label{detentes_mono_AP_10000_024} One-fluid rarefaction wave test case with zero initial magnetic field. $n$ (left panel) and $nu_x$ (right panel) as functions of $x$ at time $ t = 2 \times 10^{-4} $ with $ \lambda =1 $, $ \lambda = 10^{-2} $ and $ \lambda = 10^{-4} $ computed with the AP scheme on a $ 10^{4} $ discretization cells.
}
\end{figure}

\begin{figure}[hbtp]
 \begin{minipage}[c]{.46\linewidth}
 \includegraphics[scale=0.4]{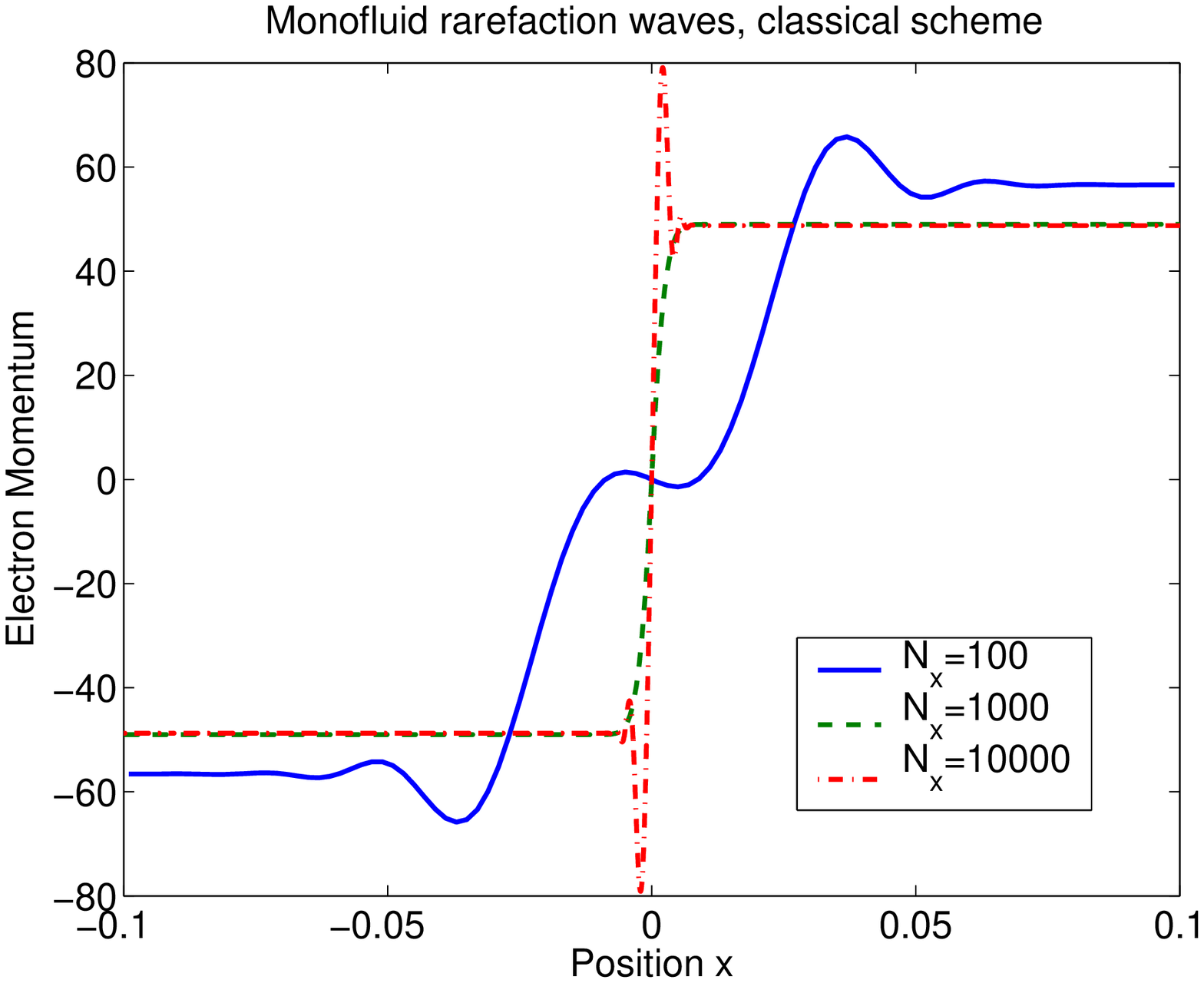}
 \end{minipage}
 \begin{minipage}[c]{.46\linewidth}
  \includegraphics[scale=0.4]{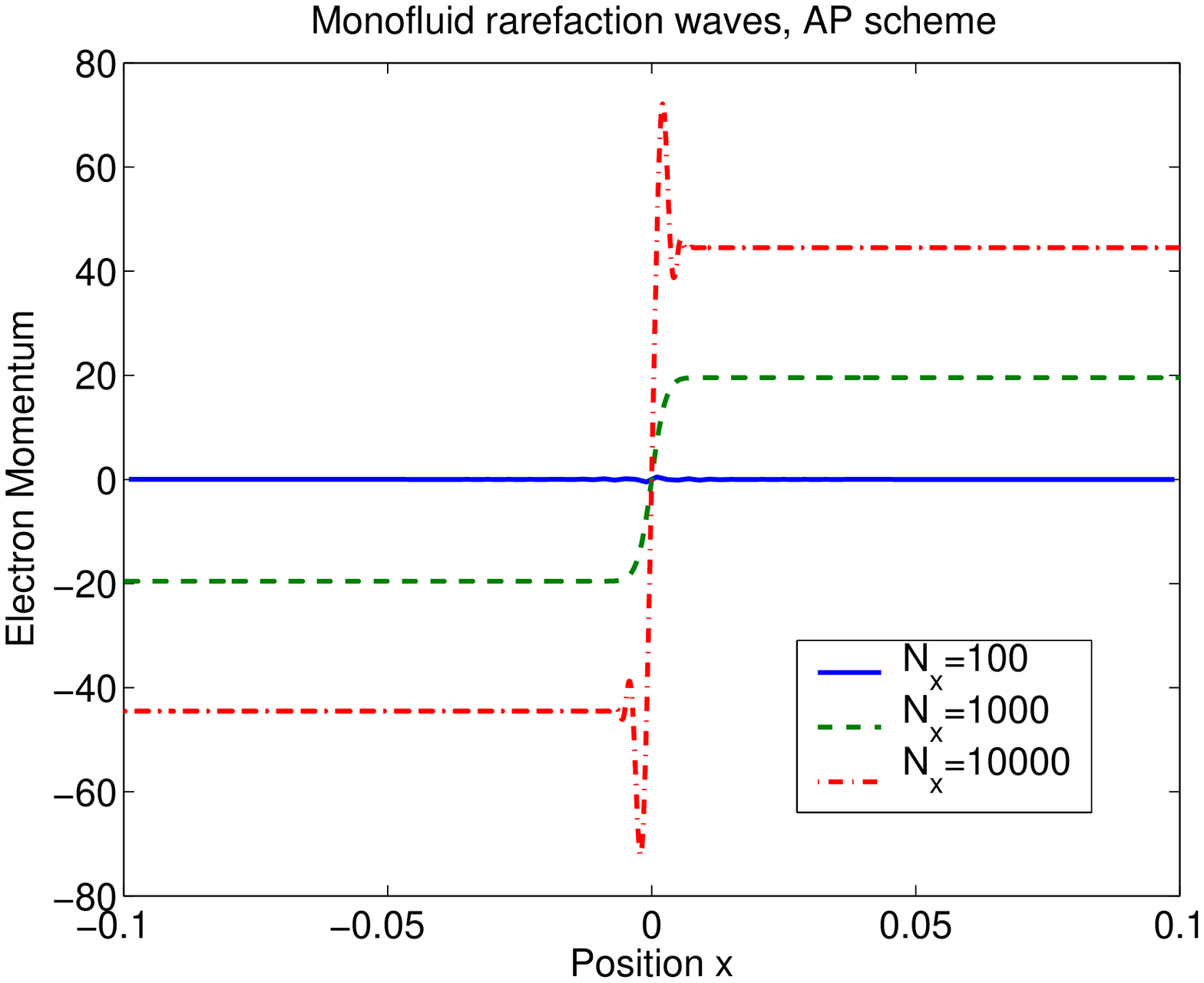}
 \end{minipage}
 \caption{\label{detentes_mono_multx_04} One-fluid rarefaction wave test case with zero initial magnetic field. $nu_x$ as a function of $x$ at time $ t = 5 \times 10^{-4} $ with $ \lambda = 10^{-4} $  for the classical scheme (left panel) and the AP-scheme (right panel).
$\Delta x / \lambda$ respectively equals $ 0.2 $, $ 2 $ and $ 20 $ for $ N_{x} = 10000 $, $ N_{x} = 1000 $ and  $ N_{x} = 100 $ discretization cells.}
\end{figure}

The initial velocities in this test case are $ u_{L} = -100 $ and $ u_{R} = 100 $. In this cases, a low density region appears at the center of the simulation domain, surrounded by two outgoing rarefaction waves. The conclusions that can be drawn from this test-case are similar as for the previous test-case.  Figs. \ref{detentes_mono_clas_10000_024} and \ref{detentes_mono_AP_10000_024} display the density (left panel) and momentum (right panel) as a function of space at time $t = 2 \times 10^{-4}$ in the cases of the classical and AP schemes respectively, for three values of $\lambda$: $\lambda = 1$, $\lambda = 10^{-2}$, $\lambda = 10^{-4}$, and for $N_x = 10^4$ cells. In this high resolution case both schemes provide the same result. For $\lambda = 1$, the results are close to those of a simulation of the Euler equations without coupling to the Lorentz force. By contrast, when $\lambda = 10^{-4}$, the density is close to a uniform one but some oscillations are visible near the origin and still generate a large amplitude in the momentum variation. However, if the ratio $\Delta x/\lambda$ is varied from values less than unity to large values, we observe that the AP-scheme converges to the quasi-neutral solution. Fig. \ref{detentes_mono_multx_04} displays momentum as a function of space in the case of the classical scheme (left panel) and the AP-scheme (right panel), for $\lambda = 10^{-4}$ and when the ratio $\Delta x/\lambda$ is varied from $0,2$ (i.e. with $N_x = 10^4$ cells) to $2$ ($N_x = 10^3$ cells) and finally $20$ ($N_x = 10^2$ cells). In the last case, the momentum computed from the AP-scheme vanishes uniformly, in accordance with the quasi-neutral limit, while that predicted by the classical still has $O(1)$ magnitude. In the intermediate case, the magnitude of the momentum predicted by AP-scheme is in between that obtained in the two extreme cases. 

\subsubsection{Two-fluid outgoing shock waves; zero initial magnetic field}
\label{subsubsec_num_twofluid_shock}

\begin{figure}[hbtp]
 \begin{minipage}[c]{.46\linewidth}
 \includegraphics[scale=0.4]{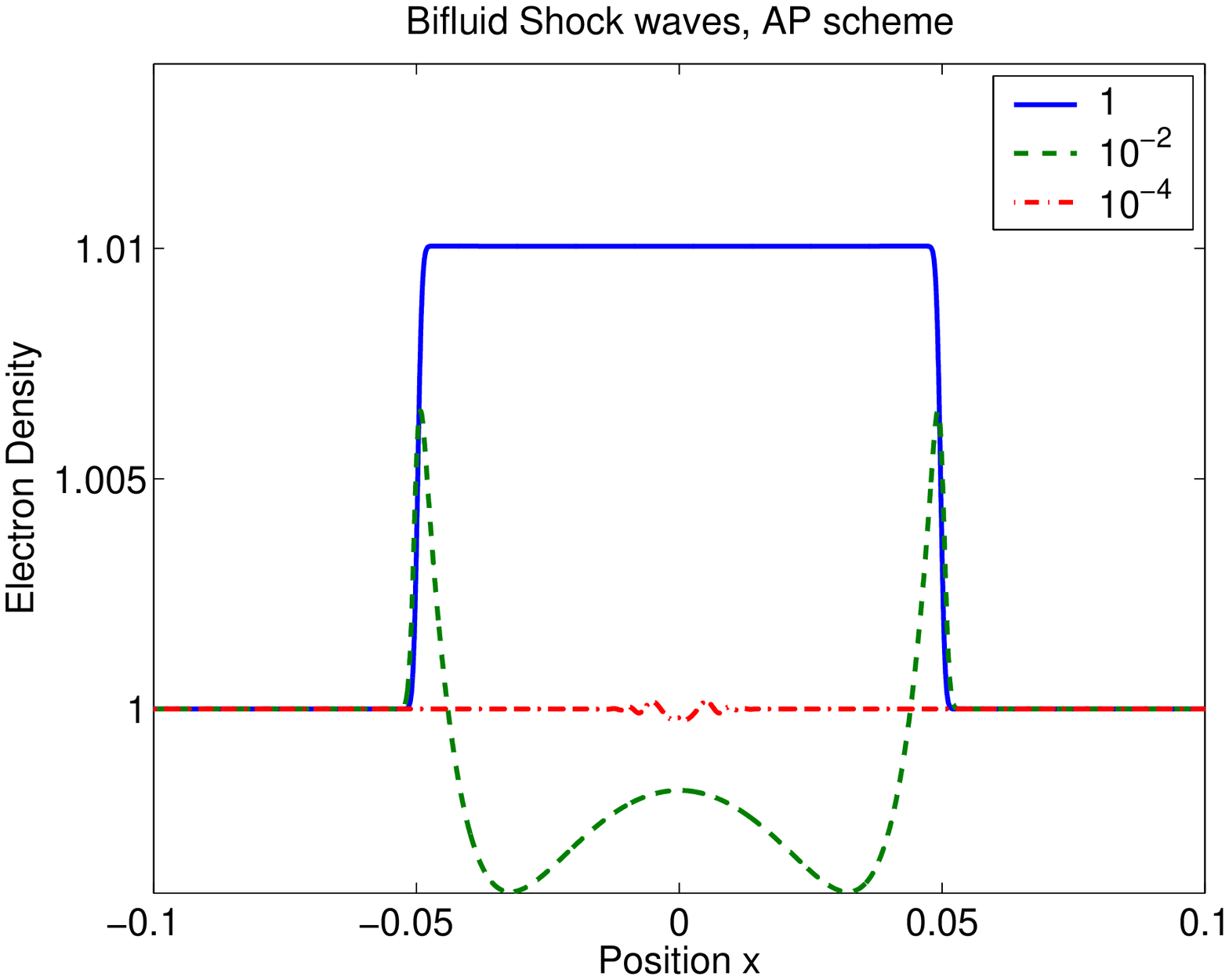}
 \end{minipage}
 \begin{minipage}[c]{.46\linewidth}
  \includegraphics[scale=0.4]{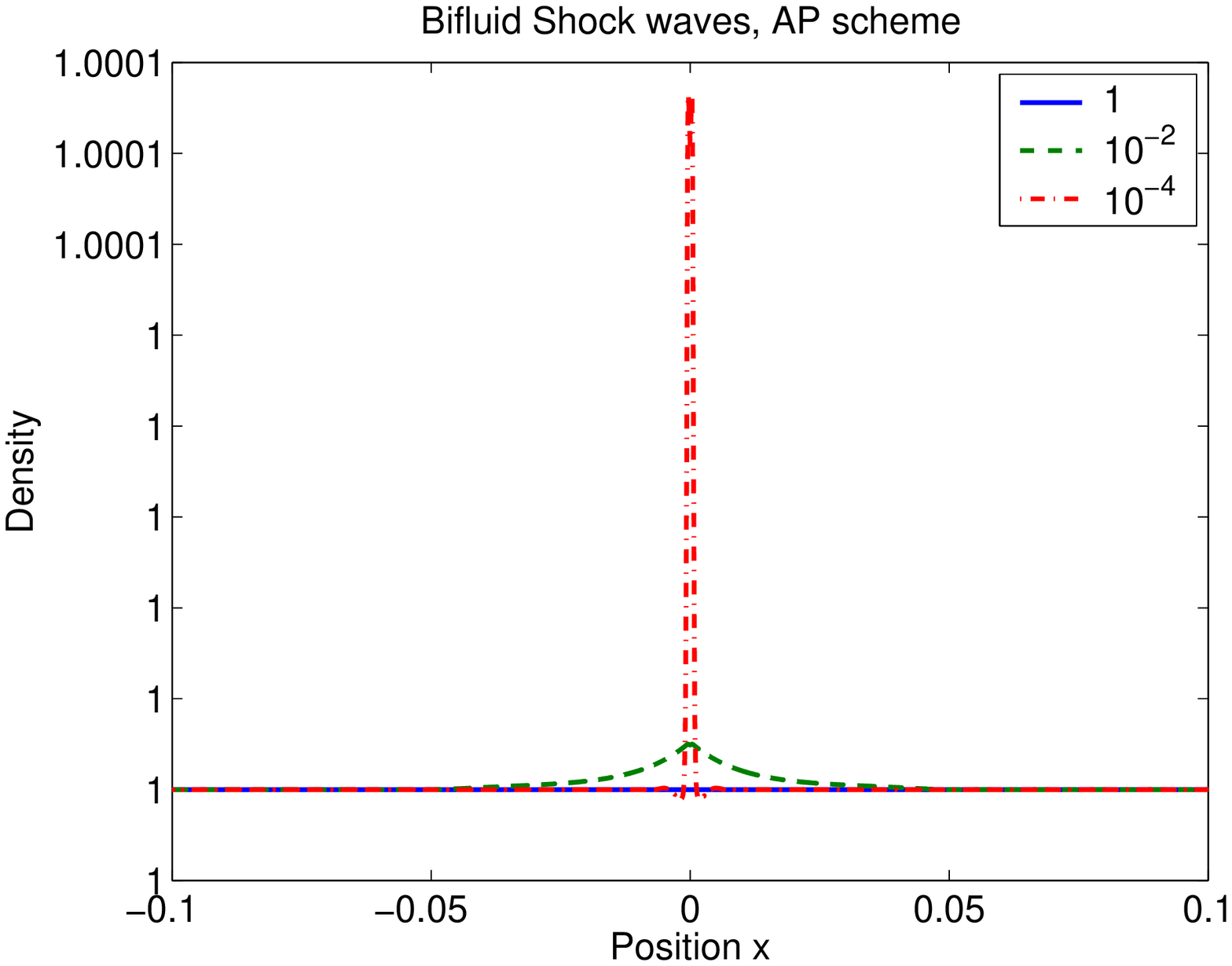}
 \end{minipage}
 \caption{\label{chocs_bif_10000_024_dens} Two-fluid shock wave test case with zero initial magnetic field. $n_e$ (left panel) and $n_i$ (right panel) as functions of $x$ at time $t=5\times10^{-4}$ for $ \lambda = 1 $, $ \lambda = 10^{-2} $ and $ \lambda = 10^{-4} $ computed with the AP-scheme on $ N_x = 10^4 $ space cells.}
\end{figure}

\begin{figure}[hbtp]
 \begin{minipage}[c]{.46\linewidth}
 \includegraphics[scale=0.4]{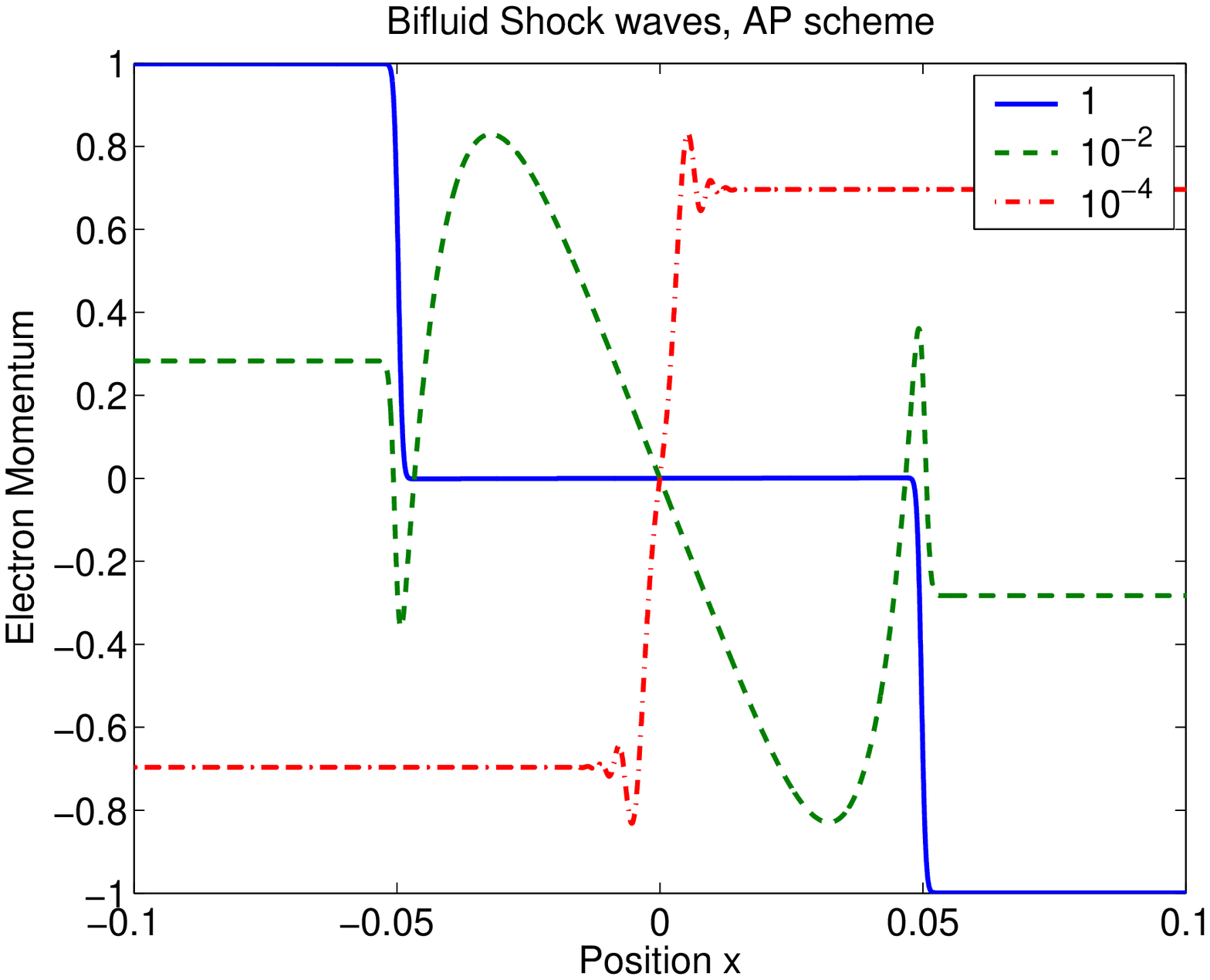}
 \end{minipage}
 \begin{minipage}[c]{.46\linewidth}
  \includegraphics[scale=0.4]{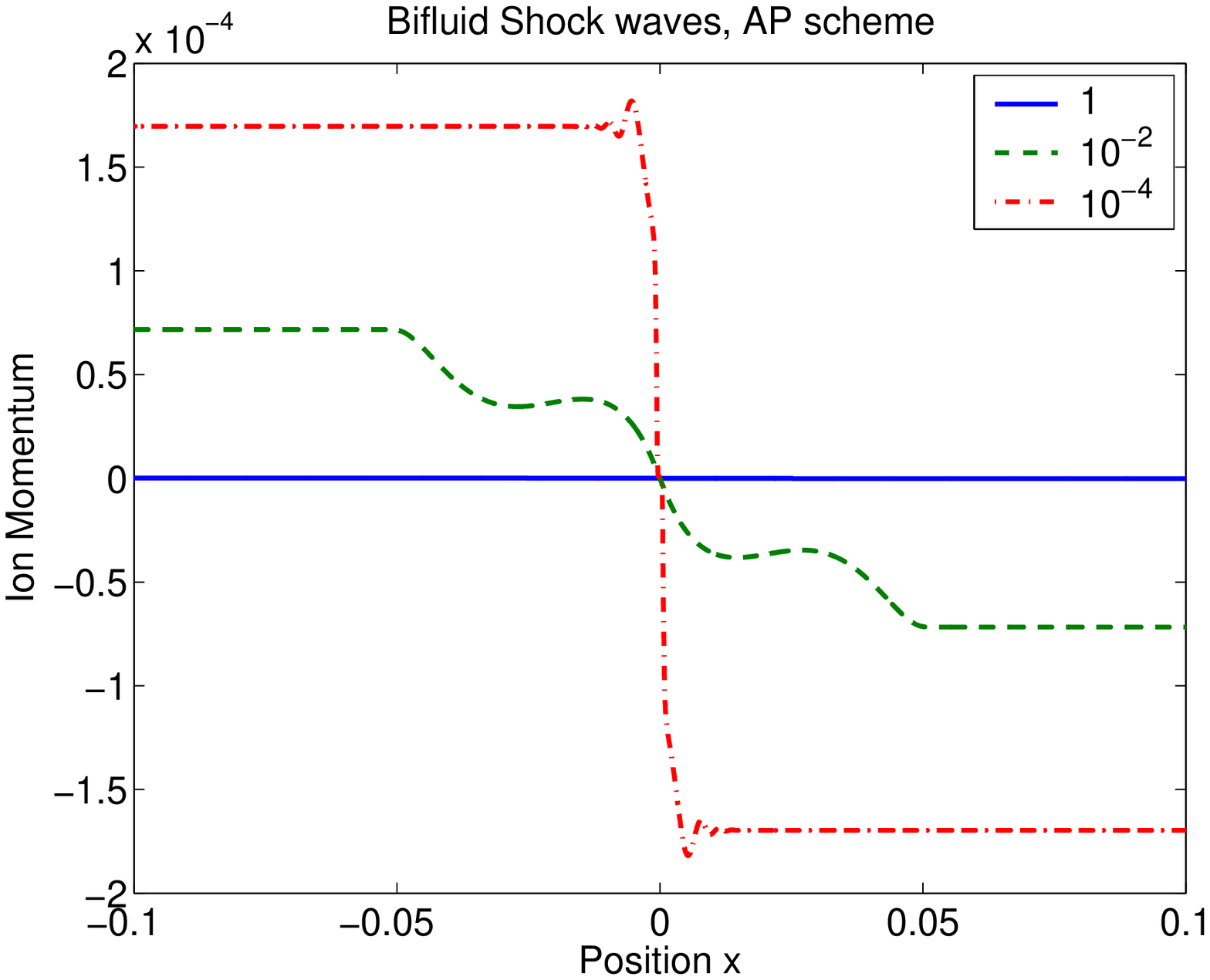}
 \end{minipage}
 \caption{\label{chocs_bif_10000_024_momx} Two-fluid shock wave test case with zero initial magnetic field.  $n_e u_{ex}$ (left panel) and $n_i u_{ix}$ (right panel) as functions of $x$ at time $t=5\times10^{-4}$ for $ \lambda = 1 $, $ \lambda = 10^{-2} $ and $ \lambda = 10^{-4} $ computed with the AP-scheme on $ N_x = 10^4 $ space cells.}
\end{figure}

\begin{figure}[hbtp]
\begin{minipage}[c]{.46\linewidth}
\psfrag{varepsilon ne}{$\varepsilon(n_e)$}
\psfrag{Delta x}{$\Delta x$}
\includegraphics[scale=0.4]{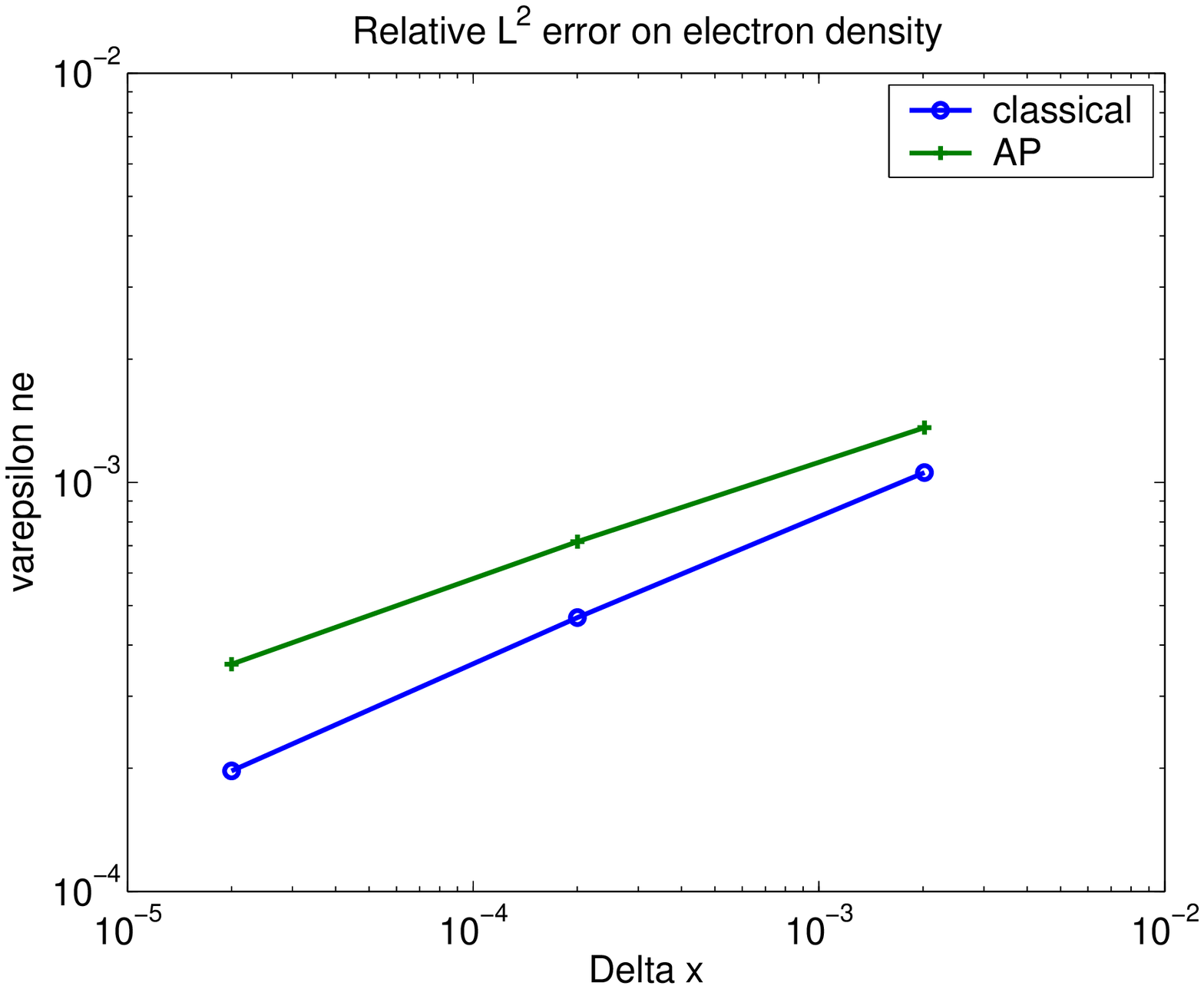}
\end{minipage}
\begin{minipage}[c]{.46\linewidth}
\psfrag{varepsilon ni}{$\varepsilon(n_i)$}
\psfrag{Delta x}{$\Delta x$}
\includegraphics[scale=0.4]{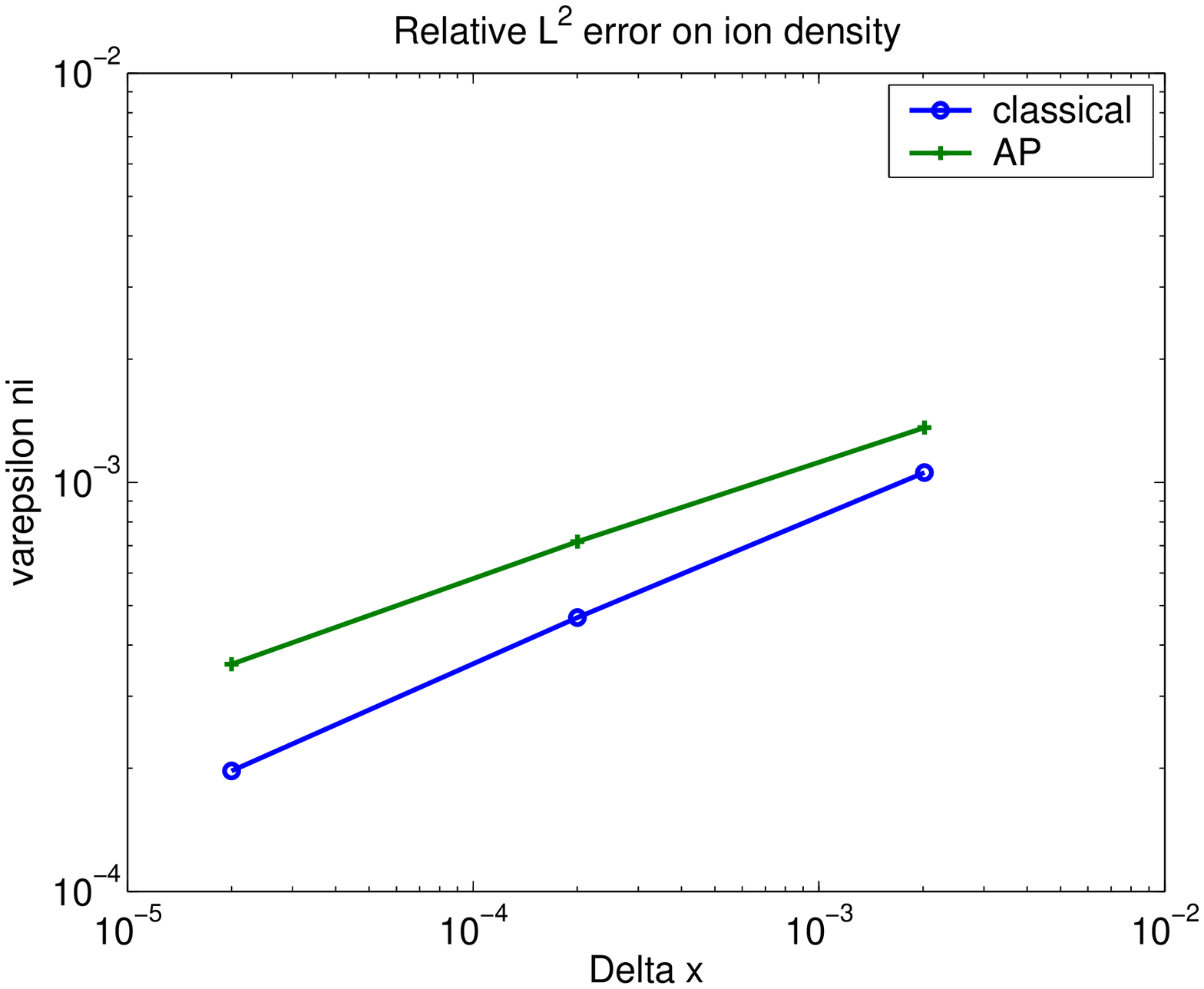}
\end{minipage}
\caption{\label{chocs_bif_error_dens_L1} Two-fluid shock wave test case with zero initial magnetic field.  Relative $L^1$ errors on $n_e$ (left panel) and $n_i$ (right panel) as functions of $\Delta x$ at time $t=5\times10^{-4}$ for both the classical and AP- schemes and with $\lambda = 1$.}
\end{figure}

We now consider a two-fluid model consisting of electrons and ions. By contrast to the one-fluid case, where only electrons are mobile, both ion and electrons are susceptible to bet set into motion. In this section, we investigate the ability of the classical and AP- schemes to describe the setting of the ions in motion. We restrict to the case of the outgoing shock waves with zero initial magnetic field. The initial electron density and velocity are taken equal to the one-fluid case, while the initial ion density is uniform equal to $1$ and the initial ion velocity is uniform equal to $0$. Figs. \ref{chocs_bif_10000_024_dens} and \ref{chocs_bif_10000_024_momx} respectively display the densities and the momenta as a function of space, at a given time, for the AP schemes. The left panels are for the electron quantities, and the right panels, for the ion ones. Three values of $\lambda$ are used: $\lambda = 1$, $\lambda = 10^{-2}$, $\lambda = 10^{-4}$. We can see that the AP-scheme provides physically meaningful results. When $\lambda = 1$, the electromagnetic coupling between the electrons and ions is weak. The ions stay immobile with uniform density while the electrons exhibit outgoing shock waves as if there would be absolutely no coupling to the Lorentz force. By contrast, in the case $\lambda = 10^{-4}$, the electron density converges to a uniform density equal to one, apart from a small oscillation near the origin, and a comparable oscillation of the ion density (the ion density scale is magnified and appears larger than the electron one, but the order of magnitudes are actually similar). The ions are set in motion in opposite directions to the electrons as they should and the ratio of the ion to electron momentum scales like the mass ratio as they should (since the densities are almost the same). In the case $\lambda = 10^{-2}$, an intermediate situation is observed. In the resolved case, a convergence study can be performed with respect to a reference solution, computed in the same way as described in section \ref{subsubsec_num_onefluid_shock}. Fig. \ref{chocs_bif_error_dens_L1} shows the relative $L^1$ errors obtained on the electron and ion densities in the case $\lambda = 1$. We can see that both scheme are convergent. The convergence rate of the AP-scheme seems a little bit slower than that of the classical scheme and the magnitude of the error a bit larger. However, this slightly lower precision is little price to pay for the AP-character which guarantees a proper behavior of the scheme in the small Debye length regime. 

\subsubsection{One-fluid outgoing shock waves; non-zero magnetic field}
\label{subsubsec_num_onefluid_shock_nonzeroB}

This test-case is similar to the one-fluid outgoing shock wave test-case of section \ref{subsubsec_num_onefluid_shock}, but the magnetic field $B_z$ at initial time is taken non-zero. This magnetic field generates a non-zero $y$-component of the electric field $E_y$ which sets the plasma into motion in this direction and consequently, generates a non-zero $y$-component of the velocity $u_y$. These components become larger as $\lambda$ is decreased. The magnitude of the dimensionless magnetic field $B_z$ at initial time is taken equal to $0.2$. Such a value generates a $y$-component of the electron momentum $n u_y$ which is of the same order of magnitude as its $x$-component $n u_x$ when the dimensionless Debye length $ \lambda = 10^{-4} $. As in the zero magnetic field case, the quasi-neutral limit simply provides a uniform density equal to $1$ and zero velocity in both components $u_x = u_y = 0$, while both components of the electric field vanish $E_x = E_y = 0$ and the magnetic field is uniform and equal to its value at time $0$: $B_z = B_z|_{t=0}$. As this Riemann problem is intended to mimic a whole space problem, we choose transparent boundary conditions, which in this simple 1D example, coincide with homogeneous Silver-M\"uller boundary conditions. However, transparent boundary conditions suppose that there are no electromagnetic sources outside the domain under consideration. In the present case, when the acoustic waves generated by the Riemann initial data escape the domain, they produce electromagnetic field sources outside the domain which are not accounted for by the homogeneous Silver-M\"uller boundary conditions. To bypass this problem, we enlarge the domain to the interval $ [-0.2,0.2] $ (i.e. twice the size of the domain of the zero magnetic field case) and we observe the results only on the domain $ [-0.1,0.1] $ and for times shorter than the time needed for the perturbations generated by the boundary conditions to reach this subdomain. 

\begin{figure}[hbtp]
 \begin{minipage}[c]{.46\linewidth}
 \includegraphics[scale=0.4]{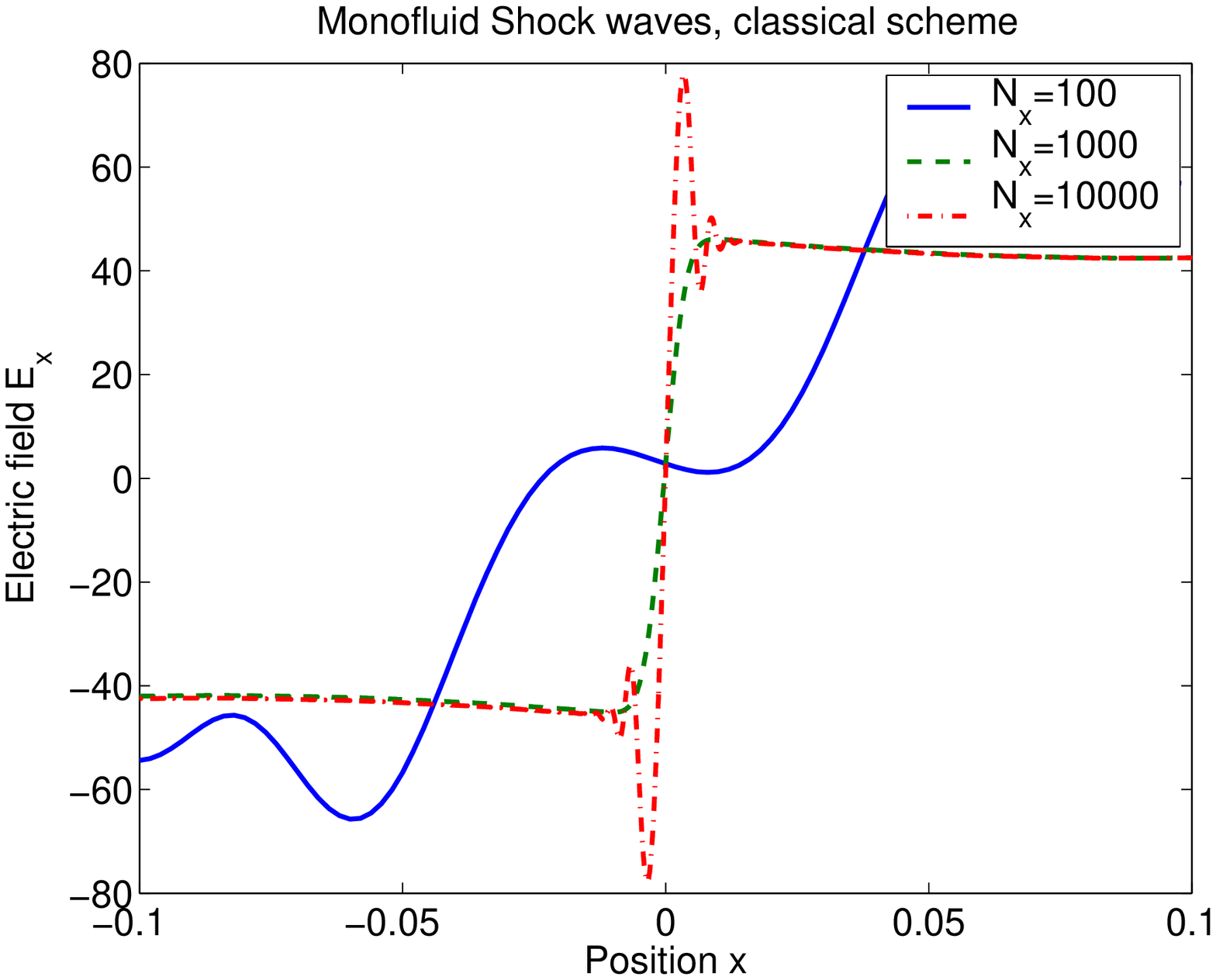}
 \end{minipage}
 \begin{minipage}[c]{.46\linewidth}
  \includegraphics[scale=0.4]{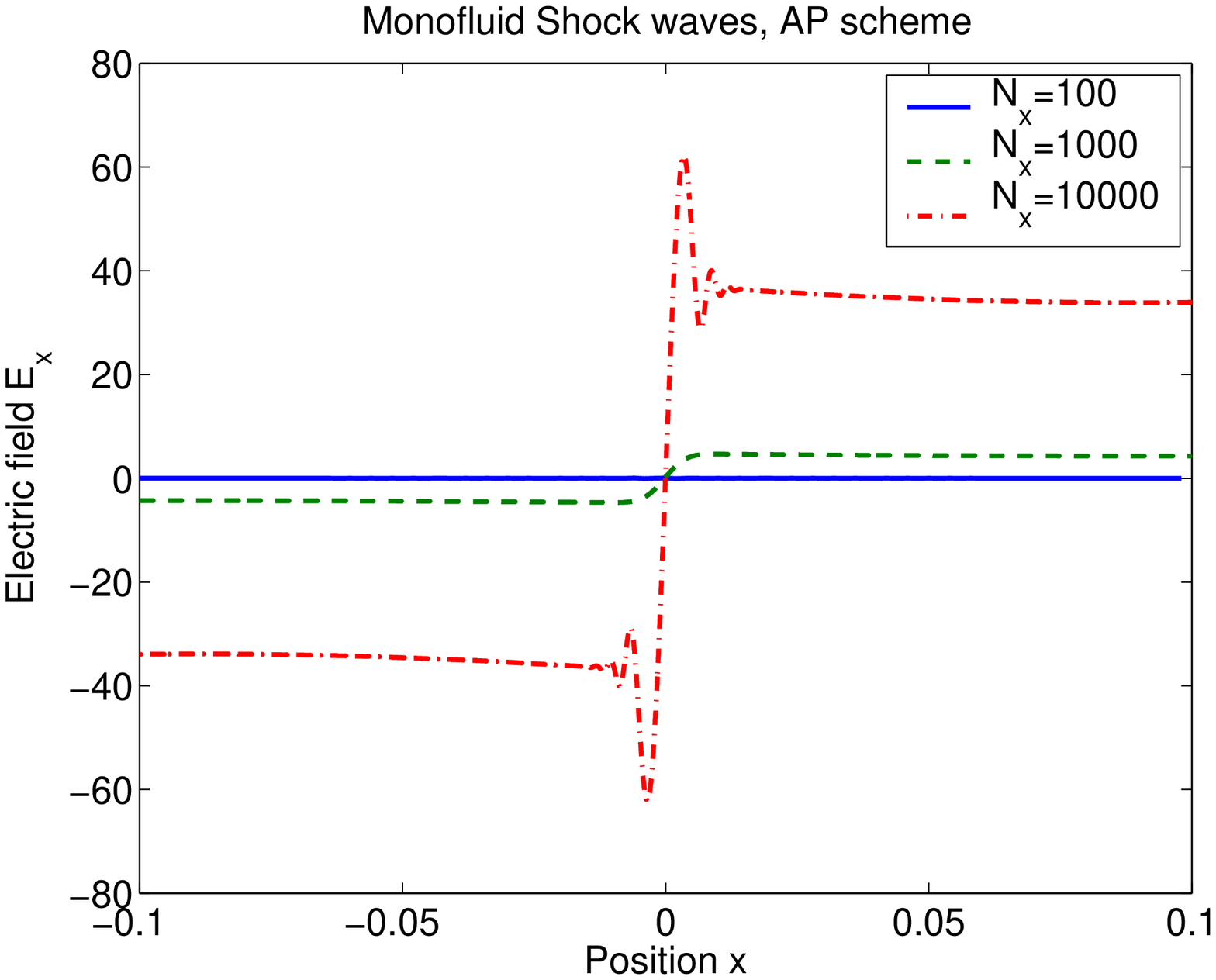}
 \end{minipage}
 \caption{\label{em_chocs_multx_04_Ex} One-fluid shock wave test case with non-zero initial magnetic field. $ E_x $  as a function of $x$ at time $ t = 5 \times 10^{-4} $ with $ \lambda = 10^{-4} $  for the classical scheme (left panel) and AP-scheme (right panel).
$\Delta x / \lambda$ respectively equals $ 0.2 $, $ 2 $ and $ 20 $ for $ N_{x} = 10000 $, $ N_{x} = 1000 $ and  $ N_{x} = 100 $ discretization cells.
}
\end{figure}

Fig. \ref{em_chocs_multx_04_Ex} displays $E_x$ as a function of space at time $t=5 \times 10^{-4}$ for the classical scheme (left panel) and the AP-scheme (right panel) in the case $\lambda = 10^{-4}$ and for $N_x = 10^2$, $N_x = 10^3$ and $N_x = 10^4$ space cells. The results are close to those obtained in the zero-magnetic field case. We observe that, as $\Delta x / \lambda$ increases from $0.2$ (in the case $N_x = 10^4$) to $20$ (in the case $N_x = 10^2$), the AP-scheme correctly captures that the magnitude of the electron momentum gradually decreases from an $O(1)$ value to $0$, as predicted by the quasi-neutral limit. By contrast, the momentum produced by the classical scheme remains $O(1)$ whatever large $\Delta x / \lambda$ becomes. 

\begin{figure}[hbtp]
 \begin{minipage}[c]{.46\linewidth}
 \includegraphics[scale=0.4]{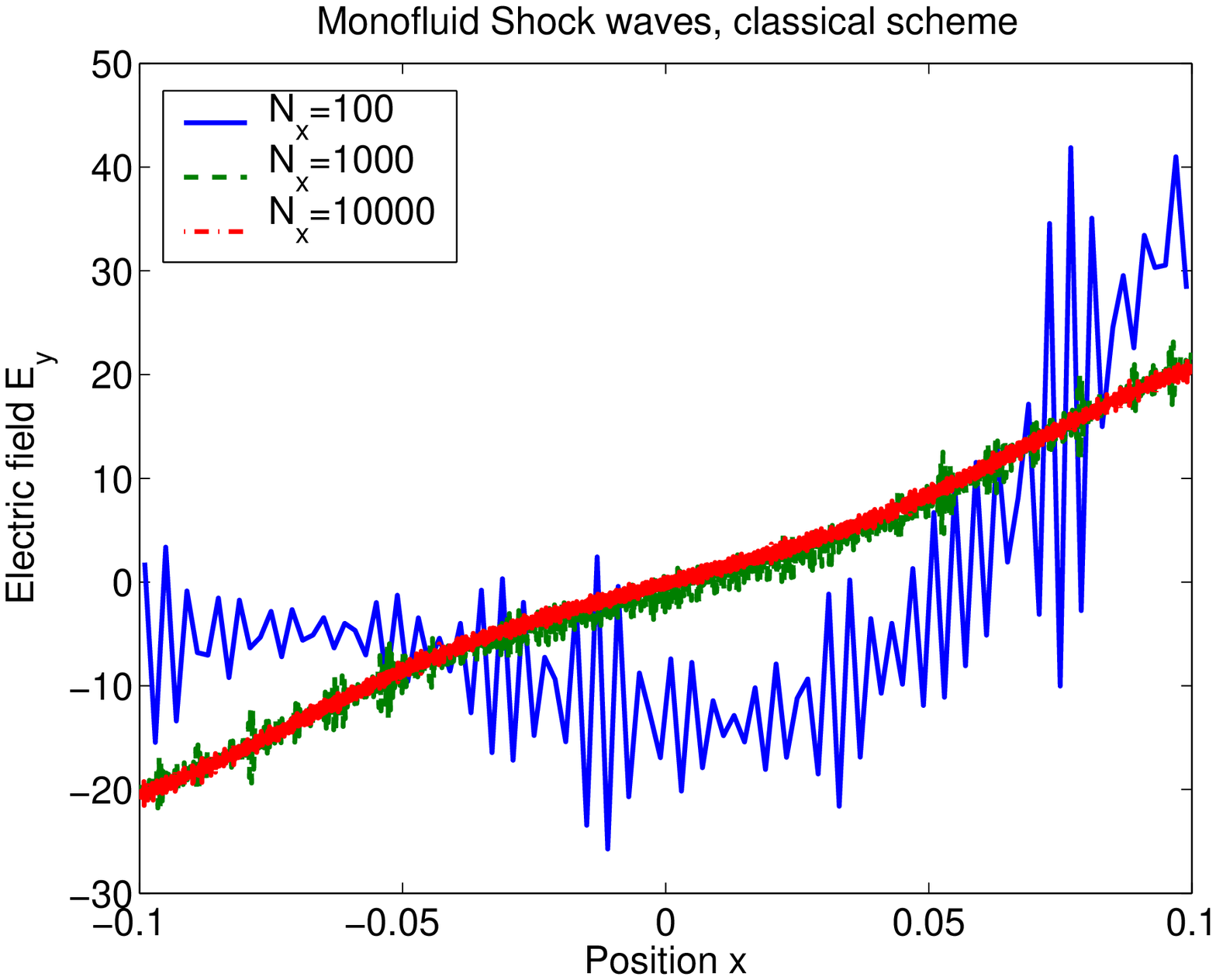}
 \end{minipage}
 \begin{minipage}[c]{.46\linewidth}
  \includegraphics[scale=0.4]{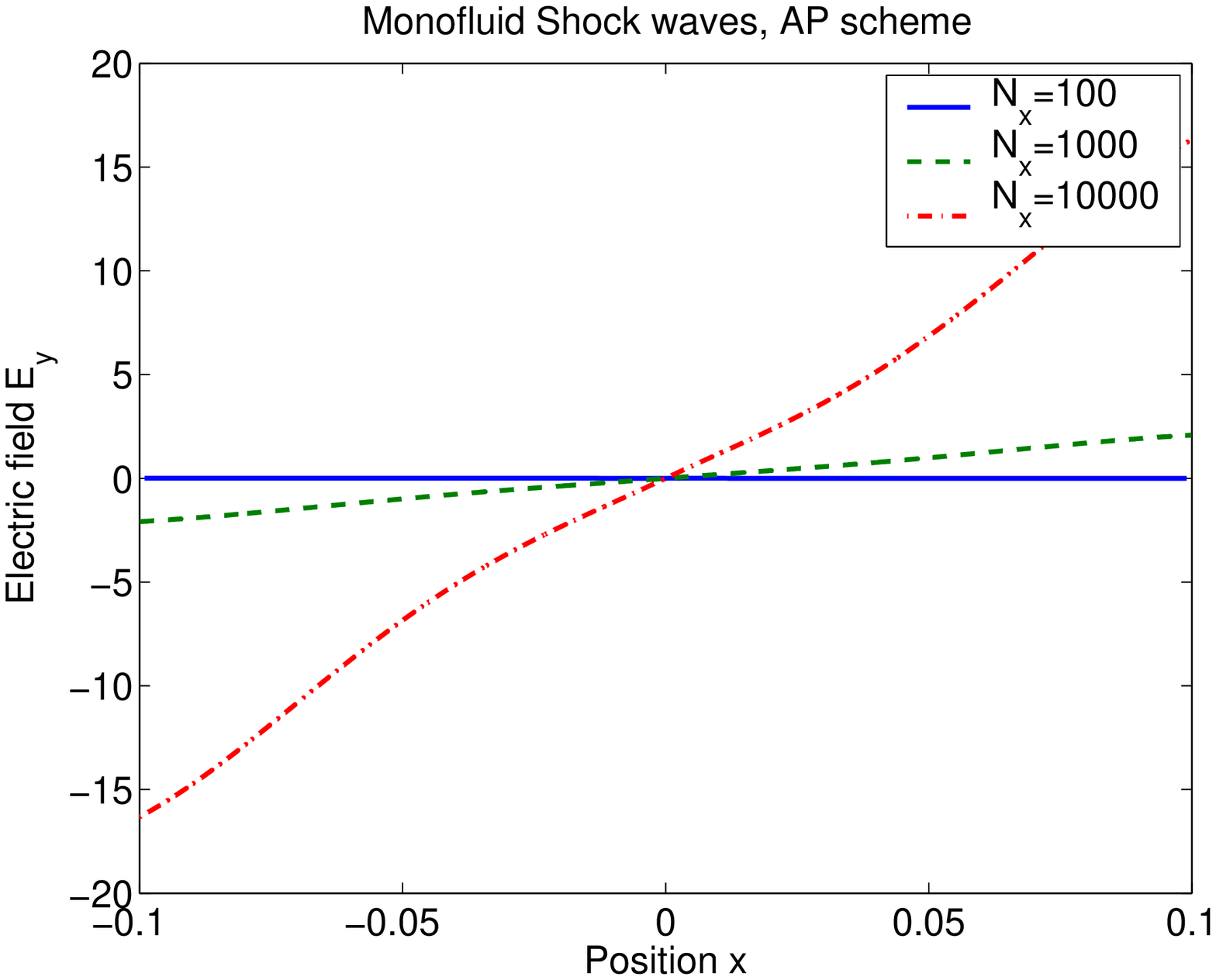}
 \end{minipage}
 \caption{\label{em_chocs_multx_04_Ey} One-fluid shock wave test case with non-zero initial magnetic field. $ E_y $  as a function of $x$ at time $ t = 5 \times 10^{-4} $ with $ \lambda = 10^{-4} $  for the classical scheme (left panel) and AP-scheme (right panel).
$\Delta x / \lambda$ respectively equals $ 0.2 $, $ 2 $ and $ 20 $ for $ N_{x} = 10000 $, $ N_{x} = 1000 $ and  $ N_{x} = 100 $ discretization cells.
}
\end{figure}

\begin{figure}[hbtp]
 \begin{minipage}[c]{.46\linewidth}
 \includegraphics[scale=0.4]{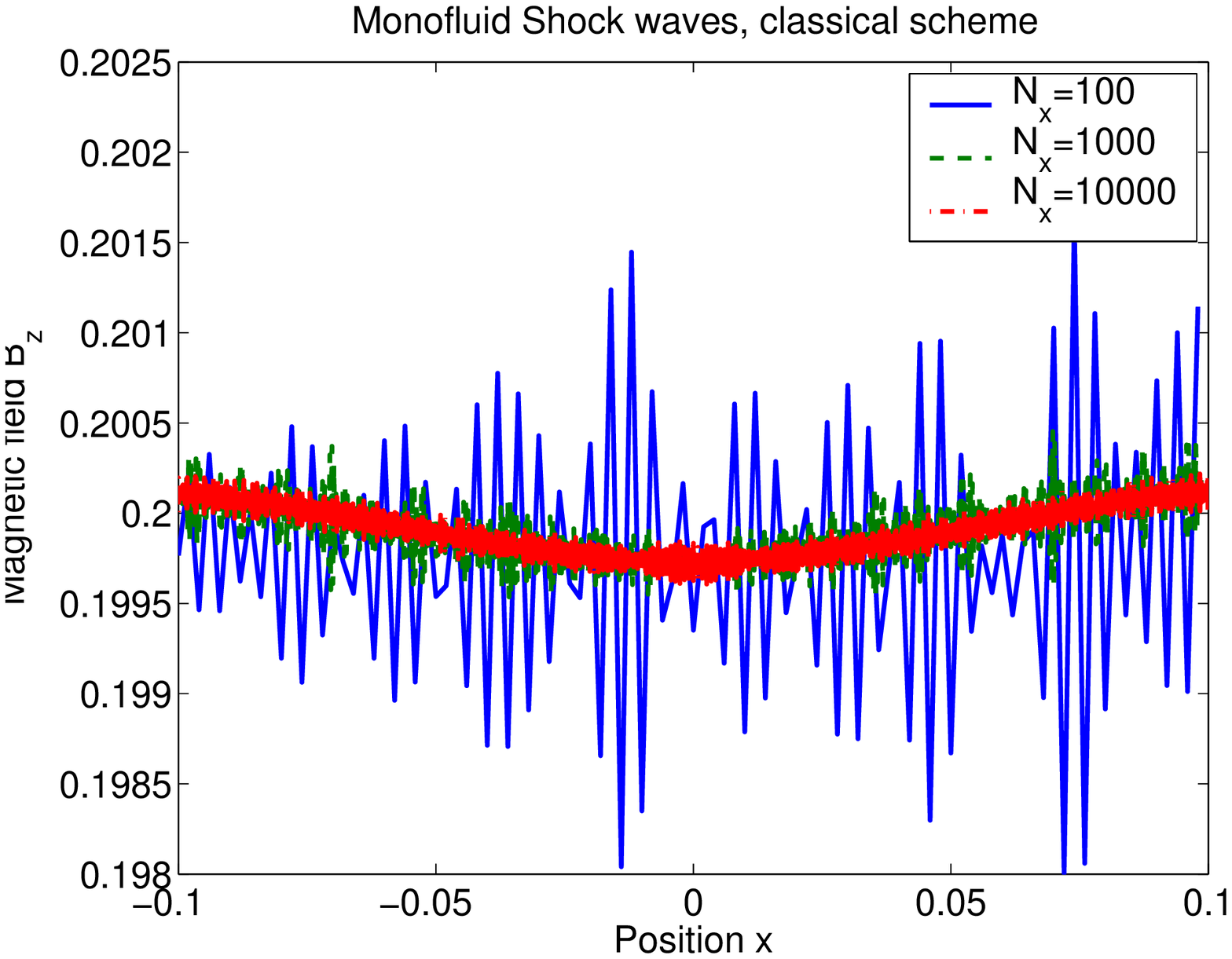}
 \end{minipage}
 \begin{minipage}[c]{.46\linewidth}
  \includegraphics[scale=0.4]{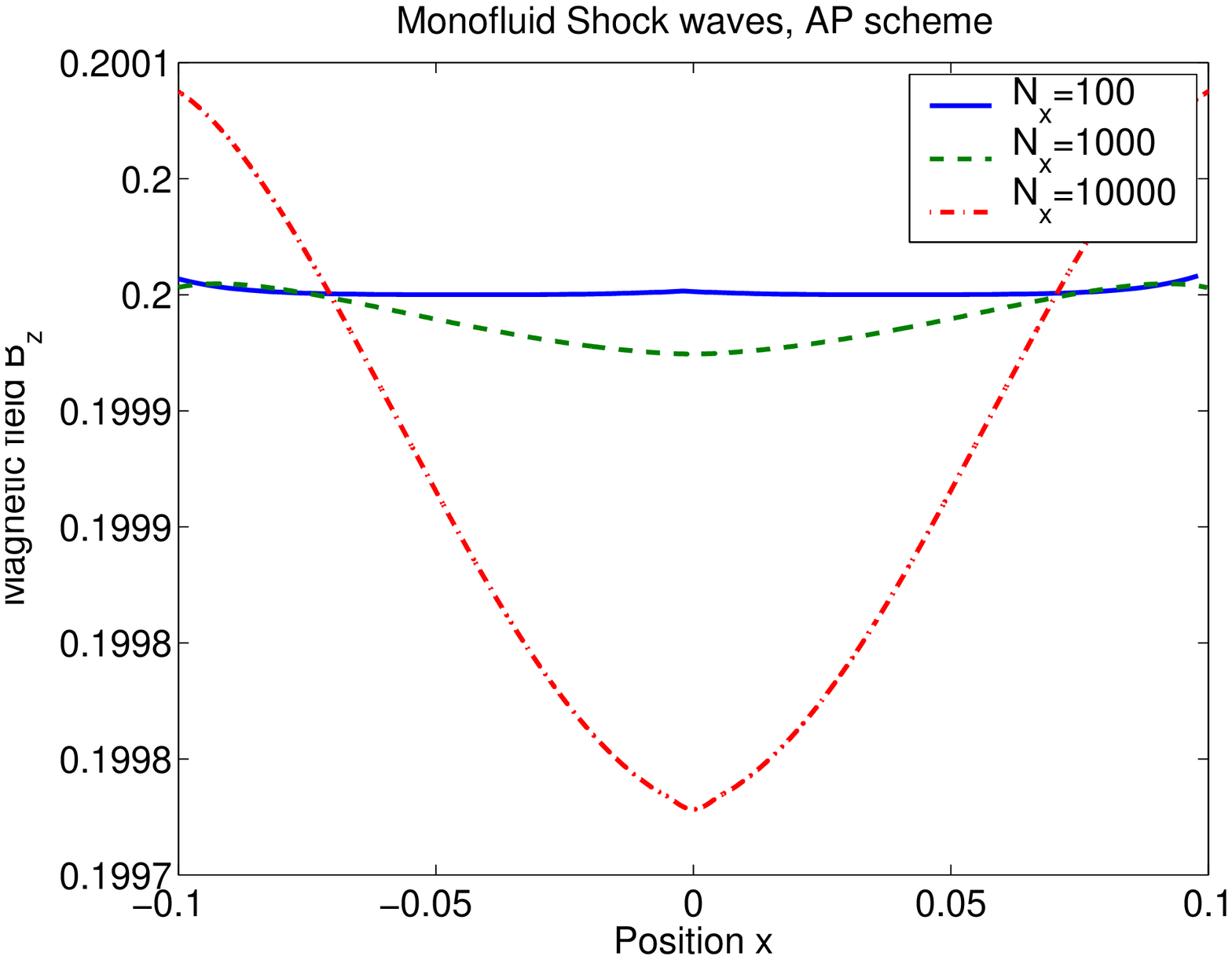}
 \end{minipage}
 \caption{\label{em_chocs_multx_04_Bz} One-fluid shock wave test case with non-zero initial magnetic field. $ B_z $  as a function of $x$ at time $ t = 5 \times 10^{-4} $ with $ \lambda = 10^{-4} $  for the classical scheme (left panel) and AP-scheme (right panel).
$\Delta x / \lambda$ respectively equals $ 0.2 $, $ 2 $ and $ 20 $ for $ N_{x} = 10000 $, $ N_{x} = 1000 $ and  $ N_{x} = 100 $ discretization cells.
}
\end{figure}

Figs. \ref{em_chocs_multx_04_Ey} and \ref{em_chocs_multx_04_Bz} display $E_y$ and $B_z$ as functions of $x$ in the same conditions (left panel: classical scheme, right panel: AP-scheme). As $\Delta x / \lambda$ increases from $0.2$ to $20$, the approximations of $E_y$ and $B_z$ given by the AP-scheme tend respectively to zero and to a constant value equal to $B_z|_{t=0}$, as predicted by the quasi-neutral limit. By contrast, the approximations of $E_y$ and $B_z$ given by the classical scheme exhibit strong oscillations with increasing amplitudes as $\Delta x / \lambda$ increases. These approximations are neither the correct solutions for the finite $\lambda$ problem, nor for the limit quasi-neutral problem. 

\begin{figure}[hbtp]
 \begin{minipage}[c]{.46\linewidth}
 \psfrag{varepsilon ne}{$\varepsilon(n_e)$}
 \psfrag{Delta x}{$\Delta x$}
 \includegraphics[scale=0.4]{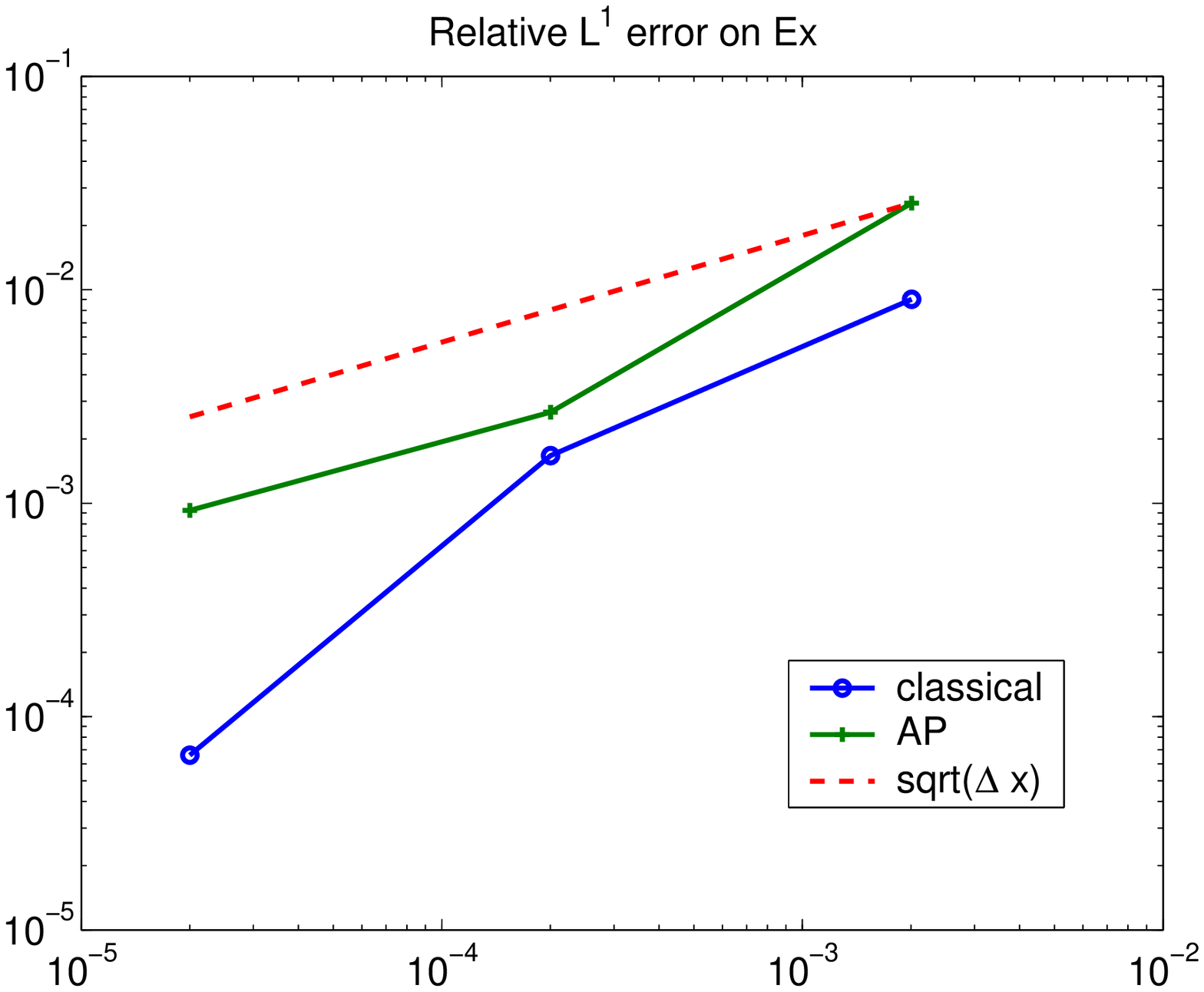}
 \end{minipage}
 \begin{minipage}[c]{.46\linewidth}
 \psfrag{varepsilon ni}{$\varepsilon(n_i)$}
 \psfrag{Delta x}{$\Delta x$}
  \includegraphics[scale=0.4]{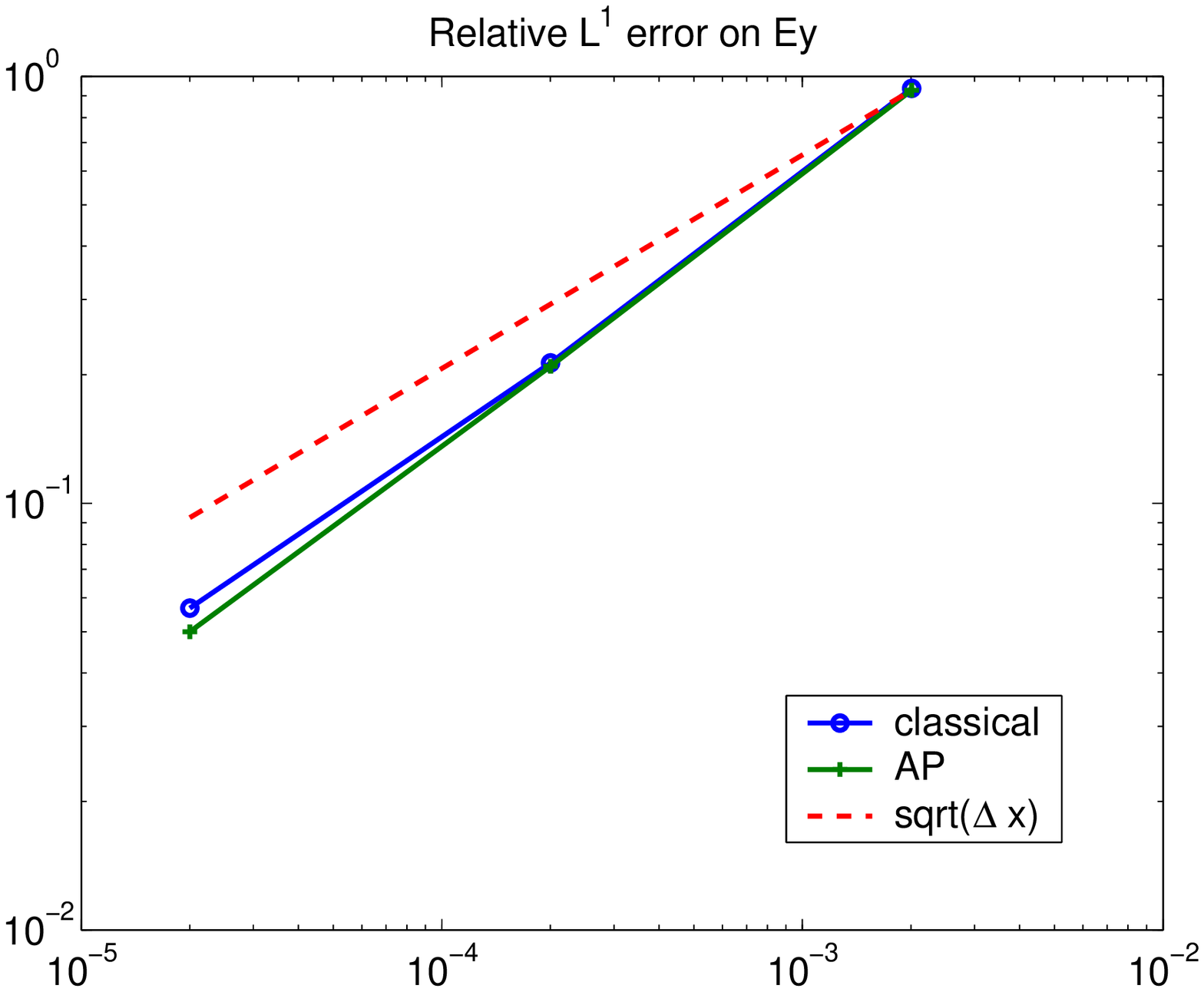}
 \end{minipage}
 \caption{\label{em_chocs_error_L1} One-fluid shock wave test case with non-zero initial magnetic field. Relative $L^1$ error on $ E_x $ (left panel) and $ E_y $ (right panel) as functions of $\Delta x$ at time $ t = 5\times 10^{-4} $ for both the classical and AP- schemes and with $\lambda = 1$.}
\end{figure}

Finally, in the case where $\lambda$ is not too small, a convergence study can  be performed. Fig. \ref{em_chocs_error_L1} displays the relative errors in $ L^{1} $ norm on $ E_x $ (left panel) and $ E_y $  (right panel) computed with the classical and AP schemes as a function of mesh size.

\subsection{Plasma opening switch}
\label{subsec_pos}

Plasma Opening switches (POS) are devices used in pulsed power systems to deliver large currents in short times compared to the rising time of generators.
A POS device consist of a transmission line (usually a coaxial transmission line) filled with a quasi-neutral plasma. The plasma short-circuits the two electrodes of the transmission line and prevents power to be delivered to the load. However, simultaneously, the electromagnetic wave gradually erodes the plasma by separating the ions and the electrons. Once a gap has been formed in the plasma, the electromagnetic wave can cross it and the tail of the power pulse can be transmitted to the load. This time-contraction enables the generation of very high power pulses. 

A preliminary validation of the AP-scheme can be performed on a reduced one-dimensional model of the 2D model such as in \cite{Deluzet_pos}.
The computational domain extends over $ 2 \times 10^{-1} m $, which is twice the length of the region filled by the plasma $ 10^{-1} m $. The plasma is located in the middle of the domain. Transparent (Silver-M\"uller) boundary conditions are imposed at the domain boundaries, to avoid including the generator and the load in the simulation. Indeed, part of the incident wave is reflected back to the load as long as the plasma short-circuits the transmission line. It is therefore necessary that the boundary conditions allow these reflected waves to escape the domain. A similar phenomenon prevails at the other end of the transmission line in the opening phase of the device. The quasi-neutral plasma is at rest at initial time. The initial densities of the ion and electron fluids inside the plasma region are equal to $1$ in dimensionless units and their velocities are both equal to $0$. Outside the plasma region, there is vacuum, i.e. initial densities are $ 0 $.

\begin{figure}[hbtp]
 \begin{center}
 \psfrag{elec}{\tiny{Electric field}}
 \psfrag{time}{\tiny{time}}
 \psfrag{Ey_inc}{\mbox{\tiny $E_y^{\text{inc}}$}}
 \psfrag{t^inc}{\mbox{\tiny $ t^{\text{inc}} $ }}
 \psfrag{incpro}{\tiny{Incident wave profile}}
 \includegraphics[width=5cm]{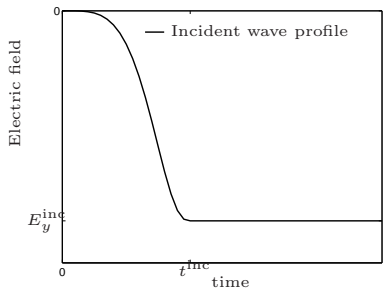}
 \caption{Incident electromagnetic wave profile as a function of time.}
 \label{Fig:incident_profile}
 \end{center}
\end{figure}

We assume a smooth transition profile for the plasma density between these two areas. The incident electromagnetic wave is supposed to be a Transverse Electromagnetic Mode, characterized by a rising time $ t^{\text{inc}} = 10^{-8} s $ and an amplitude for the electric component $ E_{y}^{\text{inc}} = -1.8 \times 10^{8} V$ (see Fig. \ref{Fig:incident_profile}). At time $ t = 0 $ the wave starts from the left side of the computational domain. The  final simulation time is $ t = 2.5 \times 10^{-9} s $. This time is long enough to allow for observation of the wave impact on the plasma and the resulting plasma motion. Unfortunately, the simple one-dimensional setting does not allow for the observation of the POS opening, as this phenomenon is related to plasma motion in transverse direction to the transmission line, which is not accounted for here. However the results from the AP-scheme shown below are consistent with the expected physical phenomena.

Two subsets of test-cases are performed. First, a low density POS is considered, with an initial density of $ 10^{16} m^{-3} $. Second, a higher density POS with an initial density of $ 10^{18} m^{-3} $ is simulated. For both the low and high density POS the temperatures of the ion and electron fluids is approximately $ 5 $ eV i.e. $ 58 \times 10^{3} K $, and a carbon plasma ($C^{+}$) is considered.
The low density POS allows for a fast penetration of the electromagnetic wave in the plasma, whereas the high density POS acts like a barrier reflecting the wave which has more difficulties to cross it. For both the low and high density POS, the one- and two-fluid models will be used.

\subsubsection{Low density POS; one-fluid model}
\label{subsubsec_pos_mono_low_dens}

In this case, the order of magnitude of the scaled Debye length in the plasma is $\lambda = 10^{-3} $. Then, a grid such that $ 10 \Delta x \leq \lambda $ is made of $ 10^{4} $ cells.
This grid is fine enough to resolve the small space and time scales (i.e. the Debye length and electron plasma period). These conditions ensure that the classical scheme is stable and accurate enough (given the computational time constraints) to build a reference solution. We denote $ \Delta x^{\text{ref}} $ and $ \Delta t^{\text{ref}} $ the space and time steps used for these computations. The electron plasma period is $ \tau_{p} = 10^{-10} s $ .
However, the most severe time constraint in this problem arises from the CFL condition for the Maxwell equations due to the explicitness of the classical scheme.
The reference time step suitable in these conditions is $ \Delta t^{\text{ref}} = 10^{-14} s $.
Then, $ 10^{4} $ time steps are needed to to obtain results at time $ t = 2.5 \times 10^{-9} s $. These reference results are used to check the accuracy of both the classical and reformulated scheme.

\begin{figure}[hbtp]
 \begin{minipage}[c]{.46\linewidth}
 \includegraphics[scale=0.4]{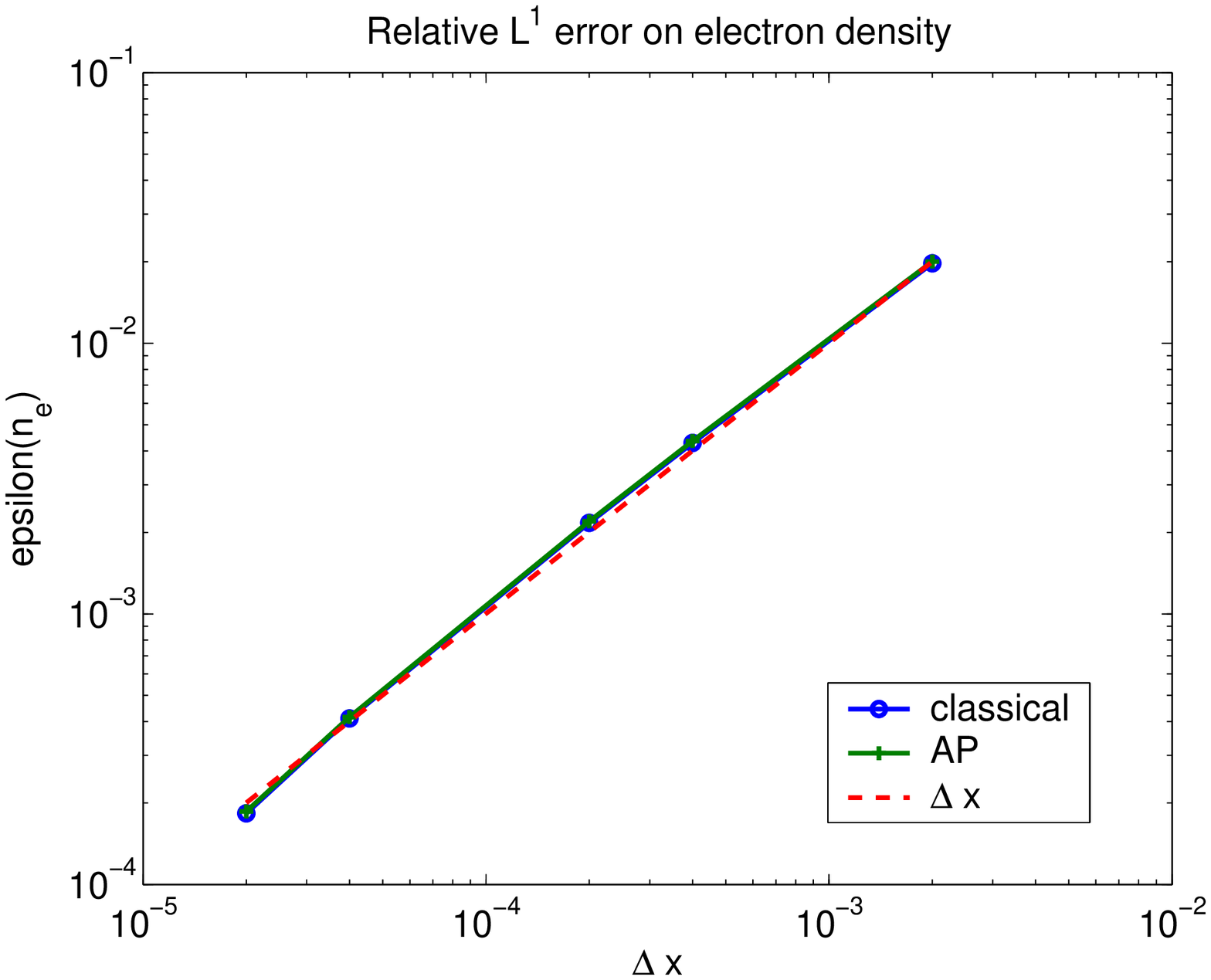}
 \end{minipage}
 \begin{minipage}[c]{.46\linewidth}
  \includegraphics[scale=0.4]{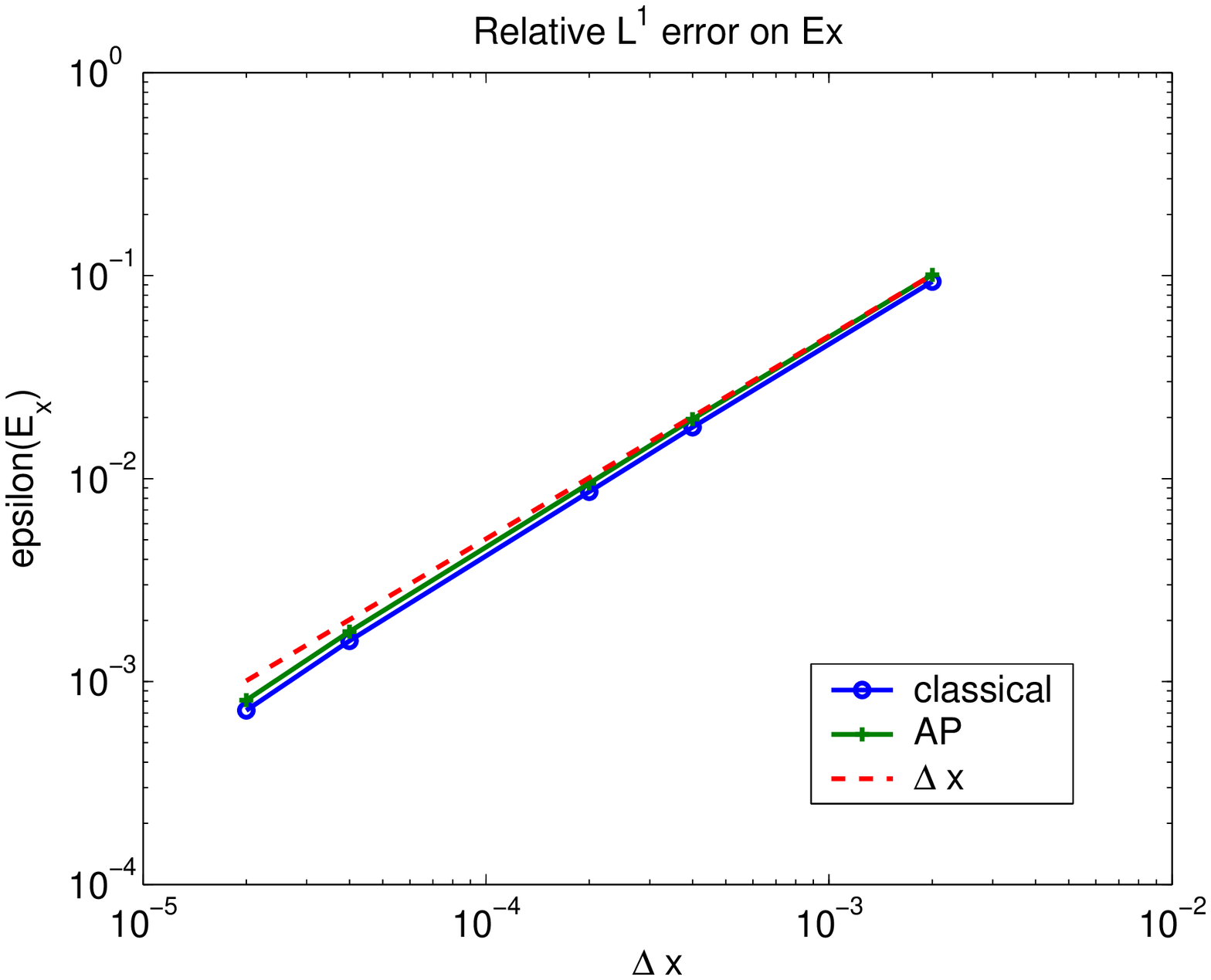}
 \end{minipage}
 \caption{\label{mono_ld_err} Low density POS; one-fluid model. $L^1$ relative error on $n$ (left panel) and $ E_x $ (right panel) as function of $\Delta s$ at time $ t = 2.5  \text{ns} $.
}
\end{figure}

First a convergence test is realized by comparison with the reference solution. The 
numerical errors for $n$ and $ E_x $ are recorded for the classical and AP- schemes with meshes consisting of $ 100 $, $ 500 $, $ 1000 $, $ 5000 $ and $ 10000 $ cells.
Figure \ref{mono_ld_err} compares the relative $L^1$ error as a function of $\Delta x$ on $n$ and $E_x$ between the two schemes. Both show exactly the same error. Moreover, the slope of the error confirms that in the case of smooth solution both numerical schemes are first order in space. Indeed, both curves are very close to the theoretical error plot (dashed line with a $ \Delta x $ slope). In this context the classical scheme ensures stable computations even if the space step is much larger than the Debye length. Its time-step however must is bounded by the CFL condition for the Maxwell equations. The level of time-implicitness in the AP-scheme ensures stability regardless of the time-step as long as it satisfies the CFL condition {\bf of the hydrodynamic equations}. Since both the fluid and acoustic velocities are much smaller than the speed of light, this provides an enormous gain in the allowed value of the time-step. 

In the following simulations, the classical scheme is used with two parameter choices.  The first choice allows for the computation of the reference solution, as explained above. The second choice is space under-resolved but time-resolved. It uses a larger mesh size than the Debye length, namely $ \Delta x = 10 \lambda$ but a time step which resolves the CFL condition of the Maxwell equations, the fastest time-scale in these conditions as mentioned above. We will refer to this situation as 'under-resolved classical scheme'. The AP-scheme will be run in a both time and space under-resolved situation. The mesh size will be the same as for the under-resolved classical scheme but the time-step will be hundred times the time-step of the  under-resolved classical scheme. 

\begin{figure}[hbtp]
 \begin{minipage}[c]{.46\linewidth}
 \includegraphics[scale=0.4]{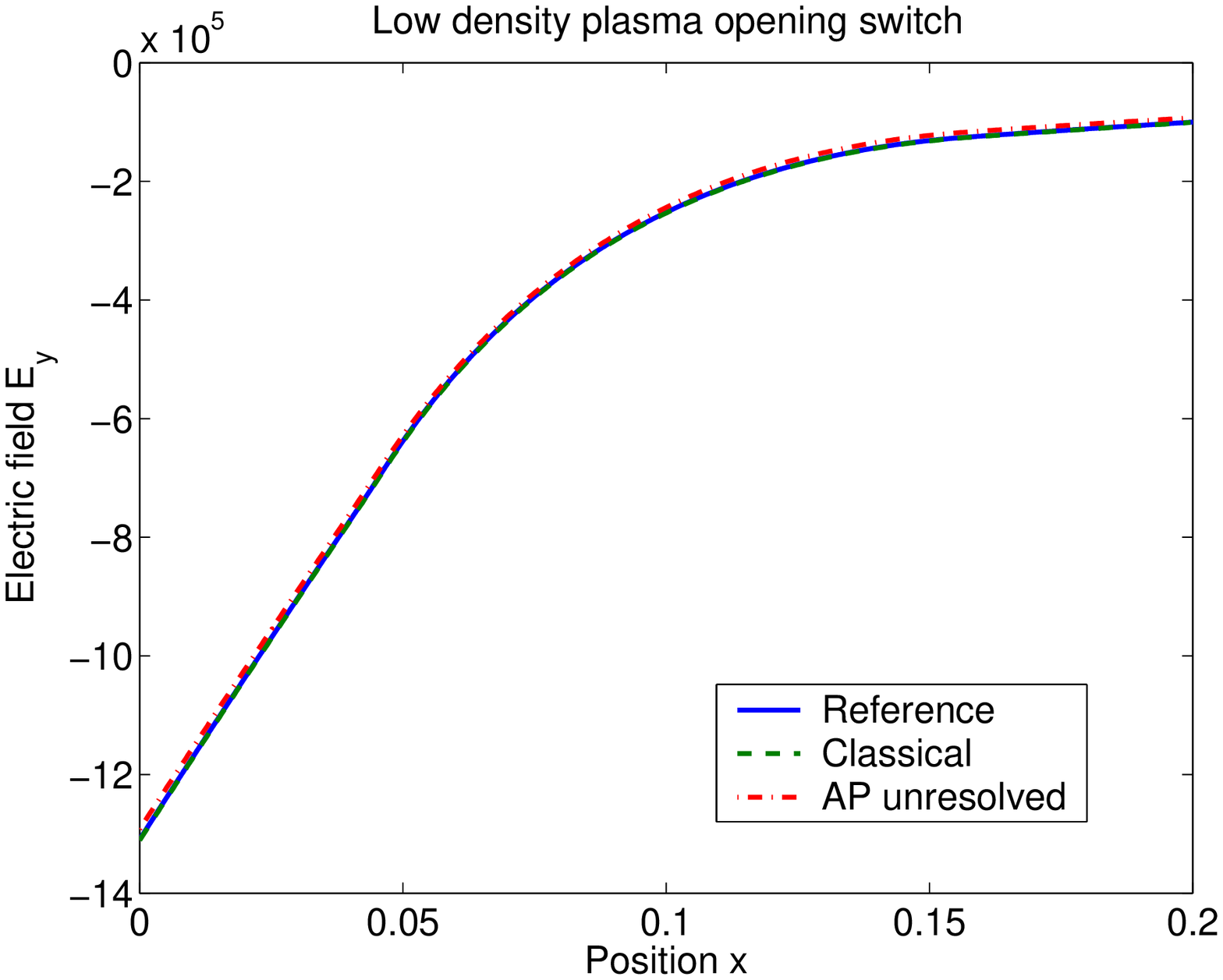}
 \end{minipage}
 \begin{minipage}[c]{.46\linewidth}
  \includegraphics[scale=0.4]{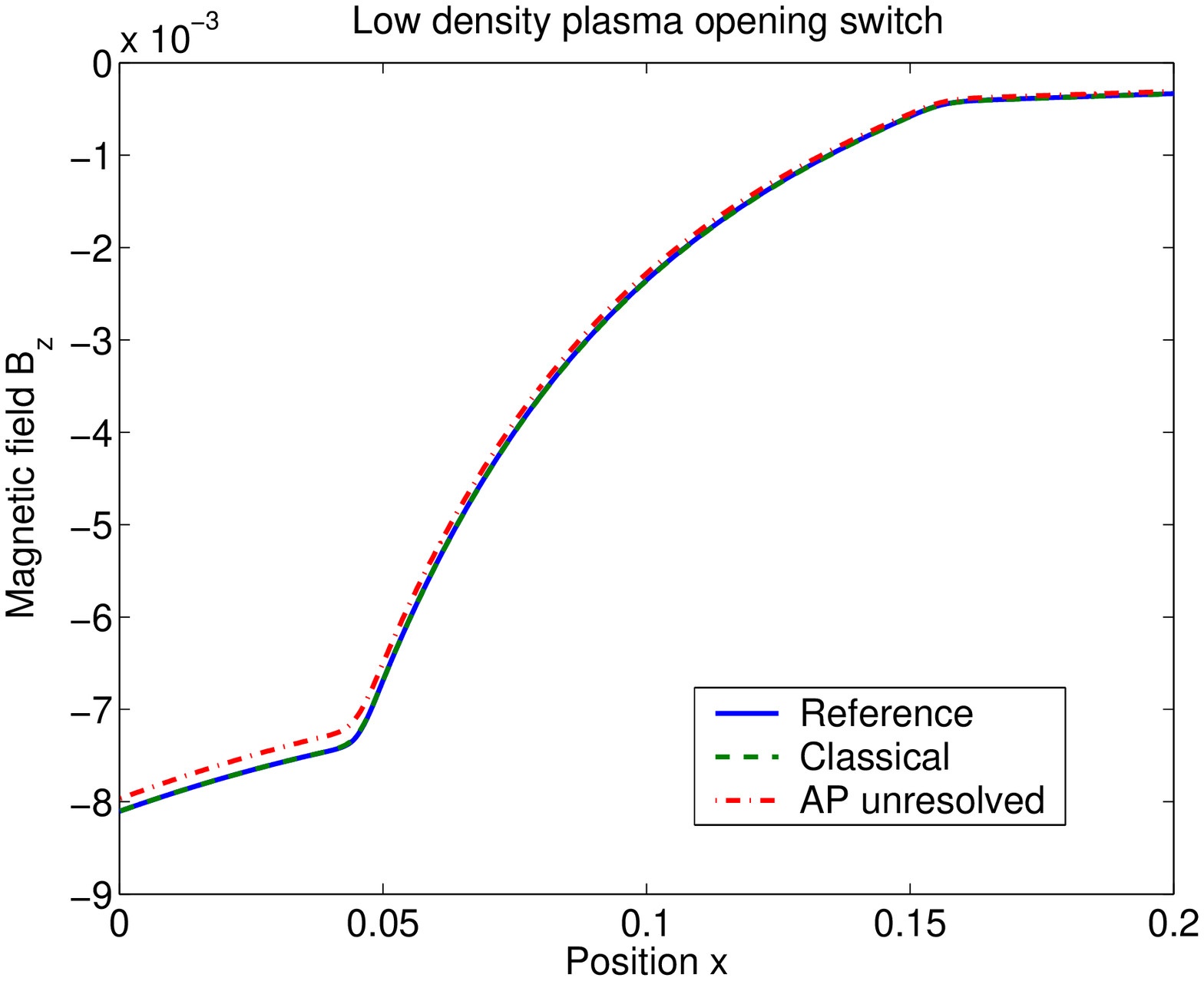}
 \end{minipage}
 \caption{\label{mono_ld_EyBz} Low density POS; one-fluid model. $ E_y $ (left panel) and  $ B_z $ (right panel) as functions of $x$ at time $ t = 2.5  \text{ns} $ with the reference, under-resolved classical and under-resolved AP- schemes.
}
\end{figure}

\begin{figure}[hbtp]
 \begin{minipage}[c]{.46\linewidth}
 \includegraphics[scale=0.4]{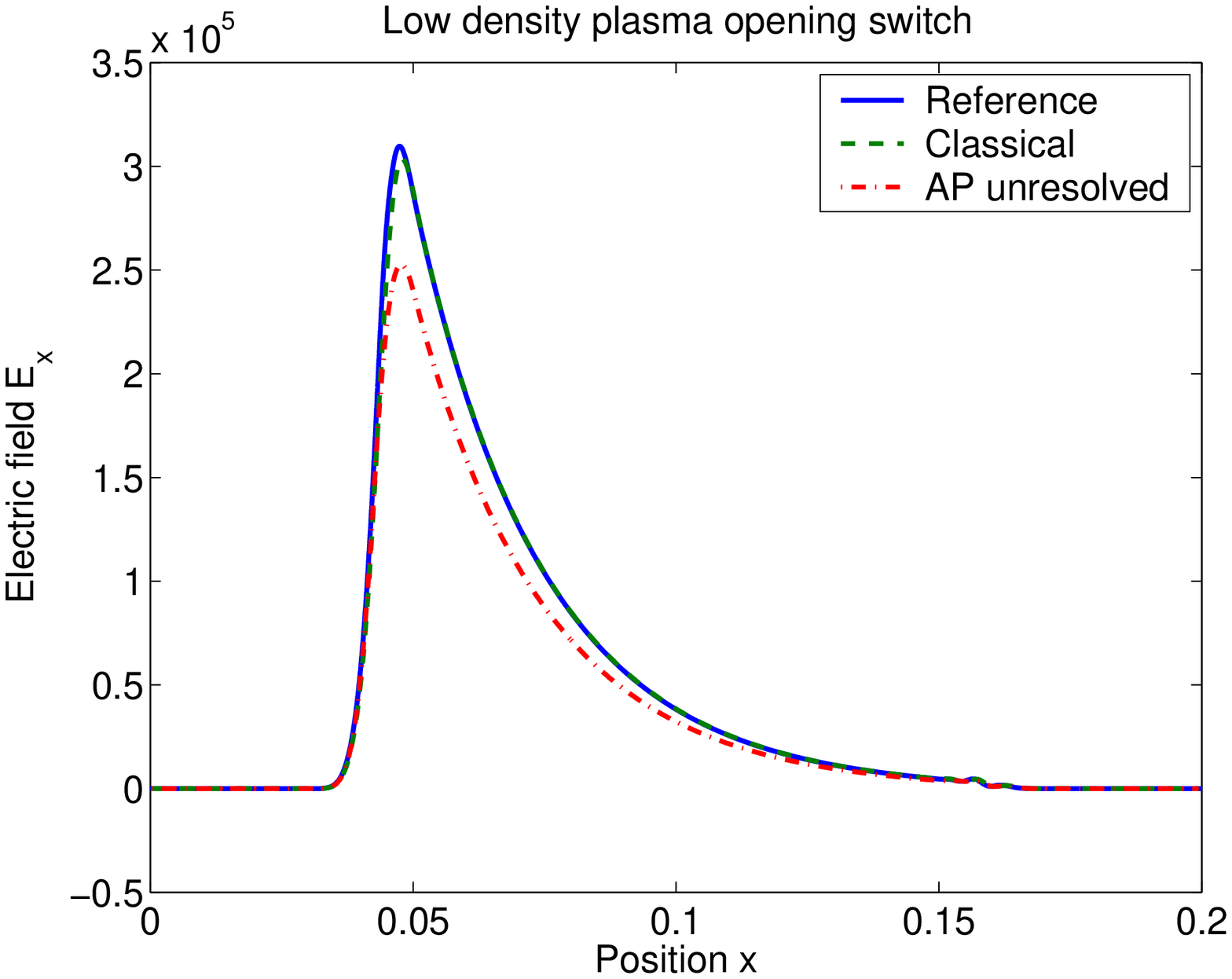}
 \end{minipage}
 \begin{minipage}[c]{.46\linewidth}
  \includegraphics[scale=0.4]{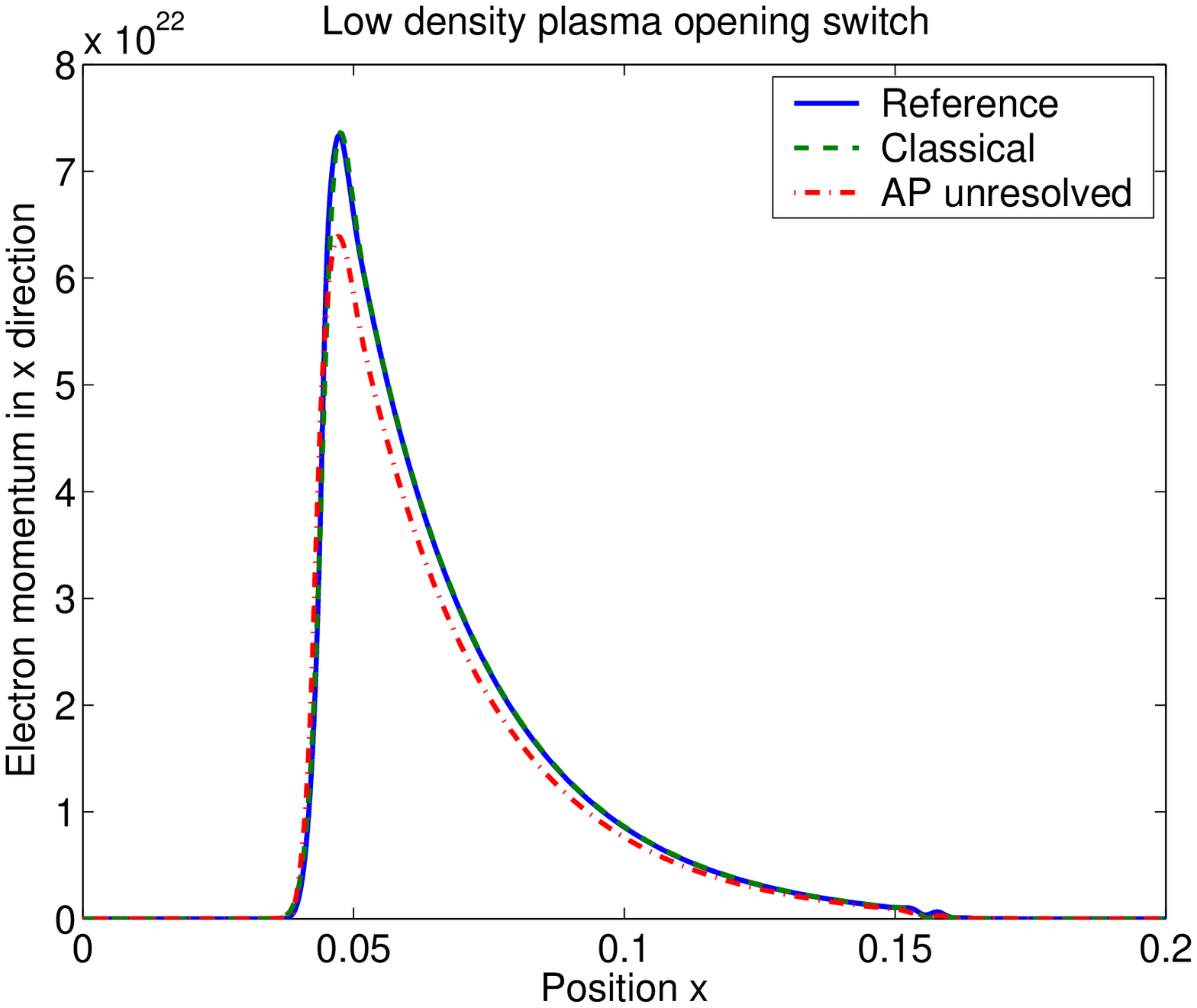}
 \end{minipage}
 \caption{\label{mono_ld_Exmomx} Low density POS; one-fluid model. $ E_x $ (left panel) and  $ n u_x $ (right panel) as functions of $x$ at time $ t = 2.5  \text{ns} $ with the reference, under-resolved classical and under-resolved AP- schemes.
}
\end{figure}

Fig. \ref{mono_ld_EyBz} displays $E_y$ (left panel) and $B_z$ (right panel) as functions of $x$ at time $t = 2.5$ ns, for both the reference, under-resolved classical and under-resolved AP- schemes. Fig. \ref{mono_ld_Exmomx} displays $E_x$ and $n u_x$ in a similar fashion. on Fig. Fig. \ref{mono_ld_EyBz},  we notice that the plasma prevents the transmission of the wave, as the values of $E_y$ and $B_z$ at the right end of the plasma are almost zero. The numerical diffusion induced by the larger time-steps used for the under-resolved AP-scheme is noticeable, but still acceptable given the large gain in computational efficiency: the computing time is reduced by a factor $ 100 $.

\subsubsection{Low density POS; two-fluid model}
\label{subsubsec_pos_two_low_dens}

\begin{figure}[hbtp]
 \begin{minipage}[c]{.46\linewidth}
 \includegraphics[scale=0.4]{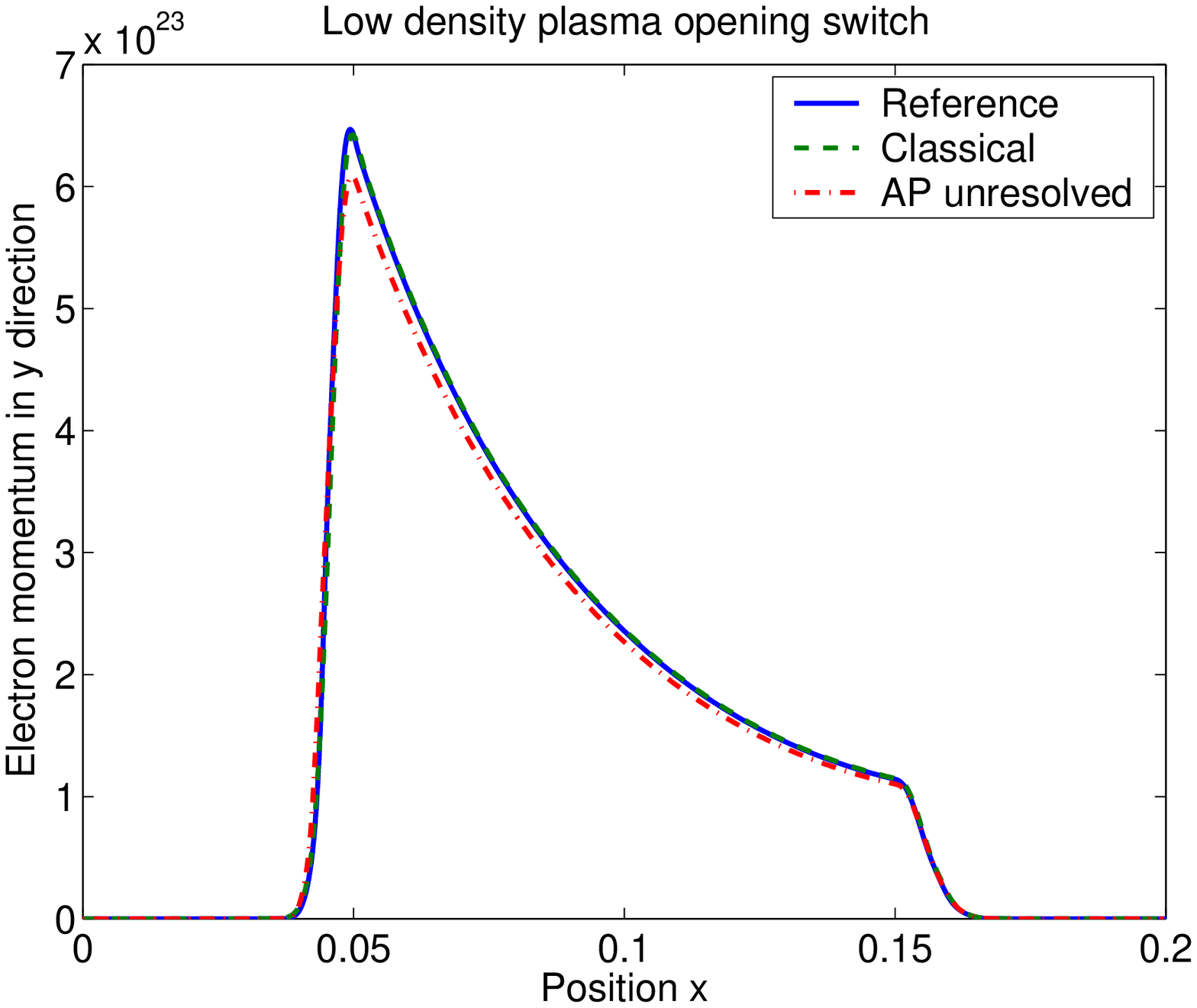}
 \end{minipage}
 \begin{minipage}[c]{.46\linewidth}
  \includegraphics[scale=0.4]{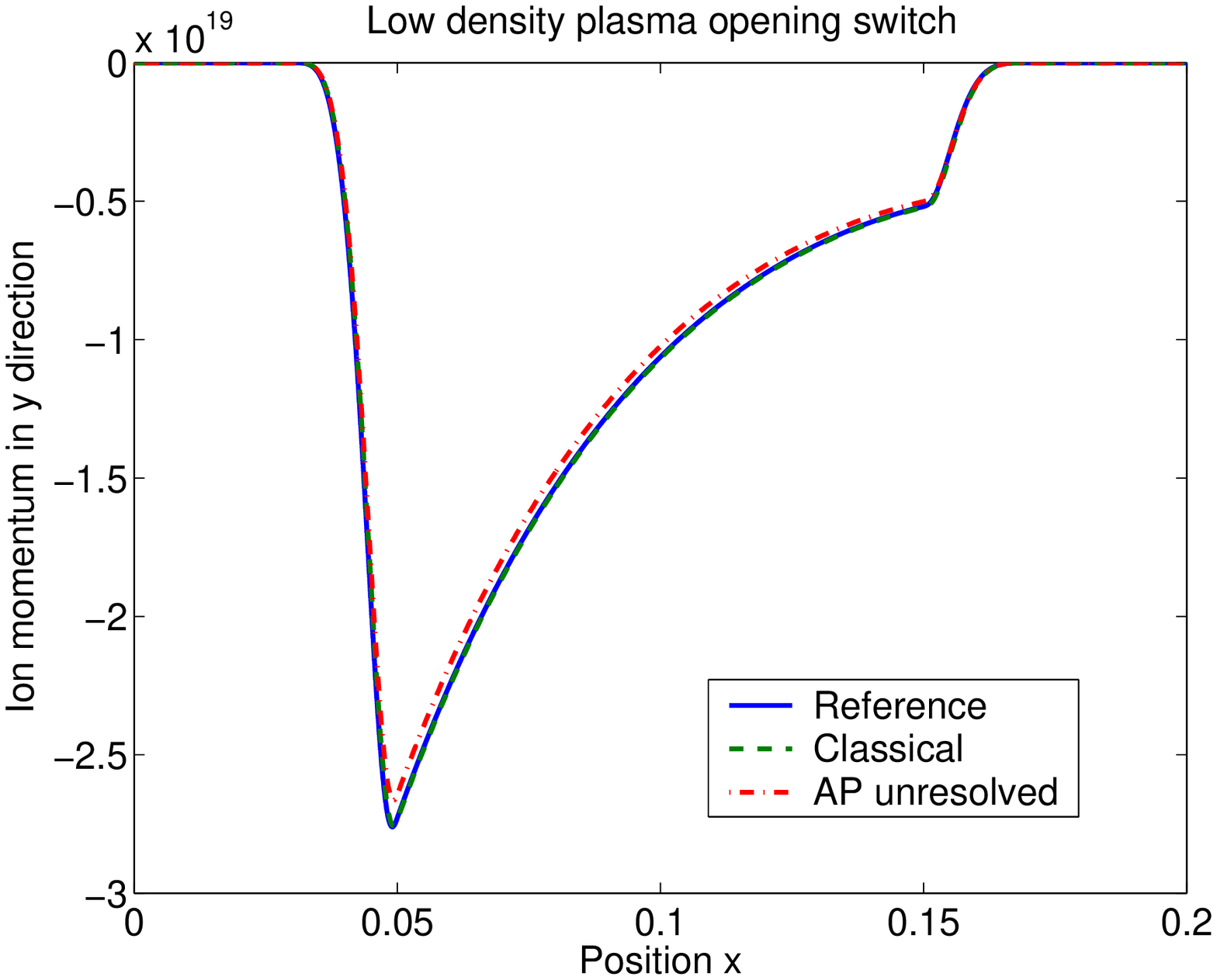}
 \end{minipage}
 \caption{\label{bif_ld_iemomy} Low density POS; two-fluid model. $n_e u_{ey}$ (left panel) and $n_i u_{iy}$ (right panel) as functions of $x$ at time $ t = 2.5  \text{ns} $ with the reference, under-resolved classical and under-resolved AP- schemes.
}
\end{figure}

Both the classical and reformulated scheme behave in a similar way in the case of two-fluid simulations, and the same conclusions hold for the numerical convergence study.
For instance, Fig.  \ref{bif_ld_iemomy} displays $n_e u_{ey}$ (left panel) and $n_i u_{iy}$ (right panel) in the same conditions as discussed for the one-fluid model. We can see that the electromagnetic wave sets electrons and ions into motion in the $y$ direction in opposite directions. With a two or three dimensional model where the extension in the $y$ direction would be bounded by the transmission line electrodes, this would induce a segregation of the electrons and ions on the different sides of the transmission line. This phenomenon induces the aperture of the POS. In the one-dimensional situation, the densities are supposed uniform in the $y$ direction and this phenomenon cannot be seen. 

\subsubsection{High density POS; one-fluid model}
\label{subsubsec_pos_high_dens}

In the case of the high density POS, the Debye length and electron plasma period in the plasma are one order of magnitude smaller. In this situation the wave cannot penetrate the plasma as fast as in the low density test case. 
The scaled Debye length is now of the order of $ 10^{-4} $. Then, a grid with a space step such that $ 10 \Delta x \leq \lambda $ (which was the ratio used for the convergence study in the low density test-case) is made of $ 10^{5} $ cells.
The computational cost induced by such a fine mesh is prohibitive. For this reason, we cannot present any convergence study in this test-case.
However the fine grid used previously for reference is such that $ \Delta x \leq \lambda$ and still can be used to generate a reference solution, to which the solution of the under-resolved classical and under-resolved AP- schemes will be compared.

\begin{figure}[hbtp]
 \begin{minipage}[c]{.46\linewidth}
 \includegraphics[scale=0.4]{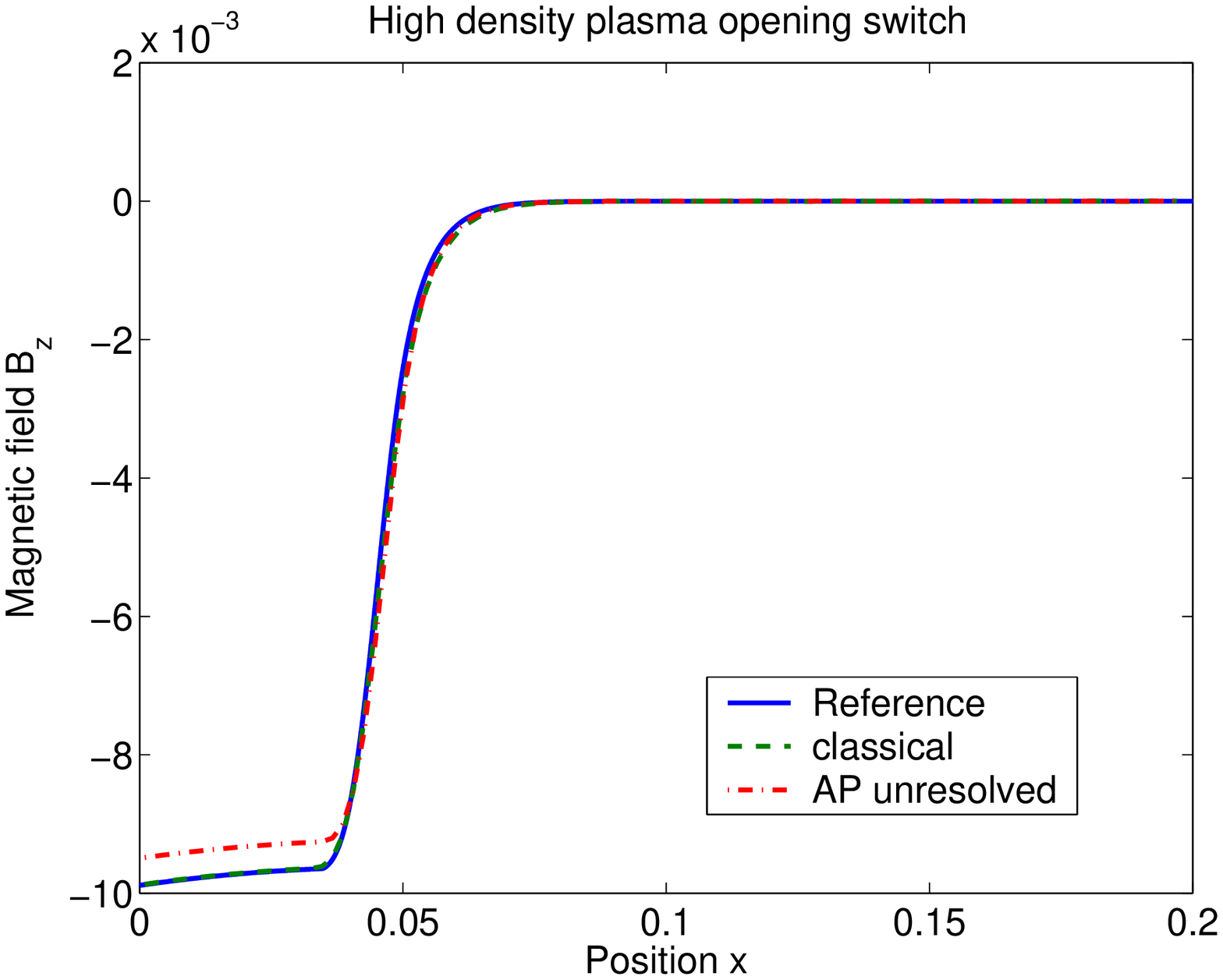}
 \end{minipage}
 \begin{minipage}[c]{.46\linewidth}
  \includegraphics[scale=0.4]{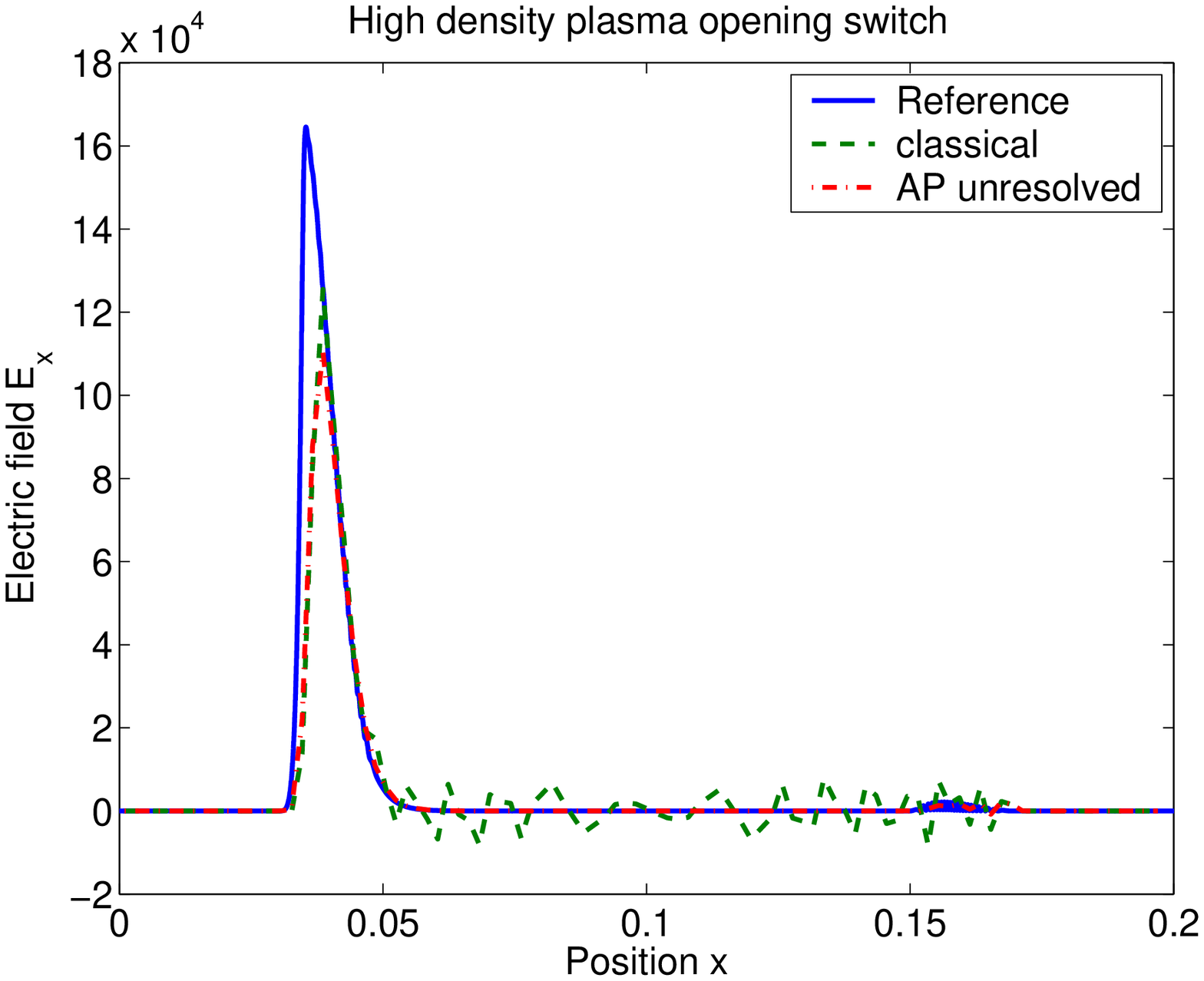}
 \end{minipage}
 \caption{\label{mono_hd_BzEx} Large density POS; one-fluid model. $B_z$ (left panel) and $E_x$ (right panel) as functions of $x$ at time $ t = 2.5  \text{ns} $ with the reference, under-resolved classical and under-resolved AP- schemes.
}
\end{figure}

Fig. \ref{mono_hd_BzEx} displays $B_z$ (left panel) and $E_x$ (right panel) as functions of $x$ at time $t = 2.5 \,$ ns, for the reference, under-resolved classical and under-resolved AP- schemes. This figure shows the plasma acting like a barrier on the magnetic field . In such a high density case, plasma waves appear at the right end of the plasma region, where an electron beam leaks outside the plasma.
The typical wave-length of these plasma waves is $O(\lambda)$, i.e. $ 10^{-4} $. Therefore, the fine grid with mesh size $ \Delta x \sim \lambda $ can resolve this scale and the reference solution is thus able to describe these waves in a satisfactory way.
By contrast, the coarse grid does not resolve these waves. Therefore, the under-resolved classical scheme is subject to instabilities generated by the impossibility of correctly describing these waves.
The under-resolved AP-scheme does not attempt to resolve these waves, but provides the correct average of the oscillation and does not suffer from any instability. We notice the slightly larger numerical diffusion of the under-resolved AP scheme, which is the counterpart of the increased time-step. Still, the use of a coarse mesh size combined with large time-steps allows for a large reduction of the computational cost : the CPU times needed to compute the reference, under-resolved classical and under-resolved AP- schemes results are respectively $ 2 \times 10^{5} \, $s , $ 20 \, $s and $\leq 1 \,$s.

\setcounter{equation}{0}
\section{Conclusion}
\label{EM:sec_conclu}

In this paper, we proposed and analyzed an Asymptotic-Preserving scheme for the Euler-Maxwell system in the quasi-neutral limit. The scheme is exposed in detail for a one-fluid plasma model where the ions are immobile and form a fixed neutralizing background. It is then extended to a two-fluid model where both ions and electrons are mobile. The analysis involves a proof of its 'Asymptotic-Preserving' character and that its linear stability condition is independent of the scaled Debye parameter when the latter tends to zero. The numerical simulations involve comparisons between the AP-scheme to a  'classical' scheme in the one- and two-fluid configurations, for two different one-dimensional test-cases: the Riemann problem and the Plasma Opening Switch device. The numerical convergence study shows that both the classical and AP-scheme are convergent to the Euler-Maxwell solution with resolved time and space discretizations. On the other hand, with under-resolved time and space discretizations, the AP scheme is consistent with the quasi-neutral Euler-Maxwell system. Additionally, the proposed spatial discretization allows for a perfect consistency with the Gauss equation. By contrast, in under-resolved situations, the classical scheme leads to spurious large amplitude oscillations and instabilities. The possibility of using large time and space discretization parameters with the AP-scheme leads to several orders of magnitude reductions in computer time and storage. Future work will pursue the validation of the methodology to multi-dimensional cases and extend it to plasma kinetic models such as the Vlasov or Fokker-Planck-Landau equations.

\end{document}